\documentclass[10pt,journal,compsoc]{IEEEtran}
\ifCLASSINFOpdf
\else
\fi
\usepackage{amsmath}
\usepackage{graphicx}
\usepackage{subfigure}
\usepackage{booktabs}
\usepackage{colortbl}
\usepackage{xcolor}
\usepackage{multirow}
\usepackage{enumitem}
\usepackage{tikz}
\usetikzlibrary{graphs, positioning,  shapes.geometric}

\usepackage{algorithm}
\usepackage{booktabs} 
\usepackage{color}
\usepackage{algpseudocode}

\usepackage{hyperref}
\usepackage{epstopdf}
\usepackage{amsmath,amssymb,mathrsfs}
\usepackage{multirow}
\newcommand{\EE}{\mathbb{E}}

\newcommand{\RR}{\mathbb{R}}

\newtheorem{lemma}{Lemma}

\newtheorem{theorem}{Theorem}

\newtheorem{definition}{Definition}
\newtheorem{assumption}{Assumption}
\newtheorem{proposition}{Proposition}
\begin{document}
%
\title{Decentralized Federated Averaging}
%
%
%

\author{Tao Sun,
        Dongsheng Li, and Bao Wang
\thanks{This work is sponsored in part by the National Key R\&D Program of China under Grant (2018YFB0204300)  and the National Natural Science Foundation of China under Grants (61932001 and 61906200).

Tao Sun and Dongsheng Li are with the
College of Computer, National University of Defense Technology,
Changsha, 410073, Hunan,  China. (e-mails: \texttt{nudtsuntao@163.com},  \texttt{dsli@nudt.edu.cn})

Bao Wang is with the Scientific Computing \& Imaging   Institute, University of Utah, USA.  (e-mail: \texttt{wangbaonj@gmail.com})

Dongsheng Li and Bao Wang the co-corresponding authors.}
}

%
%

\markboth{Journal of \LaTeX\ Class Files,~Vol.~14, No.~8, August~2015}%
{Shell \MakeLowercase{\textit{et al.}}: Bare Demo of IEEEtran.cls for IEEE Journals}
%



\maketitle

\begin{abstract}
Federated averaging (FedAvg) is a communication efficient algorithm for the distributed training with an enormous number of clients. In FedAvg, clients keep their data locally for privacy protection; a central parameter server is used to communicate between clients. This central server distributes the parameters to each client and collects the updated parameters from clients. FedAvg is mostly studied in centralized fashions, which requires massive communication between server and clients in each communication. Moreover, attacking the central server can break the whole system's privacy. In this paper, we study the decentralized FedAvg with momentum (DFedAvgM), which is implemented on clients that are connected by an undirected graph. In DFedAvgM, all clients perform stochastic gradient descent with momentum and communicate with their neighbors only. To further reduce the communication cost, we also consider the quantized DFedAvgM. We prove convergence of the (quantized) DFedAvgM under trivial assumptions; the convergence rate can be improved when the loss function satisfies the P{\L} property. Finally, we numerically verify the efficacy of DFedAvgM.
\end{abstract}


\vspace{-3mm}
\section{Introduction}
\label{sec:intro}
Federated learning (FL) is a privacy-preserving distributed machine learning (ML) paradigm \cite{mcmahan2017communication-efficient}. In FL, a central server connects with enormous clients (e.g., mobile phones, pad, etc.); the clients keep their data without sharing it with the server. In each communication round, clients receive the current global model from the server, and a small portion of clients are selected to update the global model by running stochastic gradient descent (SGD) \cite{robbins1951a} for multiple iterations using local data. The central server then aggregates these updated parameters
to obtain the updated global model. The above learning algorithm is known as federated average (FedAvg) \cite{mcmahan2017communication-efficient}. In particular, if the clients are homogeneous, FedAvg is equivalent to the local SGD \cite{zinkevich2010parallelized}. FedAvg involves multiple local SGD updates and one aggregation by the server in each communication round, which significantly reduces the communication cost between sever and clients compared to the conventional distributed training with one local SGD update and one communication.


In FL applications, large companies and government organizations usually play the role of the central server. On the one hand, since the number of clients in FL is massive, the communication cost between the server and clients can be a bottleneck
\cite{lian2017can}.
On the other hand, the updated models collected from clients encode the private information of the local data; hackers can attack the central server to break the privacy of the whole system, which remains the privacy issue as a serious concern. To this end, decentralized federated learning has been proposed \cite{lalitha2018fully,lalitha2019peer}, where all clients are connected with an undirected graph. Decentralized FL replaces the
server-clients communication in FL with clients-clients communication.

In this paper, we consider two issues about decentralized FL: 1) Although there is no expensive communication between server and clients in decentralized FL, the communication between local clients is costly when the ML model itself is large. Therefore, it is crucial to ask \emph{can we reduce the communication cost between clients?} 2) Momentum is a well-known acceleration technique for SGD \cite{sutskever2013importance}. It is natural to ask \emph{can we use SGD with momentum to improve the training of ML models in decentralized FL with theoretical convergence guarantees?}

\subsection{Our Contributions}
We answer the above questions affirmatively by proposing the decentralized FedAvg with momentum (DFedAvgM). To further reduce the communication cost between clients, we also integrate quantization with DFedAvgM. Our contributions in this paper are elaborated below in threefold.

\begin{itemize}[leftmargin=*]
\item Algorithmically, we extend FedAvg to the decentralized setting, where all clients are connected by an undirected graph. We motivate DFedAvgM from the decentralized SGD (DSGD) algorithm.
In particular, we use SGD with momentum to train ML models on each client. To reduce the communication cost between each client, we further introduce a quantized version of DFedAvgM, in which each client will send and receive a quantized model.\\

\item Theoretically, we prove the convergence of (quantized) DFedAvgM. Our theoretical results show that the convergence rate of (quantized) DFedAvgM is not inferior to that of SGD or DSGD. More specifically, we show that the convergence rates of both DFedAvgM and quantized DFedAvgM depend on the local training and the graph that connects all clients. Besides the convergence results under nonconvex assumptions, we also establish their convergence guarantee under the Polyak-{\L}ojasiewicz (P{\L}) condition, which has been widely studied in nonconvex optimization. Under the P{\L} condition, we establish a faster convergence rate for (quantized) DFedAvgM. Furthermore, we present a sufficient condition to guarantee reducing communication costs. \\

\item Empirically, we perform extensive numerical experiments on training deep neural networks (DNNs) on various datasets in both {\bf IID} and {\bf Non-IID} settings. Our results show the effectiveness of (quantized) DFedAvgM for training ML models, saving communication costs, and protecting training data's membership privacy.
\end{itemize}

\subsection{More Related Works}
We briefly review three lines of work that are most related to this paper, i.e., federated learning, decentralized training, and decentralized federated learning.

\textbf{Federated Learning}. Many variants of FedAvg have been developed with theoretical guarantees. \cite{hsu2019measuring} uses the momentum method for local clients in FedAvg.  \cite{reddi2020adaptive} proposes the adaptive FedAvg, whose central parameter server uses the adaptive learning rate ti aggregate local models. Lazy and quantizatized gradients are used to reduce communications \cite{chen2018lag,sun2019communication}.
\cite{li2019feddane} proposes a Newton-type scheme for FL.
The convergence analysis of FedAvg on heterogeneous data is discussed by \cite{khaled2019first,li2019convergence}.
The advances and open problems in FL is available in two survey papers \cite{2019Advances,li2019federated}.

\textbf{Decentralized Training}.
Decentralized algorithms are originally developed to calculate the mean of data that are distributed over multiple sensors \cite{boyd2005gossip,olfati2007consensus,schenato2007distributed,aysal2009broadcast}. Decentralized (sub)gradient descents (DGD), one of the simplest and efficient decentralized algorithms, have been studied in \cite{nedic2009distributed,chen2012fast,jakovetic2014fast,matei2011performance,yuan2016convergence}. In DGD, the convexity assumption is unnecessary \cite{zeng2018nonconvex}, which makes DGD useful for nonconvex optimization. A provably convergent DSGD is proposed
in \cite{sirb2016consensus,lan2017communication,lian2017can}. \cite{sirb2016consensus} provides the complexity result of a stochastic decentralized algorithm. \cite{lan2017communication} designs a stochastic decentralized algorithm with the dual information and provide the theoretical convergence guarantee. \cite{lian2017can} proves that DSGD outperforms SGD in communication efficiency.
Asynchronous DSGD is analyzed in \cite{lian2018a}.
DGD with momentum is proposed in \cite{sun2019non,xin2019distributed}. Quantized DSGD has been proposed in \cite{reisizadeh2018quantized}.

\textbf{Decentralized Federated Learning}. Decentralized FL is a learning paradigm  of choice when the edge devices do not trust the central server in protecting their privacy \cite{yang2019federated}. The authors in \cite{xing2020decentralized} propose a novel FL
framework without a central server for medical applications, and the new method offers a highly dynamic peer-to-peer environment. \cite{lalitha2019peer}   considers training an ML model with a connected network  whose nodes take a Bayesian-like approach by introducing a belief of the parameter space.

\subsection{Organizations}
We organize this paper as follows: in section~\ref{sec:problem}, we present a mathematical formulation of our problem and some necessary assumption. In section~\ref{sec:dfedavg}, we present the DFedAvgM and its quantized algorithms. We present the convergence of the proposed algorithm in section~\ref{sec:theory}. We provide extensive numerical verification of DFedAvgM in section~\ref{sec:numerics}. This paper ends up with concluding remarks. Technical proofs and more experimental details are provided in the appendix.

\subsection{Notation}
We denote scalars and vectors by lower case and lower case boldface letters, respectively, and matrices by upper case boldface letters. For a vector ${\bf x} = (x_1, \cdots, x_d)\in \mathbb{R}^d$, we denote its $\ell_p$ norm ($p\geq 1$) by $\|{\bf x}\|_p = {\small (\sum_{i=1}^d |x_i|^p)^{1/p}}$, and denote the $\ell_\infty$ norm of ${\bf x}$ by $\|{\bf x}\|_\infty = \max_{i=1}^d|x_i|$ and denote $\ell_2$ norm as $\|{\bf x}\|$.
For a matrix ${\bf A}$, we denote its transpose as ${\bf A}^\top$.
Given two sequences $\{a_n\}$ and $\{b_n\}$, we write $a_n=\mathcal{O}(b_n)$ if there exists  a positive constant $0<C<+\infty$ such that
$a_n \leq C b_n$, and we write $a_n=\Theta(b_n)$ if there exist two positive constants $C_1$ and $C_2$ such that $a_n \leq C_1 b_n$ and $b_n \leq C_2 a_n$. $\widetilde{\mathcal{O}}(a_n)$ hides the logarithmic factor of $a_n$. 
For a function $f({\bf x}): \mathbb{R}^d \rightarrow \mathbb{R}$, we denote its gradient as $\nabla f({\bf x})$ and its Hessian as $\nabla^2 f({\bf x})$, and denote its minimum as $\min f$. We use $\EE[\cdot]$ to denote the expectation with respect to the underlying probability space.

\section{Problem Formulation and Assumptions}\label{sec:problem}
We consider the following optimization task
\begin{align}\label{dec}
\hskip -0.3cm{\small \min_{{\bf x}\in \mathbb{R}^d} f({\bf x}):=\frac{1}{m}\sum_{i=1}^m f_i({\bf x}),~~f_i({\bf x})=\EE_{\xi\sim \mathcal{D}_i} F_i({\bf x};\xi),}
\end{align}\vspace{-1mm}
where $\mathcal{D}_i$ denotes the data distribution in the $i$-th client and $F_i({\bf x};\xi)$ is the loss function associated with the training data $\xi$. Problem \eqref{dec} models many applications in ML, which is known as empirical risk minimization (ERM).
We list several assumptions for the subsequent analysis.

\begin{assumption}\label{ass1}
The function $f_i$ is differentiable and $\nabla f_i$ is $L$-Lipschitz continuous, $\forall i \in \{1,2,\ldots,m\}$, i.e.,
$\|\nabla f_i({\bf x}) - \nabla f_i({\bf y})\| \leq L \|{\bf x} - {\bf y}\|,$
for all ${\bf x}, {\bf y} \in \mathbb{R}^d$.
\end{assumption}
The first-order Lipschitz assumption is commonly used in the ML community. Here, for simplicity, we suppose all functions enjoy the same Lipschitz constant $L$.
We can also assume that these functions have non-uniform Lipschitz constants, which does not affect our convergent analysis.

\begin{assumption}\label{ass2}
The gradient of the function $f_i$ have $\sigma_l$-bounded variance, i.e., $\mathbb{E}[\|\nabla F_i({\bf x};\xi) - \nabla f_i({\bf x})\|^2] \leq \sigma_{l}^2$ for all ${\bf x} \in \mathbb{R}^d$ $\forall i \in \{1,2,\ldots,m\}$. This paper also assumes  the (global) variance is bounded, i.e.,
$
\frac{1}{m} \sum_{i=1}^m \|\nabla f_i({\bf x}) - \nabla f({\bf x})\|^2 \leq \sigma_{g}^2
$
for all ${\bf x} \in \mathbb{R}^d$.
\end{assumption}

The uniform local variance assumption is also used for the ease of presentation, which is straightforward to generalize to non-uniform cases. The global variance assumption is used in \cite{reddi2020adaptive,li2018federated}. The constant $\sigma_{g}$ reflects the heterogeneity of the data sets $(\mathcal{D}_i)_{1\leq i\leq m}$, and
when $(\mathcal{D}_i)_{1\leq i\leq m}$ follow the same distribution,  $\sigma_{g}=0$.

\begin{assumption}\cite{ghadimi2013stochastic,lian2017can}\label{ass3}
For any $i \in \{1,2,\ldots,m\}$ and ${\bf x} \in \mathbb{R}^d$, we have $\|\nabla f_i({\bf x})\| \leq B $ for some $B>0$.
\end{assumption}

An important notion in decentralized optimization is the \textit{mixing matrix}, which is usually associated with a connected graph $\mathcal{G} = (\mathcal{V}, \mathcal{E})$ with the vertex set $\mathcal{V}=\{1,...,m\}$ and the edge set $\mathcal{E}\subseteq \mathcal{V}\times \mathcal{V}$.
Any edge $(i, l)\in\mathcal{E}$ represents a communication link between nodes  $i$
and $l$. We recall the  definition of the mixing matrix associated with the graph $\mathcal{G}$.

\begin{definition}[Mixing matrix]
\label{def:MixMat}
The mixing matrix ${\bf W} = [w_{i,j}] \in \mathbb{R}^{m\times m}$ is assumed to have the following properties:
1. (Graph) If $i\neq j$ and $(i,j) \notin {\cal E}$, then $w_{i,j} =0$, otherwise, $w_{i,j} >0$;
2. (Symmetry) ${\bf W} = {\bf W}^{\top}$;
3. (Null space property) $\mathrm{null} \{{\bf I}-{\bf W}\} = \mathrm{span}\{\bf 1\}$;
4. (Spectral property) ${\bf I} \succeq {\bf W} \succ -{\bf I}.$
\end{definition}

For a graph, the corresponding mixing matrix is not unique; given the adjacency matrix of a graph, its maximum-degree matrix and metropolis-hastings \cite{boyd2004fastest} are both mixing matrices. The symmetric property of ${\small {\bf W}}$ indicates that its eigenvalues are real and can be sorted in the non-increasing order. Let ${\small \lambda_i({\bf W})}$ denote the $i$-th largest eigenvalue of ${\small {\bf W}}$, that is, ${\small \lambda_1({\bf W})=1>\lambda_2({\bf W}) \geq \cdots \geq \lambda_m({\bf W})>-1.}$\footnote{This is based on the spectral property of mixing matrix.} The mixing matrix also serves as a probability transition matrix of a Markov chain.
A quite important constant of ${\small {\bf W}}$ is ${\small \lambda=\lambda({\bf W}):=\max\{|\lambda_2({\bf W})|,|\lambda_m({\bf W})|\}}$, which describes the speed of the Markov chain introduced by the mixing matrix converges to the stable state.

\section{Decentralized Federated Averaging}\label{sec:dfedavg}

\subsection{Decentralized FedAvg with Momentum}
We first briefly review the previous work on decentralized training, which carries out in the following fashion:
\begin{enumerate}[leftmargin=*]
\item client $i$ holds an approximate copy of the parameters ${\bf x}(i)\in \RR^d$ and calculate an unbiased estimate of $\nabla f_i:={\bf g}(i)$ at ${\bf x}(i)$.
$({\bf x}(i))_{1\leq i\leq m}$ can be non-consensus;

\item (\textit{communication}) client $i$ updates its
local
parameters ${\bf x}(i)$ as the weighted average of its neighbors:
$\tilde{{\bf x}}(i) = \sum_{l\in \mathcal{N}(i)} w_{i,l} {\bf x}(l)$;

\item (\textit{training}) client $i$ updates its parameters as ${\bf x}(i)\leftarrow \tilde{{\bf x}}(i)-\eta {\bf g}(i)$ with a learning rate $\eta>0$.
\end{enumerate}

\begin{figure*}
\centering
\subfigure[Traditional Decentralization (DSGD)]{\begin{tikzpicture}[
roundnode/.style={circle, draw=green!60, fill=green!5, very thick, minimum size=3mm},
squarednode/.style={rectangle, draw=red!60, fill=red!5, very thick, minimum size=2mm},
]
\node[squarednode]      (main)                              {Comm.};
\node[roundnode]        (r1)       [right=of main] {Train};
\node[squarednode]      (r2)       [right=of r1] {Comm.};
\node[roundnode]        (r3)       [right=of r2] {Train};
 \node[]                (r4)       [right=of r3] {\ldots};
\draw[thick,->] (main.east) -- (r1.west);
\draw[thick,->] (r1.east) -- (r2.west);
\draw[thick,->] (r2.east) -- (r3.west);
\draw[thick,->] (r3.east) -- (r4.west);
\end{tikzpicture}}

\subfigure[DFedAvgM]{ \begin{tikzpicture}[
roundnode/.style={circle, draw=green!60, fill=green!5, very thick, minimum size=3mm},
squarednode/.style={rectangle, draw=red!60, fill=red!5, very thick, minimum size=2mm},
]
\node[squarednode]      (main)                              {Comm.};
\node[roundnode]        (r1)       [right=of main] {Train};
 \node[]                (r2)       [right=of r1] {\ldots};
\node[roundnode]        (r3)       [right=of r2] {Train};
\node[squarednode]      (r4)       [right=of r3] {Comm.};
\node[roundnode]        (r5)       [right=of r4] {Train};
 \node[]                (r6)       [right=of r5] {\ldots};
\node[roundnode]        (r7)       [right=of r6] {Train};
 \node[]                (r8)       [right=of r7] {\ldots};
\draw[thick,->] (main.east) -- (r1.west);
\draw[thick,->] (r1.east) -- (r2.west);
\draw[thick,->] (r2.east) -- (r3.west);
\draw[thick,->] (r3.east) -- (r4.west);
\draw[thick,->] (r4.east) -- (r5.west);
\draw[thick,->] (r5.east) -- (r6.west);
\draw[thick,->] (r6.east) -- (r7.west);
\draw[thick,->] (r7.east) -- (r8.west);
\end{tikzpicture} }\vspace{-3mm}
 \caption{Comparison of communication and training styles of traditional decentralized stochastic gradient descent (DSGD) and the proposed decentralized federated average with momentum (DFedAvgM).
 In DSGD,
 each client will communicate with its neighbors after one single training step. In DFedAvgM, however, each client will communicate with its neighbors after multiple training iterations.}
\end{figure*}
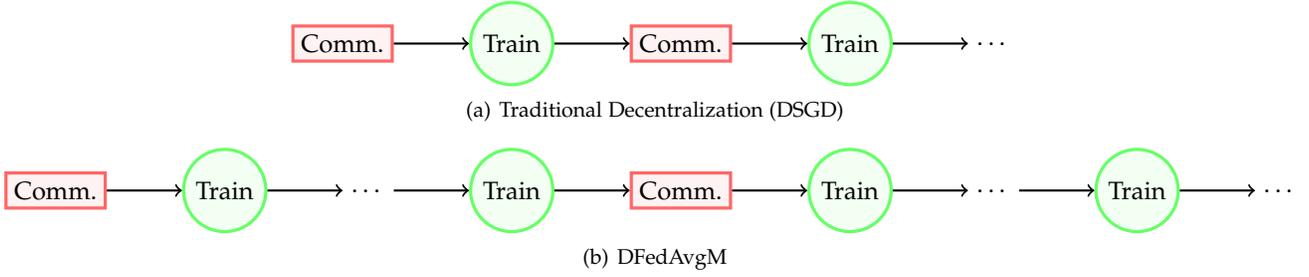

\vspace{-2mm}
\begin{algorithm}[H]
  \footnotesize
    \caption{DFedAvgM}\label{alg1}
	\begin{algorithmic}[1]
\State{\textbf{Parameters}: $\eta>0,  K\in \mathbb{Z}^+$, $0\leq \theta<1$.}
\State{\textbf{Initialization}: ${\bf x}^0=\textbf{0}$}
\For{$t=1,2,\ldots$}
\For{$i=\{1,2,\ldots,m\}$  }
\State{node $i $ performs local training \eqref{innersgd}    $K$   times and sends ~~~~~~~~~~~~~~~ ${\bf z}^t(i)={\bf y}^{t,K}(i)$ to   $\mathcal{N}(i)$ }
\State{node $i$ updates as \eqref{avera}}
\EndFor
\EndFor
	\end{algorithmic}
  \end{algorithm}\vspace{-2mm}



The traditional decentralization can be described in Figure 1 (a), in which, a {\it communication} step is needed after each \textit{training} iteration. This indicate that the above vanilla decentralized algorithm is different from FedAvg, and the later performs multiple local {\it training} step before {\it communication}.  To this end, we have to slightly modify the scheme of the decentralized algorithm. For simplicity, we consider modifying DSGD to motivate our decentralized FedAvg algorithm.
Note that when the original DGD is applied to solve \eqref{dec}, we end up with the following iteration
\vspace{-2mm}
\begin{equation}\label{dgdb}
    \begin{aligned}
{\bf x}^{t+1}(i)&=\sum_{l\in \mathcal{N}(i)} w_{i,l} {\bf x}^t(l)-\gamma {\bf g}^t(i)\\
&=\sum_{l\in \mathcal{N}(i)} w_{i,l} [ {\bf x}^t(l)-\gamma {\bf g}^t(i)],
\end{aligned}
\end{equation}
where we used the fact that $\sum_{l\in \mathcal{N}(i)} w_{i,l} =1$. In \eqref{dgdb}, if we replace ${\bf x}^t(l)$
by ${\bf x}^t(i)$,  the algorithm then iterates as
\vspace{-2mm}
\begin{align}\label{dgdb2}
{\bf x}^{t+1}(i)=\sum_{l\in \mathcal{N}(i)} w_{i,l} [ {\bf x}^t(i)-\gamma {\bf g}^t(i)].
\end{align}
In \eqref{dgdb2}, clients communicate with their neighbors after one training iteration, which is then possible to generalize to the federated optimization setting. We replace the single SGD iteration in \eqref{dgdb2} with multiple SGD with heavy-ball \cite{polyak1964some} iterations. Therefore, the DFedAvgM can be presented as follows: In each $t\in \mathbb{Z}^+$, for each client  $i\in\{1,2,\ldots,m\}$, let ${\bf y}^{t,-1}(i)={\bf y}^{t,0}(i)={\bf x}^t(i)$. The inner iteration in each node then performs as
\vspace{-2mm}\begin{equation}\label{innersgd}
 {\bf y}^{t,k+1}(i)={\bf y}^{t,k}(i)-\eta \tilde{{\bf g}}^{t,k}(i)+\theta({\bf y}^{t,k}(i)-{\bf y}^{t,k-1}(i)),
\end{equation}
where $\EE\tilde{{\bf g}}^{t,k}(i)=\nabla f_{i}({\bf y}^{t,k}(i))$.
After $K$ inner iterations in each local client, the resulting parameters $ {\bf z}^t(i)\leftarrow {\bf y}^{t,K}(i)$ is sent to its neighbors ($\mathcal{N}(i)$). Every client then updates its parameters by taking the local averaging as follows
\vspace{-2mm}{\small \begin{align}\label{avera}
 {\bf x}^{t+1}(i) = \sum_{l\in \mathcal{N}(i)} w_{i,l} {\bf z}^t(l).
\end{align}}
The procedure of DFedAvgM can be illustrated as Figure 1 (b). It is seen that DFedAvgM plays the tradeoff between local computing and communications. It is well-known that the communication costs are usually much more expensive than the computation costs \cite{li2014communication}, which indicates DFedAvgM can be more efficient than DSGD.

\vspace{-2mm}
 \begin{algorithm}[H]
\footnotesize
    \caption{Quantized DFedAvgM}\label{alg2}
	\begin{algorithmic}[1]
\State{\textbf{parameters}: $\eta>0$,  $K\in \mathbb{Z}^+$, $0\leq \theta<1$, $s$, $b$.}
\State{\textbf{initialization}: ${\bf x}^0=\textbf{0}$}
\For{$t=1,2,\ldots$}
\For{$i=\{1,2,\ldots,m\}$  }
\State{node $i $ performs local training \eqref{innersgd}    $K$  times and sends ${\bf q}^t(i)=Q[{\bf y}^{t,K}(i)-{\bf x}^t(i)]$ to   $\mathcal{N}(i)$   }
\State{node $i$ updates as \eqref{quanavera}}
\EndFor
\EndFor
	\end{algorithmic}
  \end{algorithm}

\subsection{Efficient Communication via Quantization}
In DFedAvgM, client $i$ needs to send ${\bf x}^t(i)$ to its neighbours $\mathcal{N}(i)$. Thus, when the number of neighbours $\mathcal{N}(i)$ grows, client-client communications become the major bottleneck on algorithms' efficiency. We leverage the quantization trick to reduce the communication cost \cite{alistarh2017qsgd,magnusson2019maintaining}. In particular, we consider the following quantization procedure: Given a constant $s>0$ and the limited bit number $b\in \mathbb{Z}^+$, the representable range is then $\{- 2^{b-1}s, -(2^{b-1}-1)s, \ldots, 0, s, 2s,\ldots,  (2^{b-1}-1)s\}$. For any $a\in\RR$ with $-2^{b-1}s\leq a<(2^{b-1}-1)s$,
we can find an integer $k\in\mathbb{Z}$ such that $ks\leq a <(k+1)s$ and we then use $ks$ to replace $a$.
The above quantization scheme is deterministic, which can be written as $q(a):= \lfloor \frac{a}{s}\rfloor s$ for $a\in \RR$; Besides the deterministic rule, the stochastic quantization uses the following scheme
\vspace{-1mm}$$q(a):= \left\{ \begin{array}{c}
ks,~\textrm{w.p.}~1-\frac{a-ks}{s},\\
(k+1)s,~\textrm{w.p.}~\frac{a-ks}{s}.\\
\end{array}
\right.$$
It is easy to see that the stochastic quantization is unbiased, i.e., $\EE [q(a)]=a$ for any $a\in \RR$. When $s$ is small, deterministic and stochastic quantization schemes perform very similarly. For a vector ${\bf x}\in \RR^d$ whose coordinates are all stored with $32$ bits, we consider quantizing all coordinates of $ {\bf x}=[ {x}_1 , {x}_2, \ldots, {x}_d]\in\RR^d $. The multi-dimension quantization operator is then defined as
\vspace{-1mm}\begin{align}\label{quan}
Q({\bf x}):=[q( {x}_1 ), q({x}_2), \ldots, q({x}_d)].
\end{align}
For both deterministic and stochastic quantization schemes, we have $\EE\|Q({\bf x})-{\bf x}\|^2\leq  \frac{d}{4}s^2$ if ${x}_i\in[- 2^{b-1}s, (2^{b-1}-1)s]$ for $i\in\{1,2,\ldots,d\}$. In this paper, we consider a quantization operator with the following assumption, which hold for the two quantization schemes mentioned above.

\begin{assumption}\label{assq}
The quantization operator $Q:\RR^d\rightarrow\RR^d$ satisfies   $\EE\|Q({\bf x})-{\bf x}\|^2\leq \frac{d}{4}s^2$ with $s>0$ for any ${\bf x}\in \RR^d$.
\end{assumption}

Directly quantize the parameters is feasible for sufficiently smooth loss functions, but may be impossible for DNNs. To this end, we consider quantizing the difference the difference of parameters.
Quantized DFedAvgM can be summarized as: After running \eqref{innersgd} $K$ times, client $i$ quantizes ${\small {\bf q}^t(i)\leftarrow Q({\bf y}^{t,K}(i)-{\bf x}^t(i))}$ and send it to ${\small\mathcal{N}(i)}$. After receiving
${\small[{\bf q}^t(j)]_{j\in \mathcal{N}(i)}}$, every client updates its local parameters as
\vspace{-2mm}\begin{equation}\label{quanavera}
     {\bf x}^{t+1}(i) = {\bf x}^{t}(i)+\sum_{l\in \mathcal{N}(i)} w_{i,l} {\bf q}^t(l).
\end{equation}
In each communication, client $i$ just needs to send the pair $(s, {\bf q}^t(i))$ to $\mathcal{N}(i)$, whose representation requires  $(32+db)\textrm{deg}(\mathcal{N}(i))$ bits rather than  $32d\textrm{deg}(\mathcal{N}(i))$ bits for sending the unquantized version.
If $d$ is large and $b<32$, the communications can be significantly reduced.

\section{Convergence Analysis}\label{sec:theory}
In this section, we analyze the convergence of the proposed (quantized) DFedAvgM. The convergence analysis of DFedAvgM is much more complicated than SGD, SGD with momentum, and DSGD; the technical difficulty is because ${\bf z}^t(i)-{\bf x}^t(i)$ fails to be an unbiased estimate of the gradient $\nabla f_i({\bf x}^t(i))$ after multiple iterations of SGD or SGD with momentum in each client.
In the following, we consider the convergence of the average point, which is defined as $\overline{{\bf x}^t}:={\sum_{i=1}^m {\bf x}^t(i)}/{m}$. We first present the convergence of DFedAvgM for general nonconvex objective function in the following Theorem.

\begin{theorem}[General nonconvexity]\label{th1}
Let the sequence $\{{\bf x}^{t}(i)\}_{t\geq 0}$ be generated by DFedAvgM for $i\in\{1,2\ldots,m\}$ and suppose Assumptions 1, 2 and 3 hold. Moreover, assume the stepsize $\eta$ for SGD with momentum that used for training client models satisfies
\vspace{-1mm}$$0<\eta \leq \tfrac{1}{8LK}~~\mbox{and}~~ 64L^2K^2\eta^2+64LK\eta<1,$$
where $L$ is the Lipschitz constant of $\nabla f$ and $K$ is the number of client iterations before each communication. Then
\vspace{-1mm}\begin{align*}
\min_{1\leq t\leq T} \EE\|\nabla f(\overline{{\bf x}^{t}})\|^2&\leq\frac{ 2f(\overline{{\bf x}^{1}})-2\min f}{\gamma(K,\eta)T}\\
&+\alpha(K,\eta)+\beta(K,\eta),
\end{align*}
where $T$ is the total number of communication rounds and the constants are given as
\vspace{-1mm}$$
\gamma(K,\eta):=\frac{\eta (K-\theta)}{(1-\theta)}-\frac{64(1-\theta)L^2K^4\eta^3}{K-\theta}-64LK^2\eta^2,$$
\vspace{-1mm}\begin{align*}
&\alpha(K,\eta):=\\
&{\footnotesize \frac{(\frac{ (1-\theta)L^2K^2\eta^3}{(K-\theta)}+L\eta^2)(8K\sigma_l^2+ 32K^2\sigma_g^2+\frac{64K^2\theta^2(\sigma_l^2+B^2)}{(1-\theta)^2})}{\frac{\eta (K-\theta)}{(1-\theta)}-\frac{64(1-\theta)L^2K^4\eta^3}{K-\theta}-64LK^2\eta^2}},
\end{align*}
and
\vspace{-1mm}\begin{align*}
&\beta(K,\eta,\lambda):=(\frac{64(1-\theta)L^4K^4\eta^5}{(K-\theta)}+64L^3K^2\eta^4)\times\\
&\frac{(8K\sigma_l^2+ 32K^2\sigma_g^2+32K^2B^2+\frac{64K^2\theta^2}{(1-\theta)^2}(\sigma_l^2+B^2))}{[(1-\lambda)(\frac{\eta (K-\theta)}{(1-\theta)}-\frac{64(1-\theta)L^2K^4\eta^3}{K-\theta}-64LK^2\eta^2)]}.
\end{align*}
\end{theorem}\vspace{-2mm}

To get an explicit rate on $T$ from Theorem \ref{th1}, we choose $\eta=\Theta({1}/{LK\sqrt{T}})$. As $T$ is large enough and $64L^2K^2\eta^2+64LK\eta<1$. Then, $\gamma(K,\eta)=\Theta({1}/{((1-\theta)\sqrt{T})})$, and $\alpha(K,\eta)=\Theta(\frac{(1-\theta)\sigma_l^2+(1-\theta)K\sigma_g^2+\frac{\theta^2}{(1-\theta)}K(\sigma_l^2+B^2)}{K\sqrt{T}})$, and $\beta(K,\eta,\lambda)=\Theta(\frac{(1-\theta)(\sigma_l^2+ K\sigma_g^2+ KB^2)+\frac{\theta^2}{(1-\theta)}K(\sigma_l^2+B^2)}{(1-\lambda)KT^{3/2}})$. Based on this choice of $\eta$ and the Theorem \ref{th1}, we have the following convergence rate for DFedAvgM.

\begin{proposition}\label{proc}
As the communication round number $T$ is large enough, it holds that
\vspace{-1mm}\begin{align*}
&\min_{1\leq t\leq T} \EE\|\nabla f(\overline{{\bf x}^{t}})\|^2=\mathcal{O}\Big(\frac{(1-\theta)(f(\overline{{\bf x}^{1}})-\min f)}{\sqrt{T}}\\
&+\frac{(1-\theta)\sigma_l^2+(1-\theta)K\sigma_g^2+\frac{\theta^2}{(1-\theta)}K(\sigma_l^2+B^2)}{K\sqrt{T}}\\
&\quad+\frac{(1-\theta)(\sigma_l^2+ K\sigma_g^2+ KB^2)+\frac{\theta^2}{(1-\theta)}K(\sigma_l^2+B^2)}{(1-\lambda)KT^{3/2}}\Big).
\end{align*}
\end{proposition}

From Proposition \ref{proc}, we can see that the speed of DFedAvgM can be improved when the number of local iteration, $K$, increases.  Also, when the momentum $\theta$ is $0$ and $K$ is large enough, the bound will be dominated by $\mathcal{O}\Big(\frac{1}{\sqrt{T}}+\frac{\sigma_g^2}{\sqrt{T}}+\frac{\sigma_g^2+ B^2}{(1-\lambda)^2T^{3/2}}\Big)$, in which the local variance bound diminishes. This phenomenon coincides with our intuitive understanding: in local client, the use of large $K$ can result in a local minimizer; then the local variance bound will hurt nothing. To reach any given $\epsilon>0$ error, DFedAvgM needs $\mathcal{O}(\frac{1}{\epsilon^2})$ communication rounds, which is the same as SGD and DSGD. It is worth mentioning that the above theoretical results show that whether the momentum $\theta$ can accelerate the algorithm depends on the relation between $f(\overline{{\bf x}^{1}})-\min f+\sigma_g^2$ and $\sigma_l^2+B^2$, i.e., if $f(\overline{{\bf x}^{1}})-\min f+\sigma_g^2\gg\sigma_l^2+B^2$, as $\theta\in[0,1)$ increases, the rate improves;  if  $f(\overline{{\bf x}^{1}})-\min f+\sigma_g^2\ll\sigma_l^2+B^2$, large $\theta$ may degrade the performance of DFedAvgM.

The convergence results established above, which simply require smooth assumption on the objective functions, are quite general and somehow not sharp due to extra properties are missing. For example, recent (non)convex studies \cite{karimi2016linear,reddi2016stochastic,foster2018uniform} have exploited the algorithmic performance under the P{\L} property, which is named after Polyak and {\L}ojasiewicz \cite{polyak1963gradient,lojasiewicz1963topological}. For a smooth function $f$, we say it satisfies P{\L}-$\nu$ property provided
\begin{align}\label{PL}
\|\nabla f({\bf x})\|^2\geq 2\nu(f({\bf x})-\min f),~~\forall{\bf x}\in \textrm{dom}(f).
\end{align}
The well-known strong convexity implies P{\L} condition, but not vice verse. In the following, we present the convergence of DFedAvgM under the P{\L} condition.

\begin{theorem}[P{\L} condition]\label{th2}
Assume function $f$ satisfies the P{\L}-$\nu$ condition, the following convergence rate holds
\vspace{-1mm}\begin{align*}
&\EE f(\overline{{\bf x}^{T}})-\min f\leq [1-\nu\gamma(K,\eta)]^T(f(\overline{{\bf x}^{0}})-\min f)\\
&\quad+\frac{\alpha(K,\eta)}{2\nu}+\frac{\beta(K,\eta,\lambda)}{2\nu}.
\end{align*}
\end{theorem}

Due to the fact that $f(\overline{{\bf x}^{0}})-\min f\geq 0$, the right side is larger than $\frac{\alpha(K,\eta)}{2\nu}+\frac{\beta(K,\eta,\lambda)}{2\nu}=\mathcal{O}(\eta)$.
If we still let $\eta=\Theta(\frac{1}{LK\sqrt{T}})$, the convergence rate is at least $\mathcal{O}(1/\sqrt{T})$. But we cannot choose a very small $\eta$; otherwise, the dominated term $[1-\nu\gamma(K,\eta)]^T(f(\overline{{\bf x}^{0}})-\min f)$ will decay very slowly. If the learning rate enjoys the form as $\eta={c_1\ln^{c_3}T}/{(LKT^{c_2})}$ with $c_1,c_2>0$ \footnote{This learning rate is commonly used in the ML community.}, we can prove the following results on the optimal choices for $c_1,c_2,c_3$.

\begin{proposition}\label{optimal}
Let $\eta={c_1\ln^{c_3}T}/{(LKT^{c_2})}$ with $c_1,c_2>0$, the optimal rate of DFedAvgM is $\tilde{\mathcal{O}}({1}/{T})$, in which case $c_1={L}/{\nu}$, $c_2=1$, and $c_3=-1$, that is, $\eta={1}/{(\nu KT\ln T)}$.
\end{proposition}
This finding coincides with  existing  results of the optimal rate for SGD with strong convexity \cite{shalev2011pegasos,nemirovski2009robust}. Under the P{\L} condition, the convergence rate of DFedAvgM is improved.

Next, we provide the convergence guarantee for the quantized DFedAvgM, which is stated in the following theorem.

\begin{theorem}\label{pro1}
Let the sequence $\{{\bf x}^{t}(i)\}_{t\geq 0}$ be generated by the quantized DFedAvgM for all $i\in\{1,2\ldots,m\}$, and all the assumptions in Theorem \ref{th1} and Assumption 4 hold. Let $\eta=\Theta(\frac{1}{LK\sqrt{T}})$, as $T$ is sufficiently large, it holds that
\vspace{-1mm}\begin{align*}
&\min_{1\leq t\leq T} \EE\|\nabla f(\overline{{\bf x}^{t}})\|^2=\mathcal{O}\Big(\frac{(1-\theta)(f(\overline{{\bf x}^{1}})-\min f)}{\sqrt{T}}\\
&+\frac{(1-\theta)(\sigma_l^2+K\sigma_g^2)+\frac{\theta^2}{(1-\theta)}K(\sigma_l^2+B^2)}{K\sqrt{T}}+\\
&\frac{(1-\theta)(\sigma_l^2+ K\sigma_g^2+ KB^2)+\frac{\theta^2}{(1-\theta)}K(\sigma_l^2+B^2)}{(1-\lambda)KT^{3/2}}+\sqrt{T}s\Big).
\end{align*}
If the function $f$ further satisfies the P{\L} condition and $\eta=\frac{1}{\nu TK\ln T}$, it follows that
\vspace{-1mm}$$\EE(f(\overline{{\bf x}^{T}})-\min f)=\widetilde{\mathcal{O}}(\frac{1}{T}+Ts).$$
\end{theorem}
According to Theorem \ref{pro1}, to reach any given $\epsilon>0$ error in general case, we need to set $s=\mathcal{O}(\epsilon^2)$ and set the the number of communication round as $T=\Theta(\frac{1}{\epsilon^2})$. While with P{\L} condition,
we set $T=\Theta(\frac{1}{\epsilon})$ and $s=\mathcal{O}(\epsilon^2)$. It follows $\EE(f(\overline{{\bf x}^{T}})-\min f)=\widetilde{\mathcal{O}}(\epsilon).$ Therefore, under the P{\L} condition, the number of communication round is reduced.

In the following, we provide a sufficient condition for  communications-saving of the two quantization rules mentioned in Sec. 3.2 used in quantized DFedAvgM.
\begin{proposition}\label{com-save}
Assume we use the stochastic or deterministic quantization rule with $b$ bits using stepsize $\eta=\frac{1}{LK\sqrt{T}}$.  Assume that the parameters trained in all clients do not overflow, that is, all coordinates are contained in  $[- 2^{b-1}s, (2^{b-1}-1)s]$.  Let Assumptions 1, 2 and 3 hold. If the desired error
\vspace{-1mm}{\small\begin{align*}
\epsilon&>(1-\theta)\sqrt{3LBs}d^{\frac{1}{4}}\times\\
&\sqrt{2(f(\overline{{\bf x}^{0}})-\min f)+ \frac{8\sigma_l^2}{K}+ 32\sigma_g^2+\frac{64\theta^2(\sigma_l^2+B^2)}{(1-\theta)^2}}
\end{align*}}
and $b<\frac{128}{9}+\frac{32}{d}$, the quantized DFedAvgM can beat DFedAvgM with 32 bits in term of the required communications to reach $\epsilon$.
\end{proposition}

Proposition \ref{com-save} indicates that the superiority of the quantized DFedAvgM retains when the desired error $\epsilon$ is not smaller than $\mathcal{O}((1-\theta)\sqrt{s})$. We can also see that as $K$ increases, the guaranteed lower bound of $\epsilon$ decreases, which demonstrates the necessity of multiple local iterations. Moreover, a larger $\theta$ can also reduce the lower bound.
\section{Proofs}

\subsection{Technical Lemmas}
We define
${\bf 1}:=[1, 1, \ldots, 1 ]^{\top}\in \mathbb{R}^m$ and
\begin{equation*}
    {\bf P}:=\frac{\textbf{1}\textbf{1}^{\top}}{m}\in \mathbb{R}^{m\times m}.
\end{equation*}
For a matrix ${\bf A}$, we denote its spectral norm as $\|{\bf A}\|_{\textrm{op}}$. We also define
$
{\bf X}:= \begin{bmatrix}
    {\bf x}(1), {\bf x}(2),
    \ldots,
    {\bf x}(m)
\end{bmatrix}^{\top}\in\mathbb{R}^{m\times d}$.

\begin{lemma}\label{mi}[Lemma 4 , \cite{lian2017can}]
For any $k\in \mathbb{Z}^+$, the mixing matrix ${\bf W}\in\RR^m$ satisfies
$$\|{\bf W}^k-{\bf P}\|_{\emph{op}}\leq \lambda^k,$$
where $\lambda:=\max\{|\lambda_2|,|\lambda_m(W)|\}$.
\end{lemma}
In [Proposition 1, \cite{nedic2009distributed}], the author also proved that $\|W^k-{\bf P}\|_{\textrm{op}}\leq C\lambda^k$ for some $C>0$ that depends on the matrix.

\begin{lemma}\label{diffbound}
Assume that Assumptions 2 and 3 hold, and $0\leq \theta<1$. Let $({\bf y}^{t,k}(i))_{t\geq 0}$ be generated by the (quantized) DFedAvgM. It then follows
$$\EE\|{\bf y}^{t,k+1}(i)-{\bf y}^{t,k}(i)\|^2\leq \frac{1}{(1-\theta)^2}(2\eta^2\sigma_l^2+2\eta^2B^2)$$
when $0\leq k\leq K-1$.
\end{lemma}
\begin{lemma}\label{localbound}
Given the stepsize   $0<\eta \leq \tfrac{1}{8LK}$ and $i\in \{1,2,\ldots,m\}$ and assume $({\bf y}^{t,k}(i))_{t\geq 0}$, $({\bf x}^{t}(i))_{t\geq 0}$ are generated by the (quantized) DFedAvgM for all $i\in\{1,2\ldots,m\}$. If  Assumption 3 holds, it then follows
\begin{align*}
  &\frac{1}{m}\sum_{i=1}^m\EE\|{\bf y}^{t,k}(i) - {\bf x}^{t}(i)\|^2 \\
  &\quad\leq  C_1\eta^2+ 32K^2\eta^2\frac{\sum_{i=1}^m\EE\|\nabla f({\bf x}^t(i))\|^2}{m},
\end{align*}
where $C_1:=8K\sigma_l^2+ 32K^2\sigma_g^2+\frac{64K^2\theta^2}{(1-\theta)^2}(\sigma_l^2+B^2)$ when $0\leq k\leq K$.
\end{lemma}
With the fact that ${\bf y}^{t,K}(i)={\bf z}^{t}(i)$, Lemma \ref{localbound} also holds for ${\bf z}^t(i)$.

\begin{lemma}\label{globalbound}
Given the stepsize   $\eta>0$ and let $\{{\bf x}^{t}(i)\}_{t\geq 0}$ be generated by DFedAvgM for all $i\in\{1,2\ldots,m\}$. If Assumption 3 holds, we have the following bound
\begin{align}
  \frac{1}{m}\sum_{i=1}^m\EE\|{\bf x}^t(i)-\overline{{\bf x}^t}\|^2\leq C_2\frac{\eta^2}{1-\lambda},
\end{align}
 where $C_2:=8K\sigma_l^2+ 32K^2\sigma_g^2+\frac{64K^2\theta^2}{(1-\theta)^2}(\sigma_l^2+B^2)+32K^2B^2$.
\end{lemma}

\begin{lemma}\label{quan-globalbound}
Given the stepsize $\eta>0$ and assume $\{{\bf x}^{t}(i)\}_{t\geq 0}$ are generated by the quantized DFedAvgM for all $i\in\{1,2\ldots,m\}$. If  Assumption 3 holds, it follows that
\begin{align}
  \frac{1}{m}\sum_{i=1}^m\EE\|{\bf x}^t(i)-\overline{{\bf x}^t}\|^2\leq 2C_2\frac{\eta^2}{1-\lambda}+\frac{2ds^2}{1-\lambda}.
\end{align}
\end{lemma}

\subsection{Proof of Technical Lemmas}
\subsubsection{Proof of Lemma \ref{diffbound}}
Given any $\psi>0$, the Cauchy inequality gives us
\begin{align*}
&\EE\|{\bf y}^{t,k+1}(i)-{\bf y}^{t,k}(i)\|^2\\
&=\EE\|-\eta\tilde{{\bf g}}^k(i)+\theta({\bf y}^{t,k}(i)-{\bf y}^{t,k-1}(i))\|^2\\
&\overset{a)}{\leq} (1+\psi)\theta^2\EE\|{\bf y}^{t,k}(i)-{\bf y}^{t,k-1}(i)\|^2\\
&\qquad+(1+\frac{1}{\psi})\eta^2\EE\|\tilde{{\bf g}}^k(i) - \nabla f_i({\bf y}^{t,k}(i))+\nabla f_i({\bf y}^{t,k}(i))\|^2\\
&\leq (1+\psi)\theta^2\EE\|{\bf y}^{t,k}(i)-{\bf y}^{t,k-1}(i)\|^2\\
&\qquad+(2+\frac{2}{\psi})\eta^2\|\nabla f_i({\bf y}^{t,k}(i))\|^2\\
&\qquad+2(1+\frac{1}{\psi})\eta^2\EE\|\tilde{{\bf g}}^k(i) - \nabla f_i({\bf y}^{t,k}(i))\|^2,
\end{align*}
where $a)$ uses the Cauchy's inequality $\EE\|{\bf a}+{\bf b}\|^2\leq(1+\frac{1}{\psi})\EE\|{\bf a}\|^2+(1+\psi)\EE\|{\bf b}\|^2$ with ${\bf a}=-\eta\tilde{{\bf g}}^k(i)$ and ${\bf b}=\theta({\bf y}^{t,k}(i)-{\bf y}^{t,k-1}(i))$.
Without loss of generality, we assume $\theta\neq 0$. Let $\psi=\frac{1}{\theta}-1$, we get
\begin{align*}
&\EE\|{\bf y}^{t,k+1}(i)-{\bf y}^{t,k}(i)\|^2\\
&\qquad\leq \theta\EE\|{\bf y}^{t,k}(i)-{\bf y}^{t,k-1}(i)\|+\frac{2\eta^2\sigma_l^2}{1-\theta}+\frac{2\eta^2B^2}{1-\theta}.
\end{align*}
Using the mathematical induction, for any integer $0\leq k\leq K$, we have
\begin{align*}
&\EE\|{\bf y}^{t,k+1}(i)-{\bf y}^{t,k}(i)\|^2\\
&\quad\leq \frac{2\eta^2\sigma_l^2+2\eta^2B^2}{1-\theta}(\sum_{i=0}^{k-1}\theta^i)\leq \frac{2\eta^2\sigma_l^2+2\eta^2B^2}{(1-\theta)^2}.
\end{align*}
\subsubsection{Proof of Lemma \ref{localbound}}

Note that for any $k\in\{0,1,\ldots,K-1\}$, in node $i$ it holds
\begin{align*}
    &\EE \|{\bf y}^{t,k+1}(i) - {\bf x}^{t}(i)\|^2\\
    &= \EE\|{\bf y}^{t,k}(i)-\eta \tilde{{\bf g}}^k(i)- {\bf x}^{t}(i)+\theta({\bf y}^{t,k}(i)-{\bf y}^{t,k-1}(i))\|^2\\
    &\leq \EE\|{\bf y}^{t,k}(i)- {\bf x}^{t}(i)- \eta\Big(\tilde{{\bf g}}^k(i) - \nabla f_i({\bf y}^{t,k}(i)) + \nabla f_i({\bf y}^{t,k}(i))  \\
    &\quad - \nabla f_i({\bf x}^t(i)) + \nabla f_i({\bf x}^t(i))- \nabla f({\bf x}^t(i))+ \nabla f({\bf x}^t(i))\Big)\\
    &\quad+\theta({\bf y}^{t,k}(i)-{\bf y}^{t,k-1}(i))\|^2\leq \textrm{I}+\textrm{II},
\end{align*}
where  $\textrm{I}=(1+\frac{1}{2K-1})\EE\|{\bf y}^{t,k}(i)- {\bf x}^{t}(i)- \eta(\tilde{{\bf g}}^k(i) - \nabla f_i({\bf y}^{t,k}(i)) \|^2$ and $\textrm{II}= 2K\eta^2\EE\|\nabla f_i({\bf y}^{t,k}(i)) - \nabla f_i({\bf x}^t(i)) + \nabla f_i({\bf x}^t(i)) - \nabla f({\bf x}^t(i)) + \nabla f({\bf x}^t(i))+\theta({\bf y}^{t,k}(i)-{\bf y}^{t,k-1}(i))\|^2 $.
The unbiased expectation property of  $\tilde{{\bf g}}^k(i)$ gives us
 \begin{align*}
    \textrm{I}&=(1+\frac{1}{2K-1})\Big(\EE\|{\bf y}^{t,k}(i)- {\bf x}^{t}(i)\|^2\\
    &+ \eta^2\EE\|\tilde{{\bf g}}^k(i) - \nabla f_i({\bf y}^{t,k}(i))\|^2\Big).
 \end{align*}
On the other hand,  with Lemma \ref{diffbound}, we have the following bound
\begin{align*}
&\textrm{II}\leq 8K\eta^2\EE\|\nabla f_i({\bf y}^{t,k}(i)) - \nabla f_i({\bf x}^t(i))\| \\
&+ 8K\eta^2\EE\|\nabla f_i({\bf x}^t(i)) - \nabla f({\bf x}^t(i))\| \\
&+ 8K\eta^2\EE\|\nabla f({\bf x}^t(i))\|^2+8K\theta^2\EE\|{\bf y}^{t,k}(i)-{\bf y}^{t,k-1}(i)\|^2\\
&\leq 8L^2K\eta^2\|{\bf y}^{t,k}(i) - {\bf x}^t(i)\|^2+ 8K\eta^2\sigma_g^2\\
&+ 8K\eta^2\EE\|\nabla f({\bf x}^t(i))\|^2+\frac{16K\theta^2}{(1-\theta)^2}(\eta^2\sigma_l^2+\eta^2B^2).
\end{align*}
 Thus, we can obtain
 \begin{align*}
    &\EE \|{\bf y}^{t,k+1}(i) - {\bf x}^{t}(i)\|^2 \\
    &\leq (1+\frac{1}{2K-1}+8L^2K\eta^2)\EE\|{\bf y}^{t,k}(i)- {\bf x}^{t}(i)\|^2+2\eta^2\sigma_l^2\\
    &+ 8K\eta^2\sigma_g^2+ 8K\eta^2\EE\|\nabla f({\bf x}^t(i))\|^2+\frac{16K\theta^2}{(1-\theta)^2}(\eta^2\sigma_l^2+\eta^2B^2)\\
    &\leq (1+\frac{1}{K-1})\EE\|{\bf y}^{t,k}(i)- {\bf x}^{t}(i)\|^2+2\eta^2\sigma_l^2+ 8K\eta^2\sigma_g^2\\
    &+\frac{16K\theta^2}{(1-\theta)^2}(\eta^2\sigma_l^2+\eta^2B^2)+ 8K\eta^2\EE\|\nabla f({\bf x}^t(i))\|^2,
\end{align*}
where the last inequality depends on the selection of the stepsize.
The recursion from $j=0$  to $k$ yeilds
 \begin{align*}
    &\EE \|{\bf y}^{t,k}(i) - {\bf x}^{t}(i)\|^2\\
    &\leq \sum_{j=0}^{K-1}(1+\frac{1}{K-1})^j\Big[2\eta^2\sigma_l^2+ 8K\eta^2\sigma_g^2\\
    &\quad+\frac{16K\theta^2}{(1-\theta)^2}(\eta^2\sigma_l^2+\eta^2B^2)+ 8K\eta^2\EE\|\nabla f({\bf x}^t(i))\|^2\Big]\\
    &\leq (K-1)\Big[(1+\frac{1}{K-1})^K-1\Big]\times\Big[2\eta^2\sigma_l^2+ 8K\eta^2\sigma_g^2\\
    &\quad+\frac{16K\theta^2}{(1-\theta)^2}(\eta^2\sigma_l^2+\eta^2B^2)+ 8K\eta^2\EE\|\nabla f({\bf x}^t(i))\|^2\Big]\\
    &\leq 8K\eta^2\sigma_l^2+ 32K^2\eta^2\sigma_g^2+\frac{64K^2\theta^2}{(1-\theta)^2}(\eta^2\sigma_l^2+\eta^2B^2)\\
    &\quad+ 32K^2\eta^2\EE\|\nabla f({\bf x}^t(i))\|^2,
\end{align*}
where we used the inequality  $(1+\frac{1}{K-1})^K\leq 5$ holds for any $K\geq 1$.

\subsubsection{Proof of Lemma \ref{globalbound}}
We denote ${\bf Z}^{t}:=\begin{bmatrix}
  {\bf z}^{t}(1),  {\bf z}^{t}(2),
    \ldots,
    {\bf z}^{t}(m)
\end{bmatrix}^{\top}\in\mathbb{R}^{m\times d}$.
With these notation, we have
\begin{align}\label{xtglobal}
{\bf X}^{t+1}={\bf W}{\bf Z}^{t}={\bf W}{\bf X}^{t}-{\bf \zeta}^t,
\end{align}
where ${\bf \zeta}^t:={\bf W}{\bf X}^{t}-{\bf W}{\bf Z}^{t}$.
The iteration \eqref{xtglobal} can be rewritten as the following expression
\begin{align}\label{xtglobal2}
{\bf X}^{t}=W^{t}{\bf X}^{0}-\sum_{j=0}^{t-1}{\bf W}^{t-1-j}{\bf \zeta}^j.
\end{align}
Obviously, it follows
\begin{equation}\label{trans}
    {\bf W} {\bf P}= {\bf P} {\bf W}={\bf P}.
\end{equation}
According to Lemma \ref{mi}, it holds
$$\|{\bf W}^t-{\bf P}\|\leq \lambda^t.$$ Multiplying both sides of \eqref{xtglobal2} with  ${\bf P}$ and using \eqref{trans}, we then get
\begin{align}\label{xtglobal3}
{\bf P}{\bf X}^{t}={\bf P}{\bf X}^{0}-\sum_{j=0}^{t-1}{\bf P}{\bf \zeta}^j=-\sum_{j=0}^{t-1}{\bf P}{\bf \zeta}^j,
\end{align}
where we used initialization ${\bf X}^{0}=\textbf{0}$.
Then, we are led to
\begin{equation}\label{xtglobal4}
   \begin{aligned}
&\|{\bf X}^{t}-{\bf P}{\bf X}^{t}\|=\|\sum_{j=0}^{t-1}({\bf P}-{\bf W}^{t-1-j}){\bf \zeta}^j\|\\
&\leq \sum_{j=0}^{t-1}\|{\bf P}-{\bf W}^{t-1-j}\|_{\textrm{op}}\|{\bf \zeta}^j\|\leq \sum_{j=0}^{t-1}\lambda^{t-1-j}\|{\bf \zeta}^j\|.
\end{aligned}
\end{equation}
With Cauchy inequality,
\begin{align*}
&\EE\|{\bf X}^{t}-{\bf P}{\bf X}^{t}\|^2\leq \EE(\sum_{j=0}^{t-1}\lambda^{\frac{t-1-j}{2}}\cdot \lambda^{\frac{t-1-j}{2}}\|{\bf \zeta}^j\|)^2\\
&\leq (\sum_{j=0}^{t-1}\lambda^{t-1-j})(\sum_{j=0}^{t-1} \lambda^{t-1-j}\EE\|{\bf \zeta}^j\|^2)
\end{align*}
Direct calculation gives us
$$\EE\|{\bf \zeta}^j\|^2\leq \|{\bf W}\|^2\cdot\EE\|{\bf X}^{j}-{\bf Z}^{j}\|^2\leq \EE\|{\bf X}^{j}-{\bf Z}^{j}\|^2.$$
With Lemma \ref{localbound} and Assumption 3, for any $j$,
\begin{align*}
&\EE\|{\bf X}^{j}-{\bf Z}^{j}\|^2\\
&\leq m(8K\sigma_l^2+ 32K^2\sigma_g^2+\frac{64K\theta^2}{(1-\theta)^2}(\sigma_l^2+B^2)+32K^2B^2)\eta^2.
\end{align*}
Thus, we get
\begin{align*}
&\EE\|{\bf X}^{t}-{\bf P}{\bf X}^{t}\|^2\\
&\leq \frac{ m(8K\sigma_l^2+ 32K^2\sigma_g^2+\frac{64K\theta^2}{(1-\theta)^2}(\sigma_l^2+B^2)+32K^2B^2)\eta^2}{1-\lambda}.
\end{align*}
The fact that ${\bf X}^{t}-{\bf P}{\bf X}^{t}=\left(
                                                \begin{array}{c}
                                                  {\bf x}^t(1)- \overline{{\bf x}^t}\\
                                                   {\bf x}^t(2)- \overline{{\bf x}^t} \\
                                                  \vdots \\
                                                   {\bf x}^t(m)- \overline{{\bf x}^t} \\
                                                \end{array}
                                              \right)
$ then proves the result.

\subsubsection{Proof of Lemma \ref{quan-globalbound}}
Let  $\widetilde{{\bf Z}}^{t}:={\bf Y}^{t,K}$. Obviously, it holds
\begin{align}\label{xt-quan}
{\bf X}^{t+1}={\bf W}{\bf X}^{t}-\widetilde{{\bf \zeta}}^t,
\end{align}
where $\widetilde{{\bf \zeta}}^t={\bf W}{\bf X}^{t}-{\bf W}({\bf Q}(\widetilde{{\bf Z}}^{t}-{\bf X}^t)+{\bf X}^t).$ We just need to bound $\EE\|\widetilde{{\bf \zeta}}^t\|^2$,
\begin{align*}
&\EE\|\widetilde{{\bf \zeta}}^t\|^2\leq 2\EE\|{\bf W}{\bf X}^{t}-{\bf W}\widetilde{{\bf Z}}^{t}\|^2\\
&\qquad+2\EE\|{\bf W}\widetilde{{\bf Z}}^{t}-{\bf W}({\bf Q}(\widetilde{{\bf Z}}^{t}-{\bf X}^t)+{\bf X}^t)\|^2\\
&\leq2\EE\|{\bf X}^{t}-\widetilde{{\bf Z}}^{t}\|^2+2\EE\|{\bf W}(\widetilde{{\bf Z}}^{t}-{\bf X}^t)-{\bf W}({\bf Q}(\widetilde{{\bf Z}}^{t}-{\bf X}^t))\|^2\\
&\leq 2\EE\|{\bf X}^{t}-\widetilde{{\bf Z}}^{t}\|^2+2mds^2\\
&\leq 2m\eta^2(8K\sigma_l^2+ 32K^2\sigma_g^2+\frac{64K\theta^2}{(1-\theta)^2}(\sigma_l^2+B^2)+32K^2B^2)\\
&\qquad+2mds^2,
\end{align*}
the last inequality uses Lemma \ref{localbound}.

\subsection{Proof of Theorem \ref{th1}}
Noting that ${\bf P}{\bf X}^{t+1}={\bf P}{\bf W}{\bf Z}^{t}={\bf P}{\bf Z}^{t}$, that is also
$$\overline{{\bf x}^{t+1}}=\overline{{\bf z}^{t}},$$
 we have
\begin{align}\label{th1-pro}
\overline{{\bf x}^{t+1}}-\overline{{\bf x}^{t}}=\overline{{\bf x}^{t+1}}-\overline{{\bf z}^{t}}+\overline{{\bf z}^{t}}-\overline{{\bf x}^{t}}=\overline{{\bf z}^{t}}-\overline{{\bf x}^{t}},
\end{align}
where $\overline{{\bf z}^{t}}:=\frac{\sum_{i=1}^m {\bf z}^t(i)}{m}$. With the local scheme in each node,
\begin{align*}
&\overline{{\bf z}^{t}}-\overline{{\bf x}^{t}}=\frac{\sum_{i=1}^m ({\bf z}^t(i)-{\bf x}^{t}(i))}{m}\\
&=\frac{\sum_{i=1}^m (\sum_{k=0}^{K-1}{\bf y}^{t,k+1}(i)-{\bf y}^{t,k}(i))}{m}\\
&=-\eta\frac{\sum_{i=1}^m \sum_{k=0}^{K-1}\nabla f_i({\bf y}^{t,k}(i))}{m}+\theta(\overline{{\bf z}^{t}}-\overline{{\bf x}^{t}})\\
&\qquad+\theta\eta\frac{\sum_{i=1}^m\nabla f_i({\bf y}^{t,K-1}(i))}{m}.
\end{align*}
Thus, we get
\begin{equation}\label{thcore1}
    \begin{aligned}
&\overline{{\bf z}^{t}}-\overline{{\bf x}^{t}}=\frac{1}{1-\theta}(-\eta\frac{\sum_{i=1}^m \sum_{k=0}^{K-2}\nabla f_i({\bf y}^{t,k}(i))}{m})\\
&=\frac{1}{1-\theta}(-\eta\frac{\sum_{i=1}^m(1-\theta)\nabla f_i({\bf y}^{t,K-1}(i))}{m}).
\end{aligned}
\end{equation}
The Lipschitz continuity of $\nabla f$ gives us
\begin{align}\label{th1t1}
\EE f(\overline{{\bf x}^{t+1}})&\leq \EE f(\overline{{\bf x}^{t}})+\EE\langle\nabla f(\overline{{\bf x}^{t}}),\overline{{\bf z}^{t}}-\overline{{\bf x}^{t}}\rangle\nonumber\\
&+\frac{L}{2}\EE\|\overline{{\bf x}^{t+1}}-\overline{{\bf x}^{t}}\|^2,
\end{align}
where we used \eqref{th1-pro}.
Let $$\tilde{K}=\frac{K-\theta}{1-\theta},$$
we can derive
\begin{align*}
\small
&\EE\langle \tilde{K}\nabla f(\overline{{\bf x}^{t}}),(\overline{{\bf z}^{t}}-\overline{{\bf x}^{t}})/\tilde{K}\rangle\\
&=\EE\langle \tilde{K}\nabla f(\overline{{\bf x}^{t}}),-\eta\nabla f(\overline{{\bf x}^{t}})+\eta\nabla f(\overline{{\bf x}^{t}})+(\overline{{\bf z}^{t}}-\overline{{\bf x}^{t}})/\tilde{K}\rangle\\
&=-\eta \tilde{K}\EE\|\nabla f(\overline{{\bf x}^{t}})\|^2+\EE\langle\nabla f(\overline{{\bf x}^{t}}),\eta\nabla f(\overline{{\bf x}^{t}})+(\overline{{\bf z}^{t}}-\overline{{\bf x}^{t}})/\tilde{K}\rangle\\
&\overset{a)}{\leq} -\eta \tilde{K}\EE\|\nabla f(\overline{{\bf x}^{t}})\|^2\\
&\qquad+\eta\EE \|\nabla f(\overline{{\bf x}^{t}})\|\cdot\Big\|\frac{\sum_{i=1}^m \sum_{k=0}^{K-2}[\nabla f_i(\overline{{\bf x}^{t}})-\nabla f_i({\bf y}^{t,k}(i))]}{m}\\
&\qquad+\frac{\sum_{i=1}^m(1-\theta)[\nabla f_i(\overline{{\bf x}^{t}})-\nabla f_i({\bf y}^{t,K-1}(i))]}{m})\Big\|\\
&\leq -\eta \tilde{K}\EE\|\nabla f(\overline{{\bf x}^{t}})\|^2+\frac{\eta L}{m}\sum_{i=1}^m\sum_{k=0}^{K-1}\EE \|\nabla f(\overline{{\bf x}^{t}})\|\cdot\|\overline{{\bf x}^t} -{\bf y}^{t,k}(i)\|\\
&\leq -\eta \tilde{K}\EE\|\nabla f(\overline{{\bf x}^{t}})\|^2+\frac{\eta \tilde{K}}{2}\EE\|\nabla f(\overline{{\bf x}^{t}})\|^2\\
&\qquad+\frac{\eta L^2K^2}{2\tilde{K}}( C_1\eta^2+ 32K^2\eta^2\frac{\sum_{i=1}^m\EE\|\nabla f({\bf x}^t(i))\|^2}{m}),
\end{align*}
where $a)$ uses \eqref{thcore1}.
Similarly, we can get
\begin{align*}
&\frac{L}{2}\EE(\|\overline{{\bf x}^{t+1}}-\overline{{\bf x}^{t}}\|^2)=\frac{L}{2}\EE(\|\overline{{\bf z}^{t}}-\overline{{\bf x}^{t}}\|^2)\\
&\leq \frac{L}{2}\frac{1}{m}\sum_{i=1}^m \|{\bf z}^t(i)-{\bf x}^{t}(i)\|^2\\
&\leq \frac{L}{2}C_1\eta^2+ 16LK^2\eta^2\frac{\sum_{i=1}^m\EE\|\nabla f({\bf x}^t(i))\|^2}{m}.
\end{align*}
Thus, \eqref{th1t1} can be represented as
\begin{align*}
&\EE f(\overline{{\bf x}^{t+1}})\leq \EE f(\overline{{\bf x}^{t}})-\frac{\eta \tilde{K}}{2}\EE\|\nabla f(\overline{{\bf x}^{t}})\|^2+\frac{ L^2K^2}{2\tilde{K}} C_1\eta^3\\
&+\frac{L}{2}C_1\eta^2+(\frac{16L^2K^4\eta^3}{\tilde{K}}+16LK^2\eta^2)\frac{\sum_{i=1}^m\EE\|\nabla f({\bf x}^t(i))\|^2}{m}.
\end{align*}
Direct computation together with Lemma \ref{globalbound} gives us
\begin{align*}
&\frac{\sum_{i=1}^m\EE\|\nabla f({\bf x}^t(i))\|^2}{m}\\
&=\frac{\sum_{i=1}^m\EE\|\nabla f({\bf x}^t(i))-\nabla f(\overline{{\bf x}^{t}})+\nabla f(\overline{{\bf x}^{t}})\|^2}{m}\\
&\leq\frac{\sum_{i=1}^m2\EE\|\nabla f({\bf x}^t(i))-\nabla f(\overline{{\bf x}^{t}})\|^2+2\EE\|\nabla f(\overline{{\bf x}^{t}})\|^2}{m}\\
&\leq 2L^2\frac{\sum_{i=1}^m\|{\bf x}^t(i)-\overline{{\bf x}^{t}}\|^2}{m}+2\EE\|\nabla f(\overline{{\bf x}^{t}})\|^2\\
&\leq \frac{2L^2C_2\eta^2}{1-\lambda}+2\EE\|\nabla f(\overline{{\bf x}^{t}})\|^2.
\end{align*}
Therefore, we have
\begin{equation}\label{descent}
    \begin{aligned}
&\EE f(\overline{{\bf x}^{t+1}})\leq \EE f(\overline{{\bf x}^{t}})\\
&\quad-(\frac{\eta (K-\theta)}{2(1-\theta)}-\frac{32(1-\theta)L^2K^4\eta^3}{K-\theta}-32LK^2\eta^2)\\
&\qquad\times\EE\|\nabla f(\overline{{\bf x}^{t}})\|^2+(\frac{ (1-\theta)L^2K^2}{2(K-\theta)}\eta^3+\frac{L}{2}\eta^2)\\
&\qquad\times(8K\sigma_l^2+ 32K^2\sigma_g^2+\frac{64K^2\theta^2}{(1-\theta)^2}(\sigma_l^2+B^2))\\
&\quad+ (32(1-\theta)L^4K^4\eta^5/(K-\theta)+32L^3K^2\eta^4)/(1-\lambda)\\
&\qquad\times(8K\sigma_l^2+ 32K^2\sigma_g^2+32K^2B^2+\frac{64K^2\theta^2}{(1-\theta)^2}(\sigma_l^2+B^2)).
\end{aligned}
\end{equation}
Summing the inequality  \eqref{descent} from  $t=1$  to $T$, we then proved the result.
\subsection{Proof of Theorem \ref{th2}}
With the P{\L} condition, $$\EE\|\nabla f(\overline{{\bf x}^{t}})\|^2\geq 2\nu\EE(f(\overline{{\bf x}^{t}})-\min f).$$
We start from \eqref{descent},
\begin{equation}\label{descent2}
    \begin{aligned}
&\EE f(\overline{{\bf x}^{t+1}})\leq \EE f(\overline{{\bf x}^{t}})-\nu\gamma(K,\eta)\EE(f(\overline{{\bf x}^{t}})-\min f)\\
&\quad+\frac{\gamma(K,\eta)\alpha(K,\eta)}{2}
+\frac{\gamma(K,\eta)\beta(K,\eta,\lambda)}{2}.
\end{aligned}
\end{equation}
By defining $\xi_t:=\EE(f(\overline{{\bf x}^{t}})-\min f)$, it then follows
\begin{equation}\label{descent3}
    \begin{aligned}
\xi_{t+1}&\leq [1-\nu\gamma(K,\eta)]\xi_t\\
&+\frac{\gamma(K,\eta)\alpha(K,\eta)}{2}+\frac{\gamma(K,\eta)\beta(K,\eta,\lambda)}{2}.
\end{aligned}
\end{equation}
Thus, we are then led to
\begin{equation*}
    \begin{aligned}
&\xi_{T}\leq [1-\nu\gamma(K,\eta)]^T \xi_0
\\
&\quad+\Big(\frac{\gamma(K,\eta)\alpha(K,\eta)}{2}+\frac{\gamma(K,\eta)\beta(K,\eta,\lambda)}{2}\Big)\\
&\qquad\times(\sum_{t=0}^{T-1}[1-\nu\gamma(K,\eta)]^t)\\
&\leq [1-\nu\gamma(K,\eta)]^T \xi_0\\
&\quad+\Big(\frac{\gamma(K,\eta)\alpha(K,\eta)}{2}+\frac{\gamma(K,\eta)\beta(K,\eta,\lambda)}{2}\Big)\frac{1}{\nu\gamma(K,\eta)}\\
&= [1-\nu\gamma(K,\eta)]^T \xi_0+\frac{\alpha(K,\eta)}{2\nu}+\frac{\beta(K,\eta,\lambda)}{2\nu}.
\end{aligned}
\end{equation*}
The result is then proved.
\subsection{Proof of Proposition \ref{optimal}}
A quick calculation gives us
\begin{align*}
\left\{\begin{array}{c}
         \gamma(K,\eta)=\Theta(\frac{1}{T^{c_2}}), \\
         \frac{\alpha(K,\eta)}{2\nu}+\frac{\beta(K,\eta,\lambda)}{2\nu}=O(\frac{1}{T^{c_2}}).
       \end{array}
\right.
\end{align*}
Thus, we just need to bound the first term in Theorem \ref{th2}. As $T$ is large, $\gamma(K,\eta)\rightarrow 0$. Its  logarithm is then
\begin{align*}
T\log[1-\nu\gamma(K,\eta)]=T\log[1-\nu\gamma(K,\eta)]=\Theta(-T\nu\gamma(K,\eta)).
\end{align*}
With our setting, it follows
$$T\nu\gamma(K,\eta)\approx \frac{\nu c_1\ln^{c_3}T}{LT^{c_2-1}}.$$
Then we have
$$\EE f(\overline{{\bf x}^{T}})-\min f=\mathcal{O}(\textrm{exp}(-\frac{\nu c_1\ln^{c_3}T}{LKT^{c_2-1}})+\frac{1}{T^{c_2}}).$$
We first consider how to choose $c_2$. From the L'Hospital's rule, for any $\delta>0$, as $T\rightarrow+\infty$
$$\textrm{exp}(-\frac{1}{T^{\delta}})\rightarrow 1.$$
Thus, we need to set $c_2\leq 1$ and the fast rate is slower than $O(\frac{1}{T})$. To this end, we choose $c_1=\frac{L}{\nu}$, $c_2=1$, and $c_3=-1$.
\subsection{Proof of Theorem \ref{pro1}}
Let
$\widetilde{{\bf y}}^{t}:=\frac{\sum_{i=1}^m {\bf y}^{t,K}(i)}{m}$, in the quantized DFedAvgM, it  follows
$\overline{{\bf x}^{t+1}}-\overline{{\bf x}^{t}}=Q(\widetilde{{\bf y}}^{t}-\overline{{\bf x}^{t}}).$
With the Lipschitz continuity of $\nabla f$,
\begin{align*}
\EE f(\overline{{\bf x}^{t+1}})&\leq \EE f(\overline{{\bf x}^{t}})\\
&+\EE\langle\nabla f(\overline{{\bf x}^{t}}),Q(\widetilde{{\bf y}}^{t}-\overline{{\bf x}^{t}})\rangle+\frac{L}{2}\EE\|\overline{{\bf x}^{t+1}}-\overline{{\bf x}^{t}}\|^2,
\end{align*}
We have
\begin{align*}
&\EE\langle \nabla f(\overline{{\bf x}^{t}}),Q(\widetilde{{\bf y}}^{t}-\overline{{\bf x}^{t}})\rangle\\
&=\EE\langle \nabla f(\overline{{\bf x}^{t}}),\widetilde{{\bf y}}^{t}-\overline{{\bf x}^{t}}\rangle+\EE\langle \nabla f(\overline{{\bf x}^{t}}),\widetilde{{\bf y}}^{t}-\overline{{\bf x}^{t}}-Q(\widetilde{{\bf y}}^{t}-\overline{{\bf x}^{t}})\rangle\\
&\leq \EE\langle \nabla f(\overline{{\bf x}^{t}}),\widetilde{{\bf y}}^{t}-\overline{{\bf x}^{t}}\rangle+B\sqrt{d}s
\end{align*}
and
\begin{align*}
&\frac{L}{2}\EE\|\overline{{\bf x}^{t+1}}-\overline{{\bf x}^{t}}\|^2=\frac{L}{2}\EE\|Q(\widetilde{{\bf y}}^{t}-\overline{{\bf x}^{t}})\|^2\\
&\leq L\EE\|\widetilde{{\bf y}^{t}}-\overline{{\bf x}^{t}}\|^2+L\EE\|Q(\widetilde{{\bf y}}^{t}-\overline{{\bf x}^{t}})-(\widetilde{{\bf y}}^{t}-\overline{{\bf x}^{t}})\|^2\\
&\leq L\EE\|\widetilde{{\bf y}^{t}}-\overline{{\bf x}^{t}}\|^2+\frac{Lds^2}{m}.
\end{align*}
Note that both $\EE\langle \nabla f(\overline{{\bf x}^{t}}),\widetilde{{\bf y}}^{t}-\overline{{\bf x}^{t}}\rangle$ and $\EE\|\widetilde{{\bf y}^{t}}-\overline{{\bf x}^{t}}\|^2$ can inherit the bounds of $\EE\langle\nabla f(\overline{{\bf x}^{t}}),\overline{{\bf z}^{t}}-\overline{{\bf x}^{t}}\rangle$ and $\EE\|\overline{{\bf x}^{t+1}}-\overline{{\bf x}^{t}}\|^2$ in the proof of Theorem \ref{th1}. Thus, we obtain
\begin{align*}
&\EE f(\overline{{\bf x}^{t+1}})\leq \EE f(\overline{{\bf x}^{t}})-\frac{\eta \tilde{K}}{2}\EE\|\nabla f(\overline{{\bf x}^{t}})\|^2\\
&+\frac{ L^2K^2}{2\tilde{K}} C_1\eta^3+LC_1\eta^2+\frac{Lds^2}{m}+B\sqrt{d}s\\
&+(\frac{32L^2K^4\eta^3}{\tilde{K}}+32LK^2\eta^2)\frac{\sum_{i=1}^m\EE\|\nabla f({\bf x}^t(i))\|^2}{m}.
\end{align*}
With Lemma  \ref{quan-globalbound}, we can get
\begin{align*}
&\frac{\sum_{i=1}^m\EE\|\nabla f({\bf x}^t(i))\|^2}{m}\\
&\leq\frac{\sum_{i=1}^m2\EE\|\nabla f({\bf x}^t(i))-\nabla f(\overline{{\bf x}^{t}})\|^2+2\EE\|\nabla f(\overline{{\bf x}^{t}})\|^2}{m}\\
&\leq 2L^2\frac{\sum_{i=1}^m\|{\bf x}^t(i)-\overline{{\bf x}^{t}}\|^2}{m}+2\EE\|\nabla f(\overline{{\bf x}^{t}})\|^2\\
&\leq \frac{2L^2C_3\eta^2}{1-\lambda}+\frac{4L^2ds^2}{1-\lambda}+2\EE\|\nabla f(\overline{{\bf x}^{t}})\|^2.
\end{align*}
Combining the inequalities together,
\begin{align*}
\EE f(\overline{{\bf x}^{t+1}})&\leq \EE f(\overline{{\bf x}^{t}})+ \zeta(K,\eta,\lambda,s)\\
&-(\frac{\eta \tilde{K}}{2}-\frac{64L^2K^4\eta^3}{\tilde{K}}-64LK^2\eta^2)\EE\|\nabla f(\overline{{\bf x}^{t}})\|^2,
\end{align*}
where $\zeta(K,\eta,\lambda,s):=\frac{ L^2K^2}{2\tilde{K}} C_1\eta^3+LC_1\eta^2+(\frac{32L^2K^4\eta^3}{\tilde{K}}+32LK^2\eta^2)(\frac{2L^2C_3\eta^2}{1-\lambda}+\frac{4L^2ds^2}{1-\lambda})+\frac{Lds^2}{m}+B\sqrt{d}s.$
Given the stepsize $\eta=\Theta(\frac{1}{LK\sqrt{T}})$, we can  see that $\frac{\eta \tilde{K}}{2}-\frac{64L^2K^4\eta^3}{\tilde{K}}-64LK^2\eta^2>0$ as $T$ is large. When $s>0$ is small, $s^2=O(s)$.
And it then follows
 $\frac{\eta \tilde{K}}{2}-\frac{64L^2K^4\eta^3}{\tilde{K}}-64LK^2\eta^2=\Theta(\frac{1}{(1-\theta)\sqrt{T}})$. We now consider
 \begin{equation}\label{quancomplex}
     \begin{aligned}
     &\Big[ (\frac{32L^2K^4\eta^3}{\tilde{K}}+32LK^2\eta^2)(\frac{2L^2C_3\eta^2}{1-\lambda}+\frac{4L^2ds}{1-\lambda})\\
     &\quad+LC_1\eta^2+\frac{ L^2K^2}{2\tilde{K}}C_1\eta^3+\frac{Lds^2}{m}+B\sqrt{d}s\Big]\Big/\Big[\frac{\eta \tilde{K}}{2}\\
     &\quad-\frac{64L^2K^4\eta^3}{\tilde{K}}-64LK^2\eta^2\Big],
 \end{aligned}
 \end{equation}
which is at the order as
 $O\Big(\frac{1}{\sqrt{T}}+\frac{\sigma_l^2+K\sigma_g^2+\frac{\theta^2}{(1-\theta)^2}K(\sigma_l^2+B^2)}{K\sqrt{T}}+\frac{\sigma_l^2+ K\sigma_g^2+ KB^2+\frac{\theta^2}{(1-\theta)^2}K(\sigma_l^2+B^2)}{(1-\lambda)KT^{3/2}}+\sqrt{T}s\Big)$.
 If the function $f$ satisfies the P{\L}-$\nu$, we have
 \begin{align*}
&\EE(f(\overline{{\bf x}^{t}})-\min f)\\
&\qquad\leq [1-\nu\gamma(K,\eta)]^T (f(\overline{{\bf x}^{0}})-\min f) +\frac{\zeta(K,\eta,\lambda,s)}{\nu\gamma(K,\eta)}.
\end{align*}
When $\eta=\frac{1}{\nu TK\ln T}$, $[1-\nu\gamma(K,\eta)]^T=\widetilde{\mathcal{O}}(\frac{1}{T})$ and
$$\frac{\zeta(K,\eta,\lambda,s)}{\nu\gamma(K,\eta)}=\widetilde{\mathcal{O}}(\frac{1}{T}+Ts).$$
\subsection{Proof of Proposition \ref{com-save}}
We calculate to reach the same error, the communication costs by both algorithms. Omitting the order larger than 1 for $\eta$, from \eqref{descent}, we have
\begin{equation*}
    \begin{aligned}
&\min_{1\leq t\leq T}\EE\|\nabla f(\overline{{\bf x}^{t}})\|^2\approx \frac{2(f(\overline{{\bf x}^{0}})-\min f)}{\eta K T}\\
&\quad+ L\eta(8\sigma_l^2+ 32K\sigma_g^2+\frac{64K\theta^2}{(1-\theta)^2}(\sigma_l^2+B^2))\\
&=\frac{2(1-\theta)(f(\overline{{\bf x}^{0}})-\min f)}{ \sqrt{ T}}\\
&\quad+ \frac{8(1-\theta)\sigma_l^2+ 32(1-\theta)K\sigma_g^2+\frac{64K\theta^2}{(1-\theta)}(\sigma_l^2+B^2)}{K\sqrt{T}}.
\end{aligned}
\end{equation*}
for DFedAvgM. From  \eqref{quancomplex},
\begin{equation*}
    \begin{aligned}
&\min_{1\leq t\leq T}\EE\|\nabla f(\overline{{\bf x}^{t}})\|^2\approx \frac{2(1-\theta)(f(\overline{{\bf x}^{0}})-\min f)}{ \sqrt{ T}}\\
&\quad+ \frac{8(1-\theta)\sigma_l^2+ 32(1-\theta)K\sigma_g^2+\frac{64K\theta^2}{(1-\theta)}(\sigma_l^2+B^2)}{K\sqrt{T}}\\
&\quad+2(1-\theta)LB\sqrt{d}\sqrt{T}s.
\end{aligned}
\end{equation*}
Given the $\epsilon>0$, assume $T_{\epsilon}$ obeys
\begin{align*}
&\frac{8(1-\theta)\sigma_l^2+ 32(1-\theta)K\sigma_g^2+\frac{64K\theta^2}{(1-\theta)}(\sigma_l^2+B^2)}{K\sqrt{T_{\epsilon}}}\\
&\quad+\frac{2(1-\theta)(f(\overline{{\bf x}^{0}})-\min f)}{ \sqrt{ T_{\epsilon}}}=\epsilon.
\end{align*}
That means DFedAvgM can output an $\epsilon$ error in $T_{\epsilon}$ iterations. However, due to the error caused by the quantization,  we have to increase the iteration number for quantized DFedAvgM.
We set the iteration number as $\frac{9}{4}T_{\epsilon}$. To get the $\epsilon$ error for quantized DFedAvgM, we also need
$3(1-\theta)LB\sqrt{d}\sqrt{T_{\epsilon}}s\leq\epsilon$, which yields
\begin{align*}
&3(1-\theta)^2LB\sqrt{d}s\Big[2(f(\overline{{\bf x}^{0}})-\min f)+ \frac{8\sigma_l^2}{K}\\
&\quad+ 32\sigma_g^2+\frac{64\theta^2}{(1-\theta)^2}(\sigma_l^2+B^2)\Big]\leq \epsilon^2.
\end{align*}
The total communication cost   of DFedAvgM to reach $\epsilon$ is
$$32 d T_{\epsilon}\sum_{i=1}^m [\textrm{deg}(\mathcal{N}(i))].$$
While for quantized version, the total communication cost is
$$(32+ db) \frac{9}{4}T_{\epsilon}\sum_{i=1}^m [\textrm{deg}(\mathcal{N}(i))].$$
Thus, the communications can be reduced if
$$(32+ db) \frac{9}{4}<32 d .$$

\section{Numerical Results}\label{sec:numerics}
We apply the proposed DFedAvgM with communication quantization to train DNNs for both image classification and language modeling, where we consider a simple ring structured communication network. We aim to verify that DFedAvgM can train DNNs effectively, especially in communication efficiency. Moreover, we consider the membership privacy protection when DFedAvgM is used for training DNNs.
We apply the membership inference attack (MIA) \cite{salem2018ml} to test the efficiency of (quantized) DFedAvgM in protecting the training data's MP. In MIA, the attack model is a binary classifier \footnote{We use a multilayer perceptron with a hidden layer of 64 nodes, followed by a softmax output function as the attack model, which is adapted from \cite{salem2018ml}.}, which is to decide if a data point is in the training set of the target model. For each of the following dataset, to perform MIA we first split its training set into $D_{\rm shadow}$ and $D_{\rm target}$ with the same size. Furthermore, we split $D_{\rm shadow}$ into two halves with the same size and denote them as $D_{\rm shadow}^{\rm train}$ and $D_{\rm shadow}^{\rm out}$, and we split $D_{\rm target}$ by half into $D_{\rm target}^{\rm train}$ and $D_{\rm target}^{\rm out}$. MIA proceeds as follows: 1) train the shadow model by using $D_{\rm shadow}^{\rm train}$; 2) apply the trained shadow model to predict all data points in $D_{\rm shadow}$ and train the corresponding classification probabilities of belonging to each class. Then we take the top three classification probabilities (or two in the case of binary classification) to form the feature vector for each data point.A feature vector is tagged as1if the corresponding data point is in $D_{\rm shadow}^{\rm train}$, and $0$ otherwise. Then we train the attack model by leveraging all the labeled feature vectors; 3) train the target model by using $D_{\rm target}^{\rm train}$ and obtain feature vector for each point in $D_{\rm target}$. Finally, we leverage the attack model to decide if a data point is in $D_{\rm target}^{\rm train}$. Note the attack model we build is a binary classifier, which is to decide if a data point is in the training set of the target model.  For any data $\xi\in D_{\rm target}$, we apply the attack model to predict its probability ($p$) of belonging to the training set of the target model. Given any fixed threshold $t$ if $p\geq t$, we classify $\xi$ as a member of the training set (positive sample), and if $p<t$, we conclude that $\xi$ is not in the training set (negative sample); so we can obtain different attack results with different thresholds. We can plot the ROC curve for different threshold, and regard the area under the ROC curve (AUC) as an evaluation of the membership inference attack.The target model protects perfect membership privacy if the AUC is 0.5 (attack model performs random guess), and the higher AUC is, the less private the target model is.

\subsection{MNIST Classification}
\begin{figure}[t!]
\centering
\begin{tabular}{ccc}
\hskip -0.3cm\includegraphics[width=0.32\linewidth]{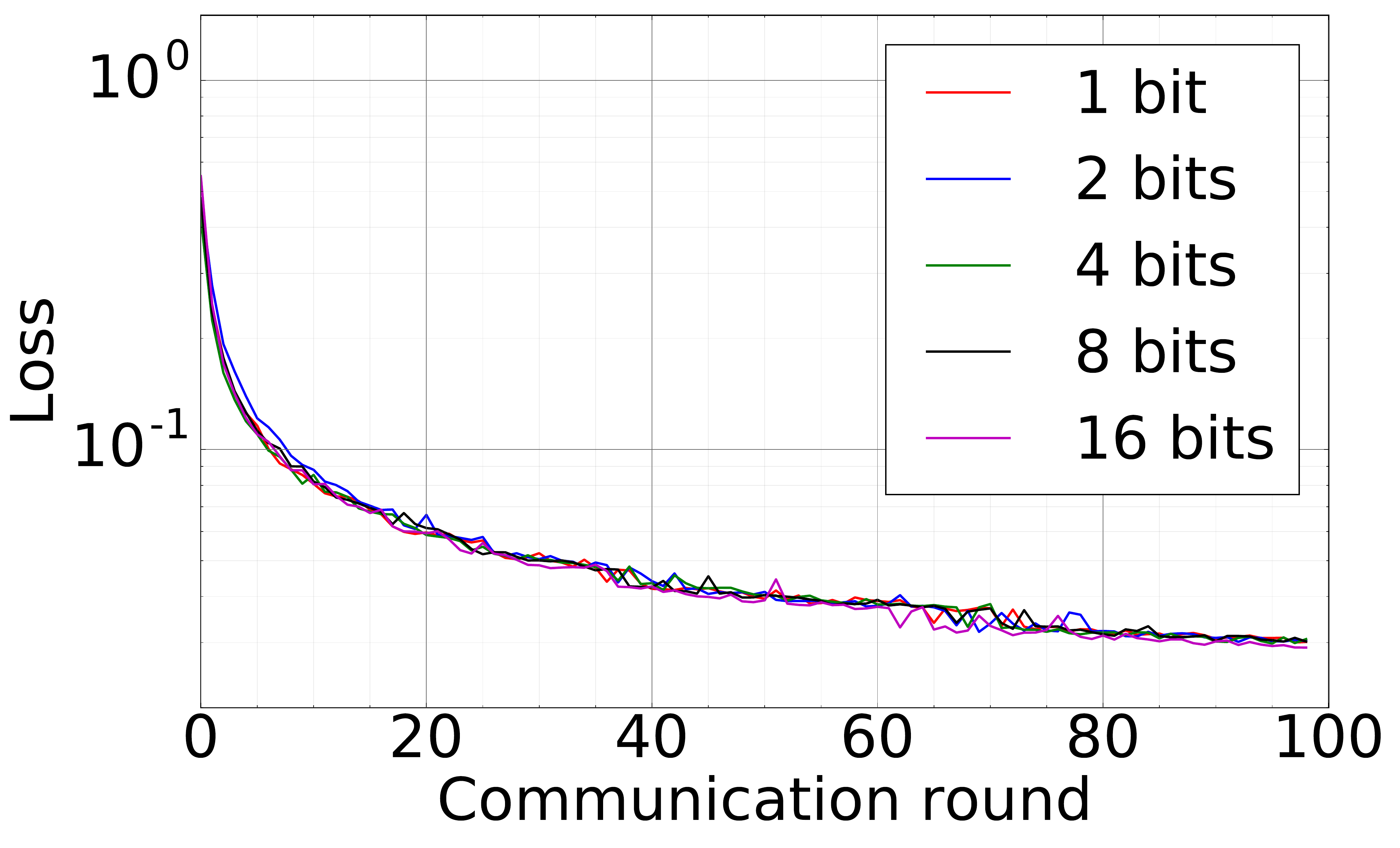}&
\hskip -0.45cm\includegraphics[width=0.32\linewidth]{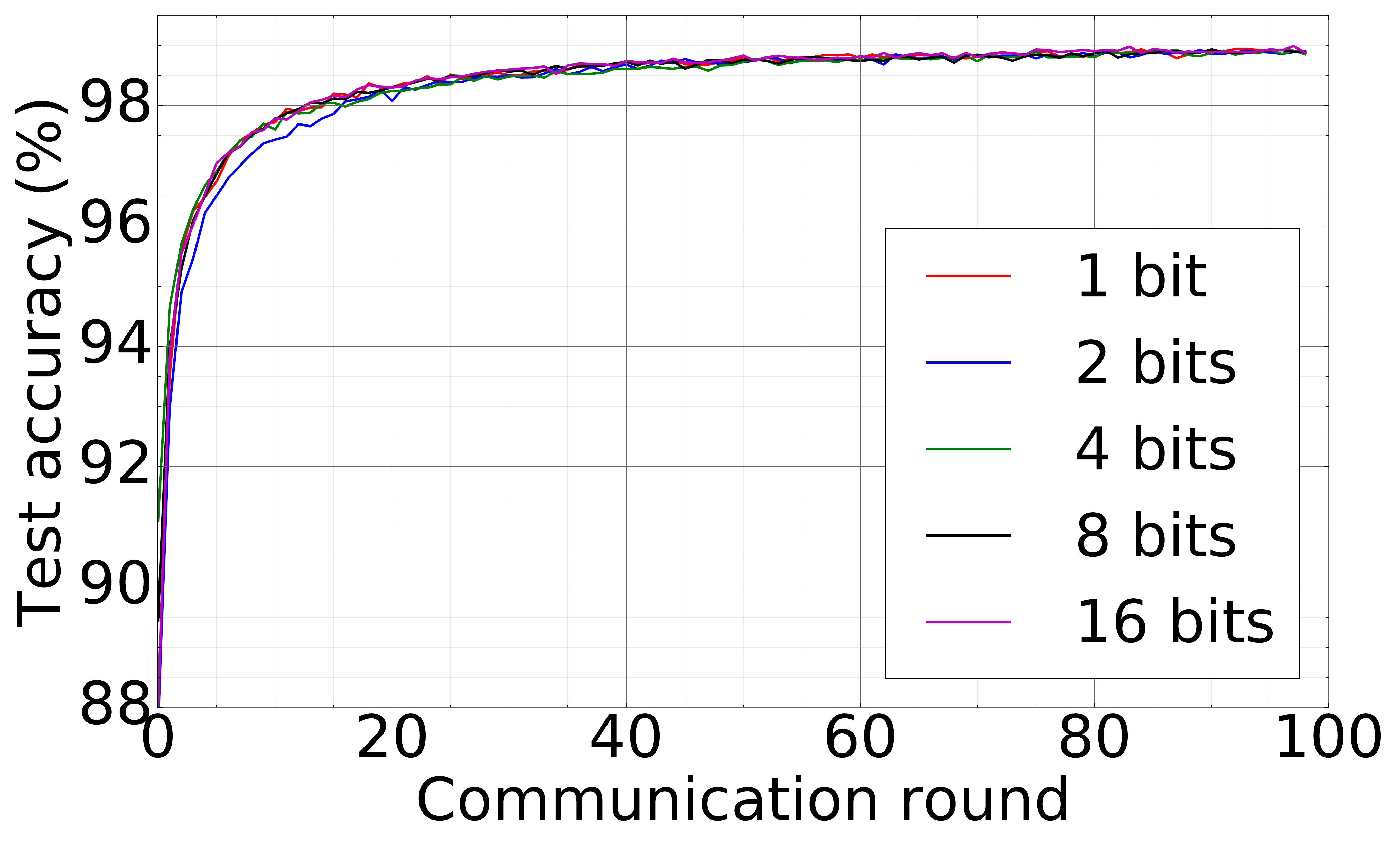}&
\hskip -0.45cm\includegraphics[width=0.32\linewidth]{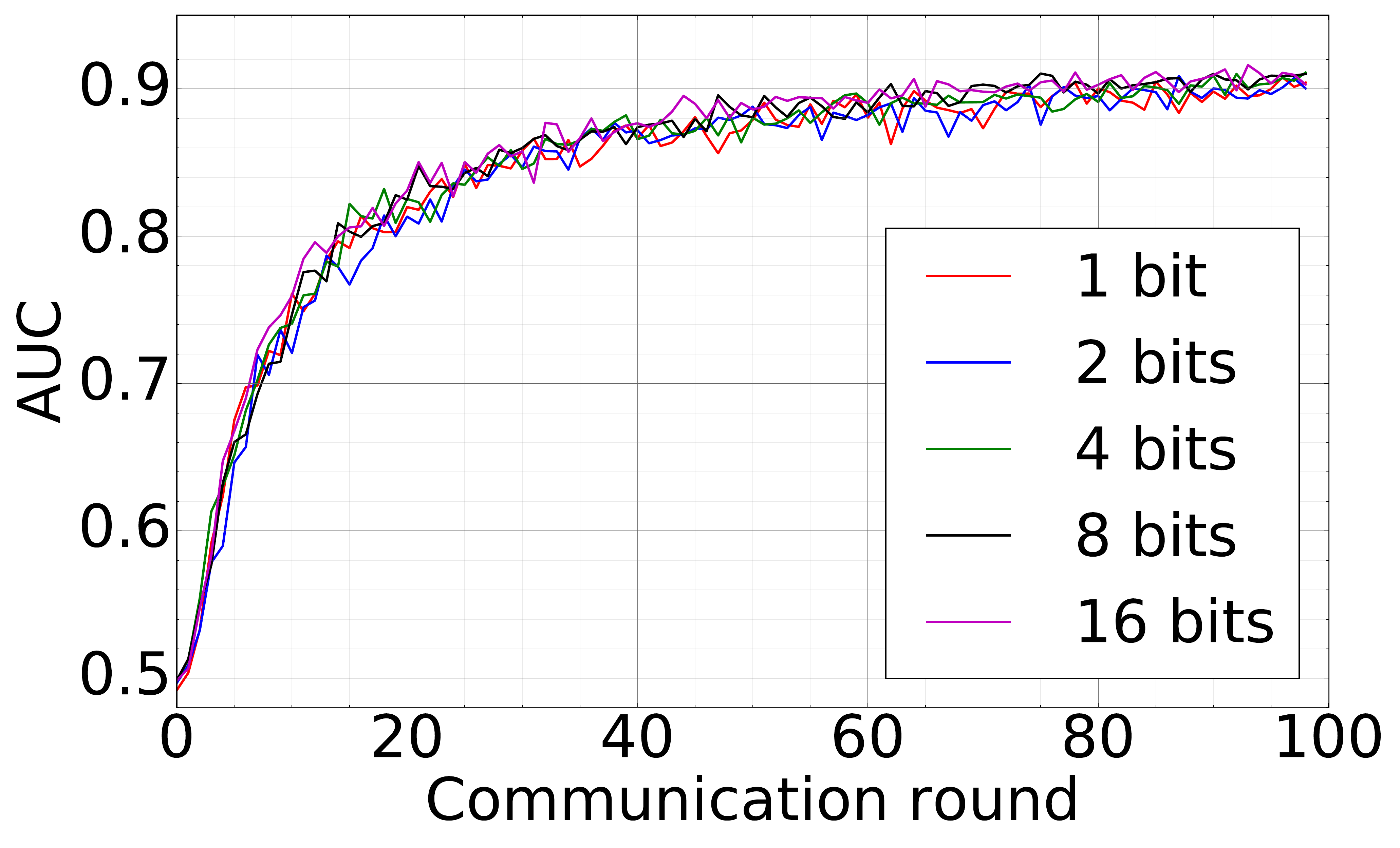}\\
{\footnotesize CR vs. Training loss} & {\footnotesize CR vs. Test acc} & {\footnotesize CR vs. AUC}\\
\hskip -0.3cm\includegraphics[width=0.32\linewidth]{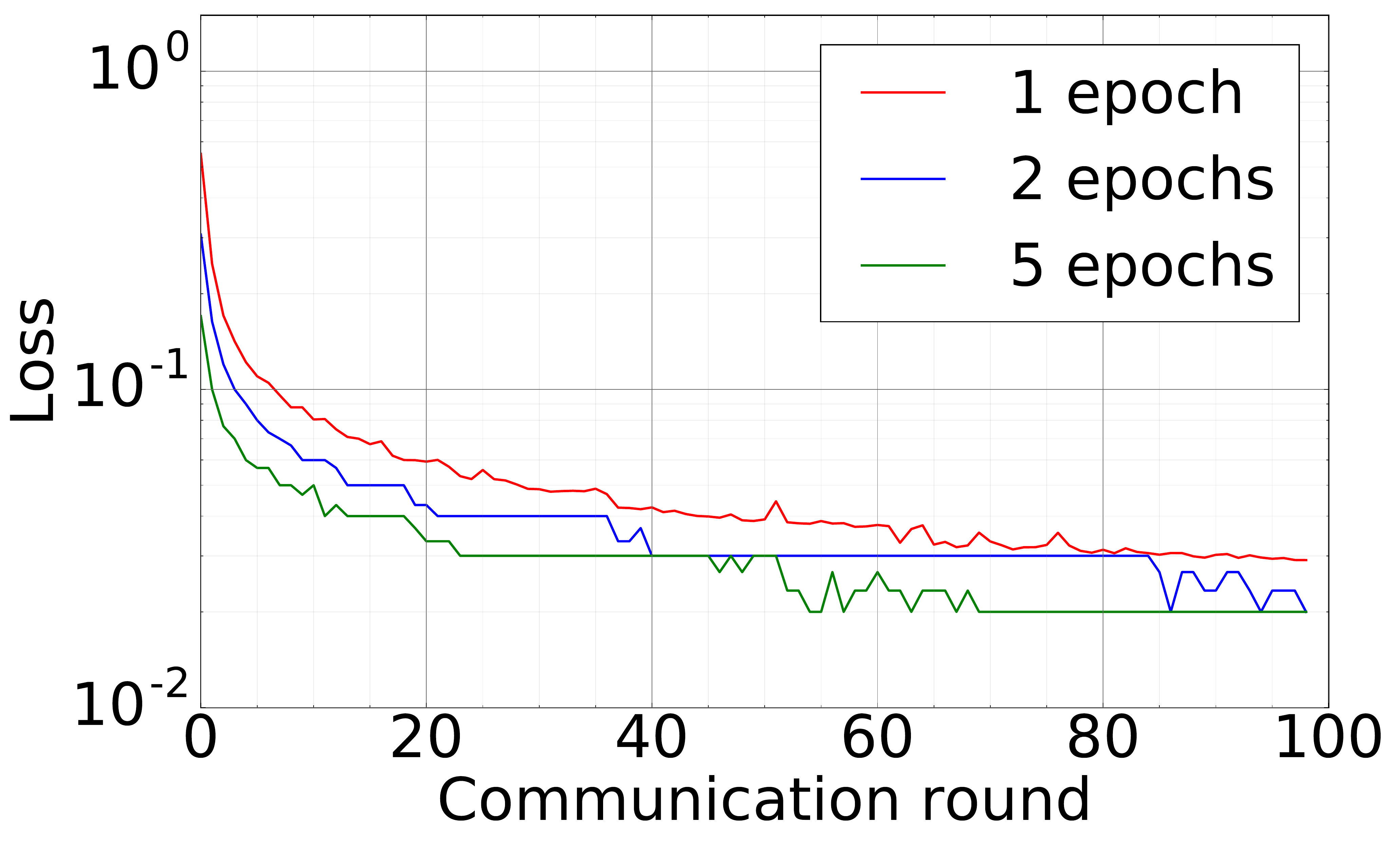}&
\hskip -0.45cm\includegraphics[width=0.32\linewidth]{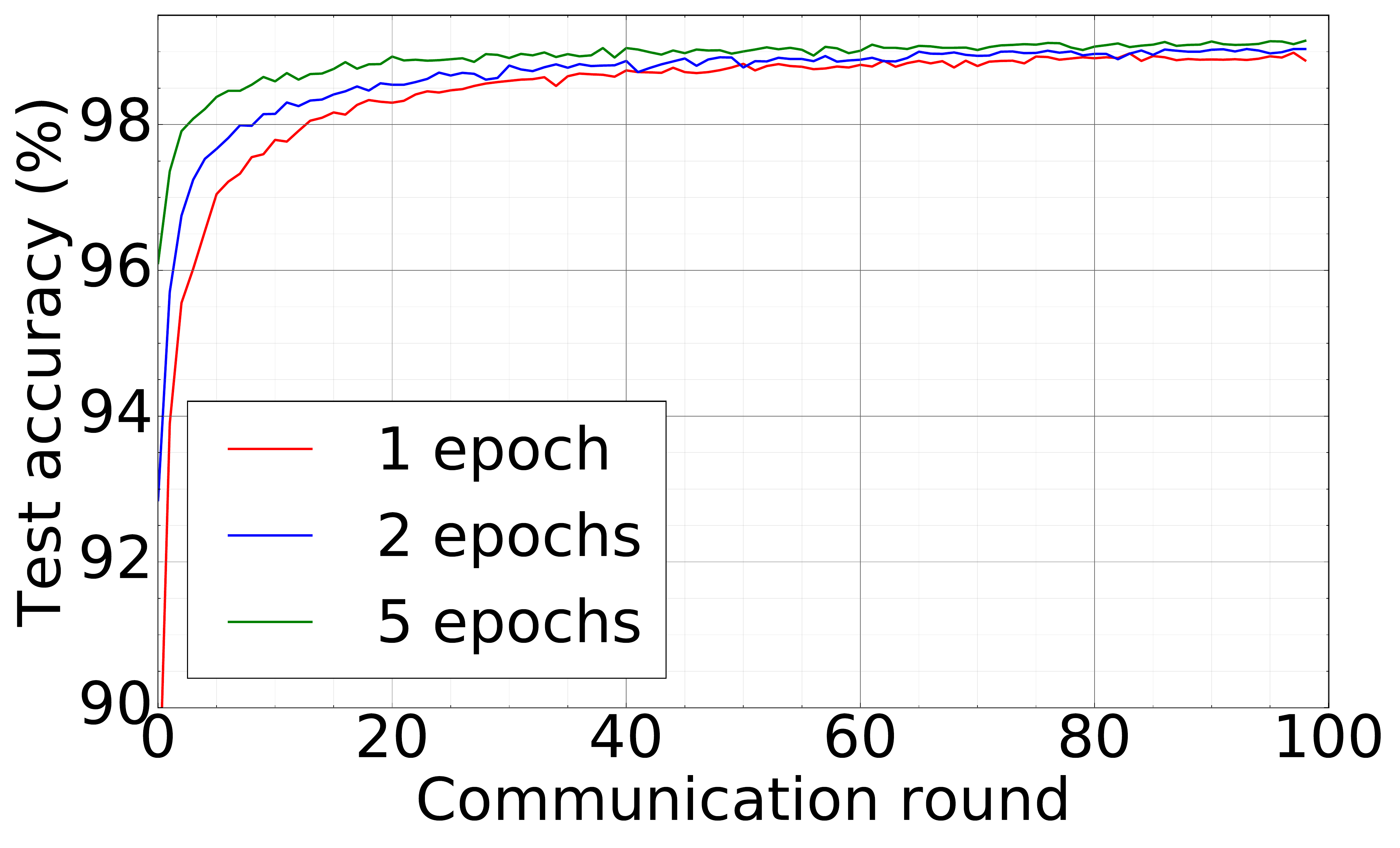}&
\hskip -0.45cm\includegraphics[width=0.32\linewidth]{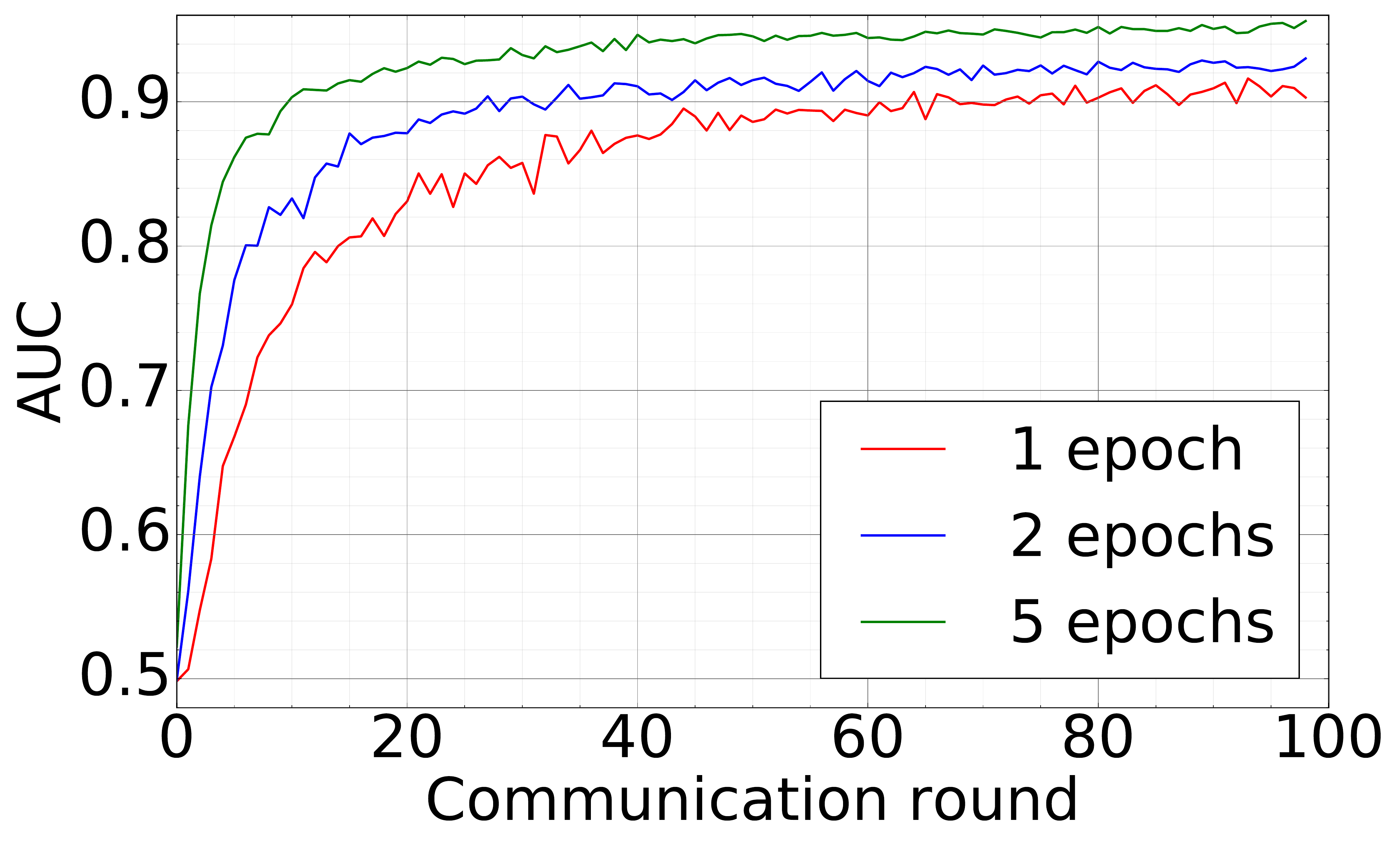}\\
{\footnotesize CR vs. Training loss} & {\footnotesize CR vs. Test acc} & {\footnotesize CR vs. AUC}\\
\end{tabular}
\vskip -0.3cm
\caption{Training CNN for IID MNIST classification with DFedAvgM using: different communication bits but fix local epoch to one (first row) and different local epochs but fix the communication bits to 16 (second row). Different quantized DFedAvgM performs almost similar, and more local epoch can accelerate training at the cost of faster privacy leakage. CR: communication round.
}
\label{fig:mnist-cnn-iid}
\end{figure}

\begin{figure}[t!]
\centering
\begin{tabular}{ccc}
\hskip -0.3cm\includegraphics[width=0.32\linewidth]{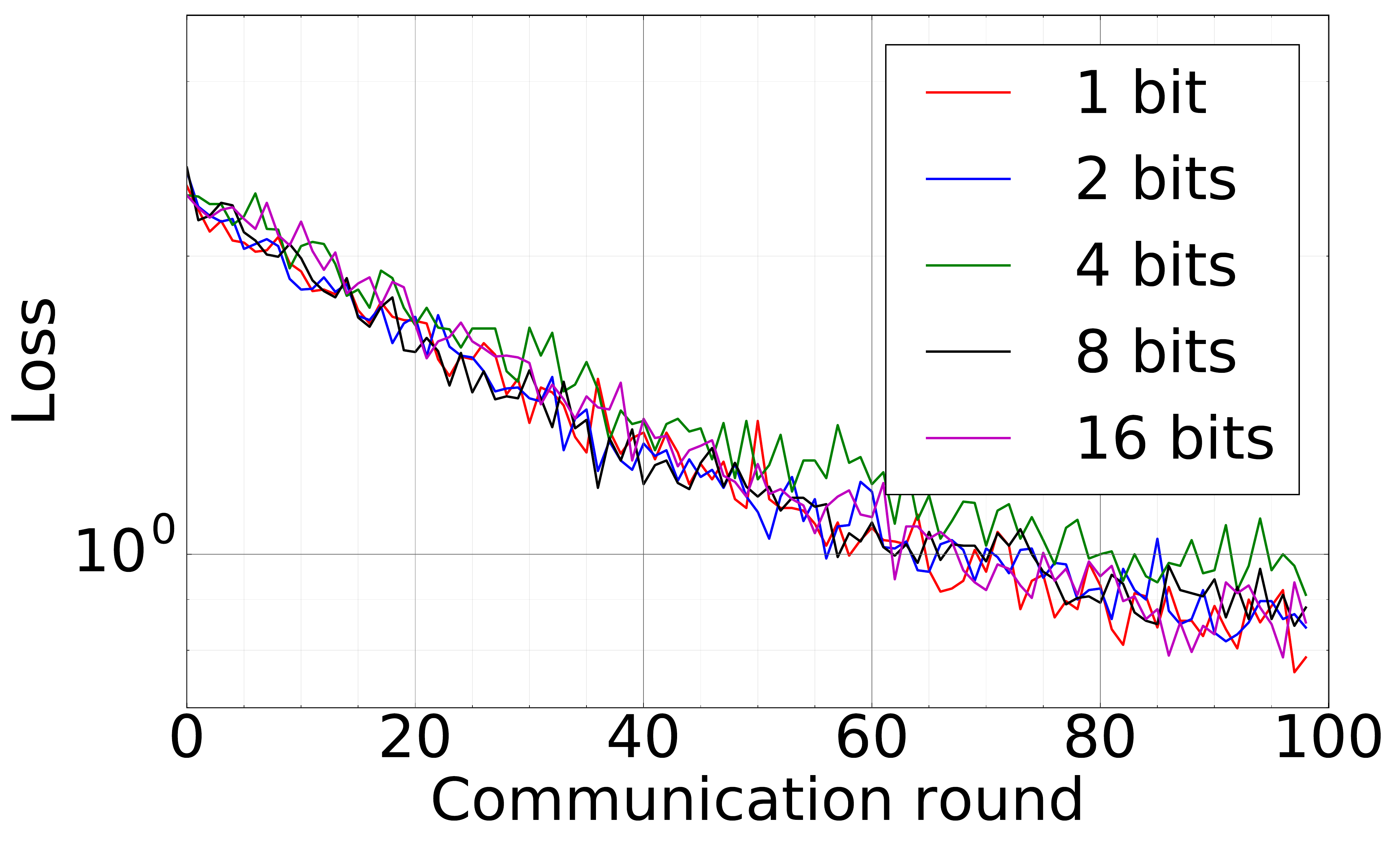}&
\hskip -0.45cm\includegraphics[width=0.32\linewidth]{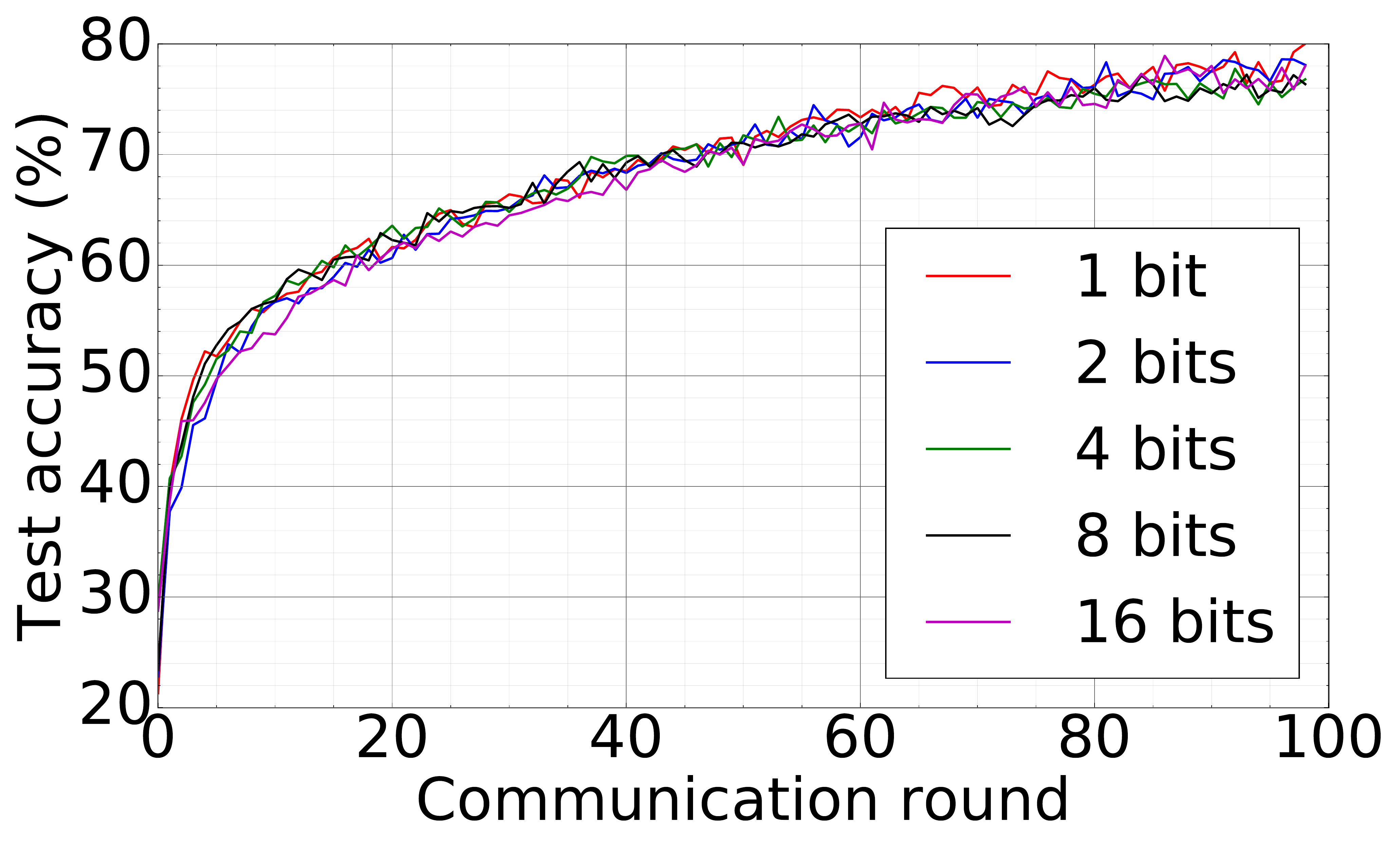}&
\hskip -0.45cm\includegraphics[width=0.32\linewidth]{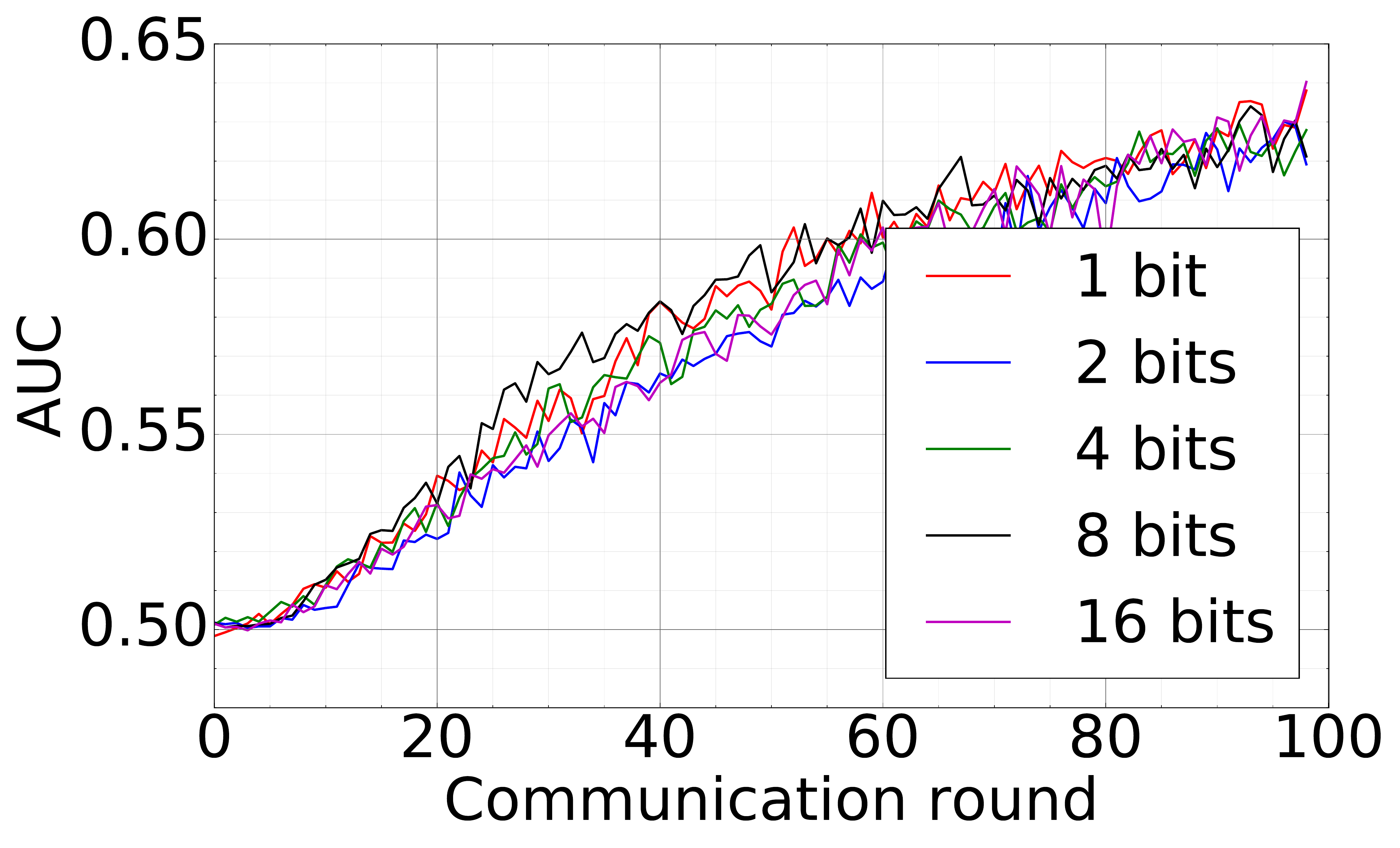}\\
{\footnotesize CR vs. Training loss} & {\footnotesize CR vs. Test acc} & {\footnotesize CR vs. AUC}\\
\hskip -0.3cm\includegraphics[width=0.32\linewidth]{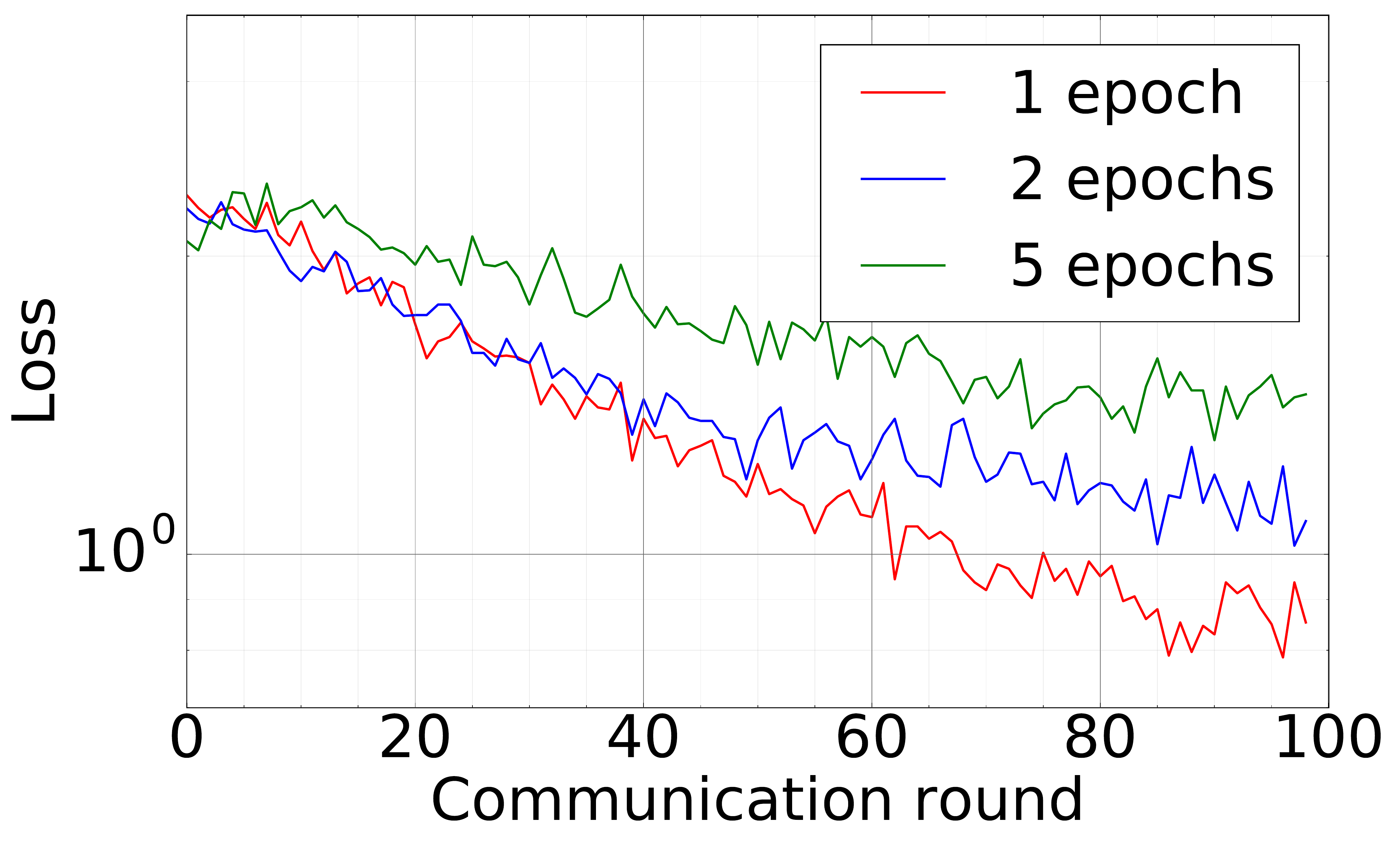}&
\hskip -0.45cm\includegraphics[width=0.32\linewidth]{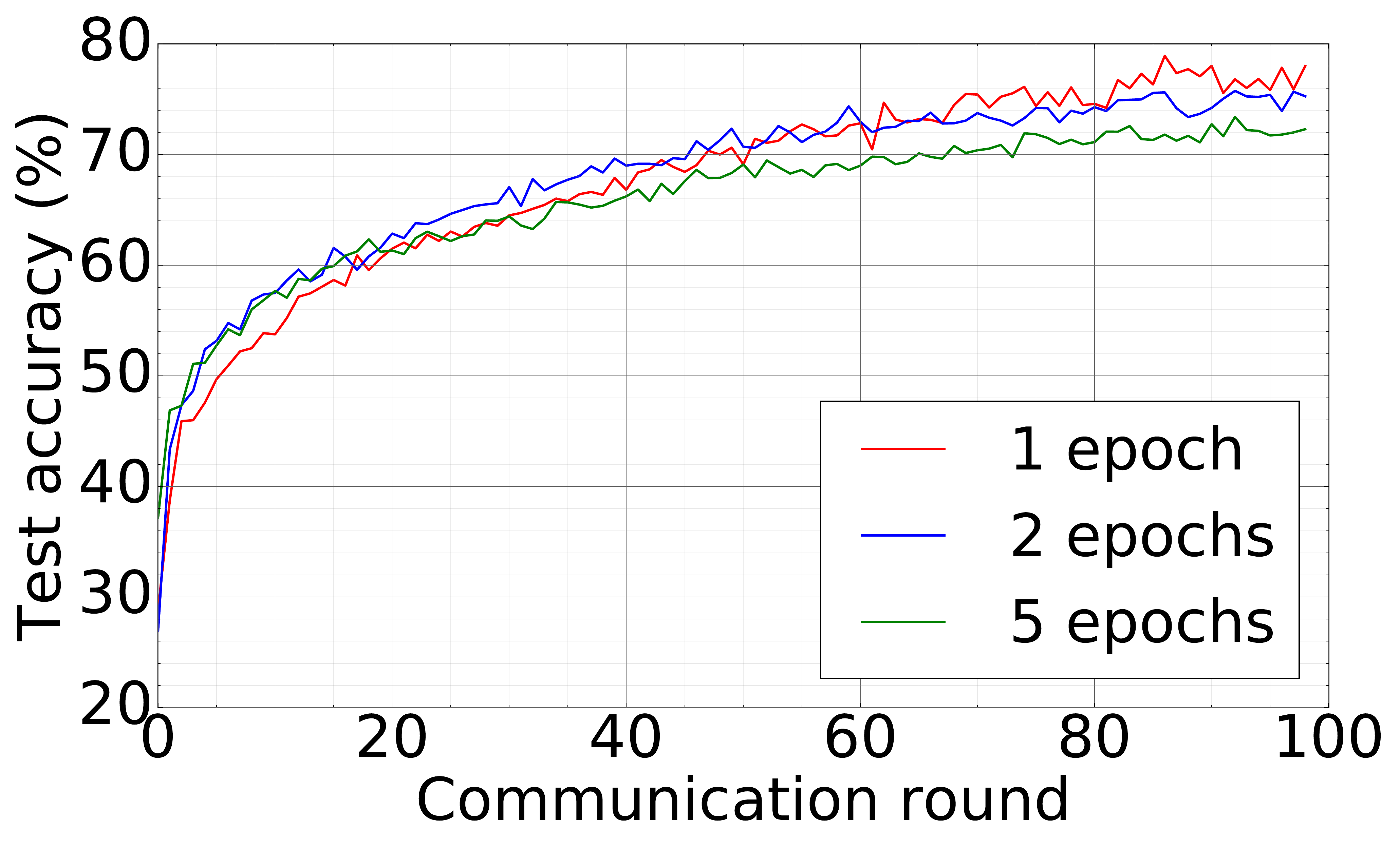}&
\hskip -0.45cm\includegraphics[width=0.32\linewidth]{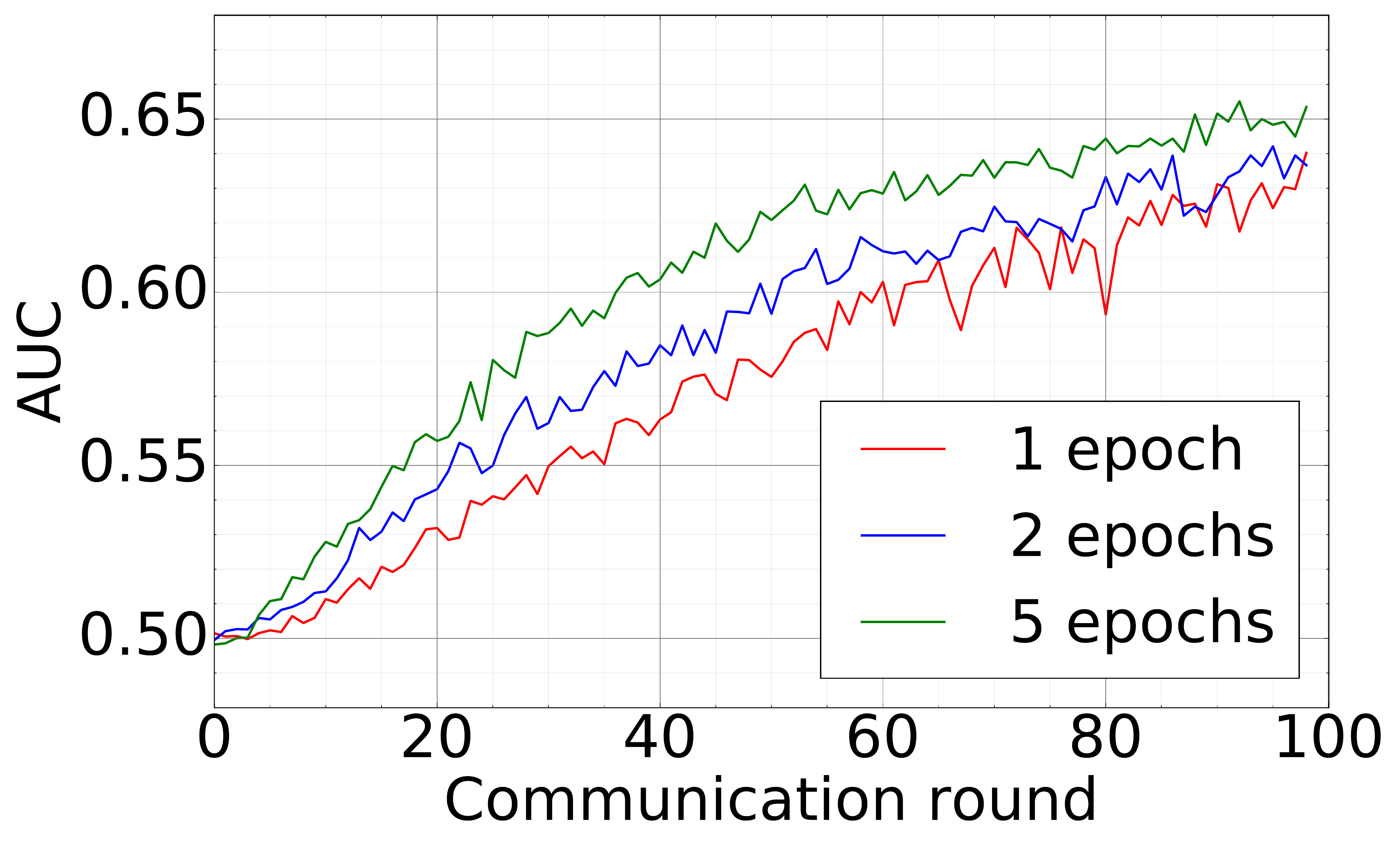}\\
{\footnotesize CR vs. Training loss} & {\footnotesize CR vs. Test acc} & {\footnotesize CR vs. AUC}\\
\end{tabular}
\vskip -0.3cm
\caption{Training CNN for Non-IID MNIST classification with DFedAvgM using: different communication bits but fix local epoch to one (first row) and different local epochs but fix the communication bits to 16 (second row). Different quantized DFedAvgM does not lead to much difference in performance. More local epoch does not help in accelerating training or protect data privacy.
}
\label{fig:mnist-cnn-noniid}
\end{figure}

\begin{figure}[t!]
\centering
\begin{tabular}{ccc}
\hskip -0.3cm\includegraphics[width=0.32\linewidth]{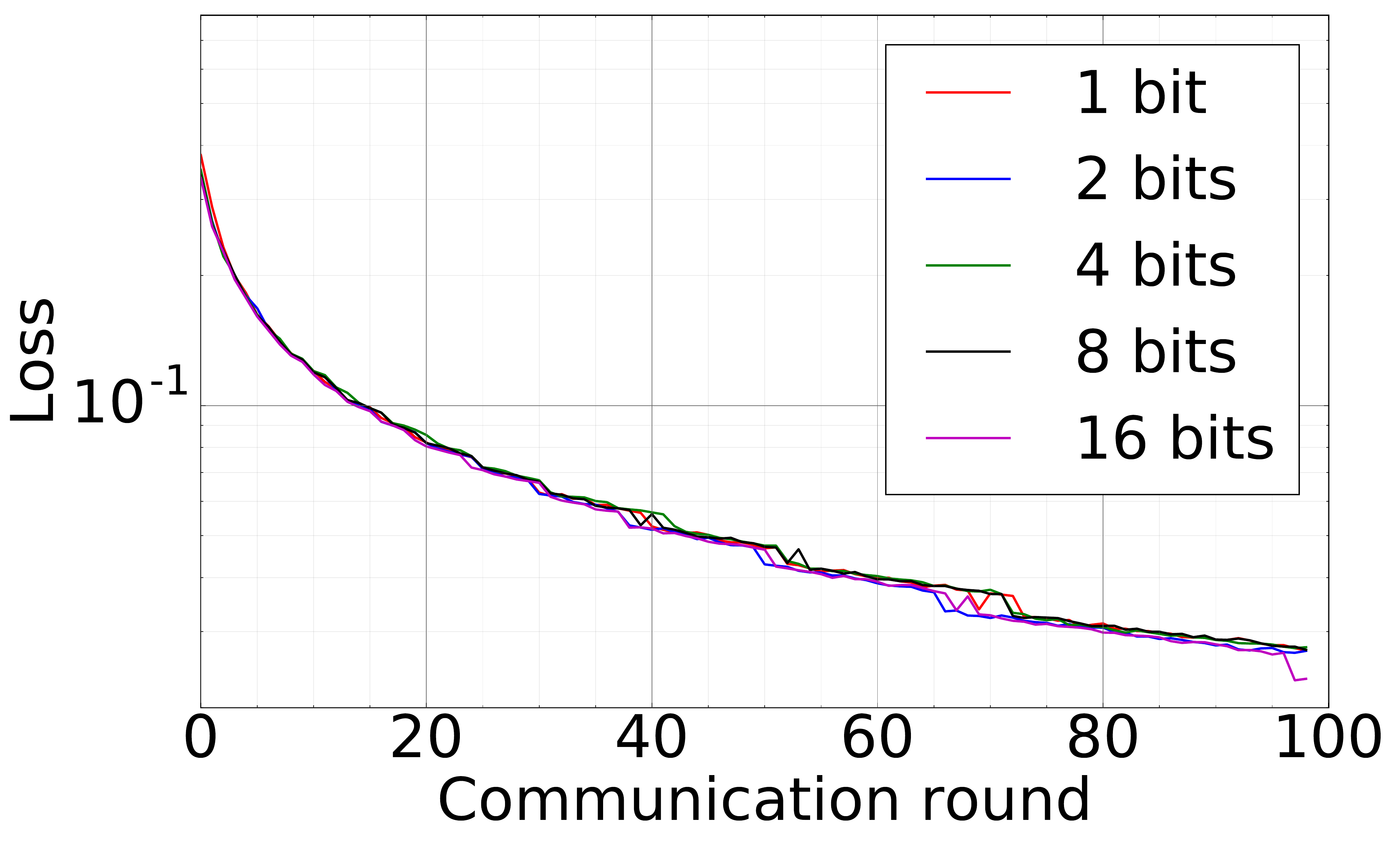}&
\hskip -0.45cm\includegraphics[width=0.32\linewidth]{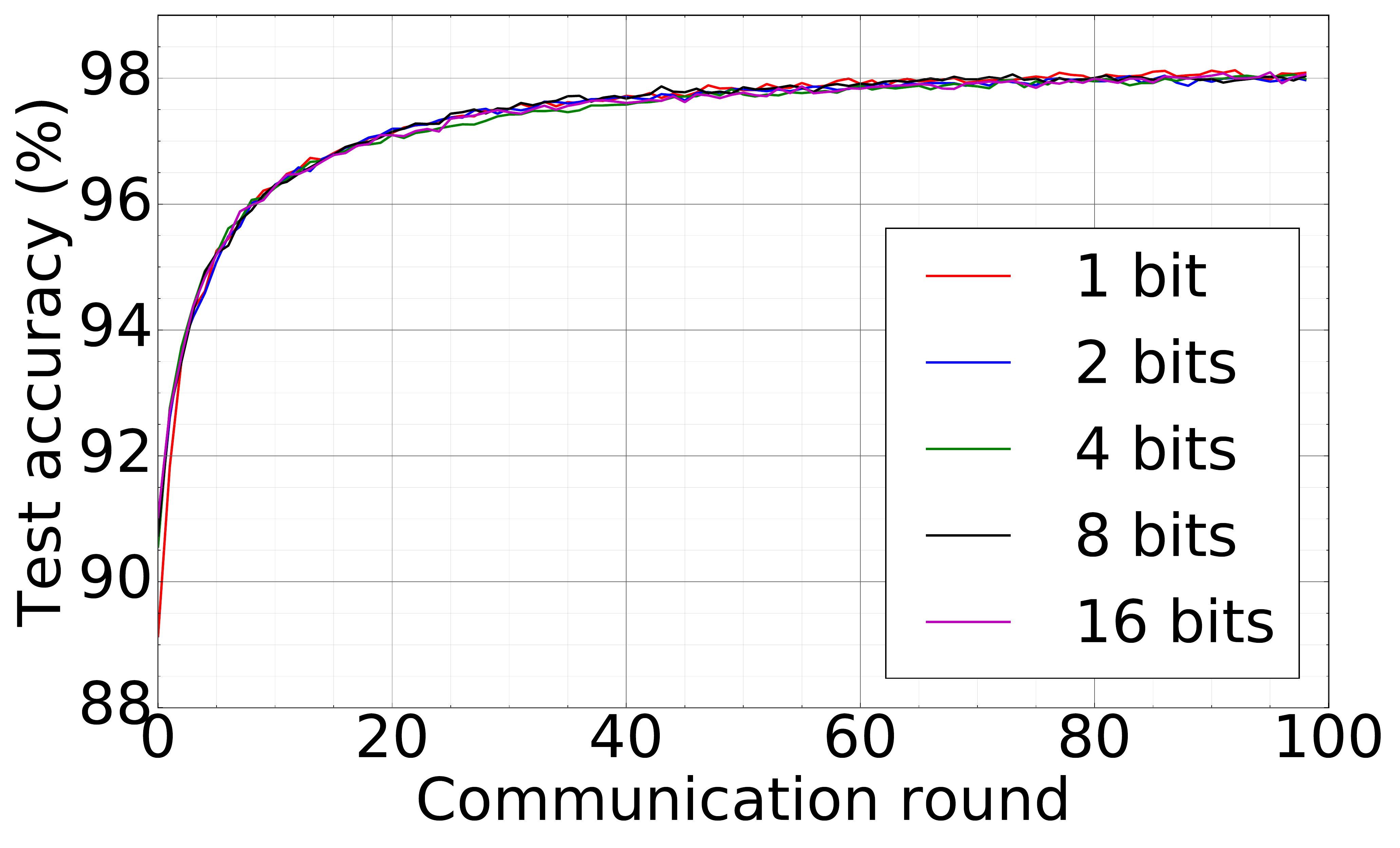}&
\hskip -0.45cm\includegraphics[width=0.32\linewidth]{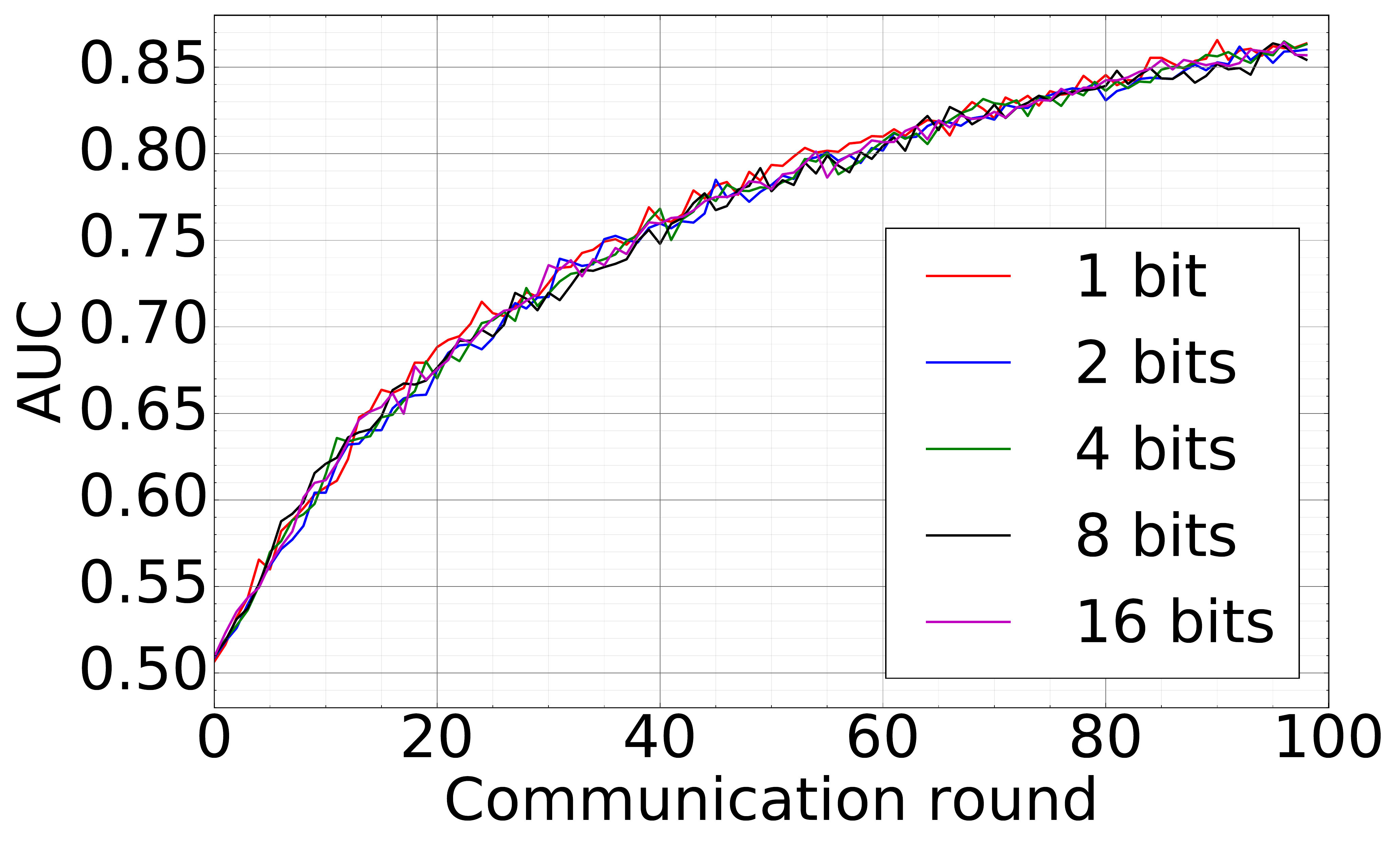}\\
{\footnotesize CR vs. Training loss} & {\footnotesize CR vs. Test acc} & {\footnotesize CR vs. AUC}\\
\hskip -0.3cm\includegraphics[width=0.32\linewidth]{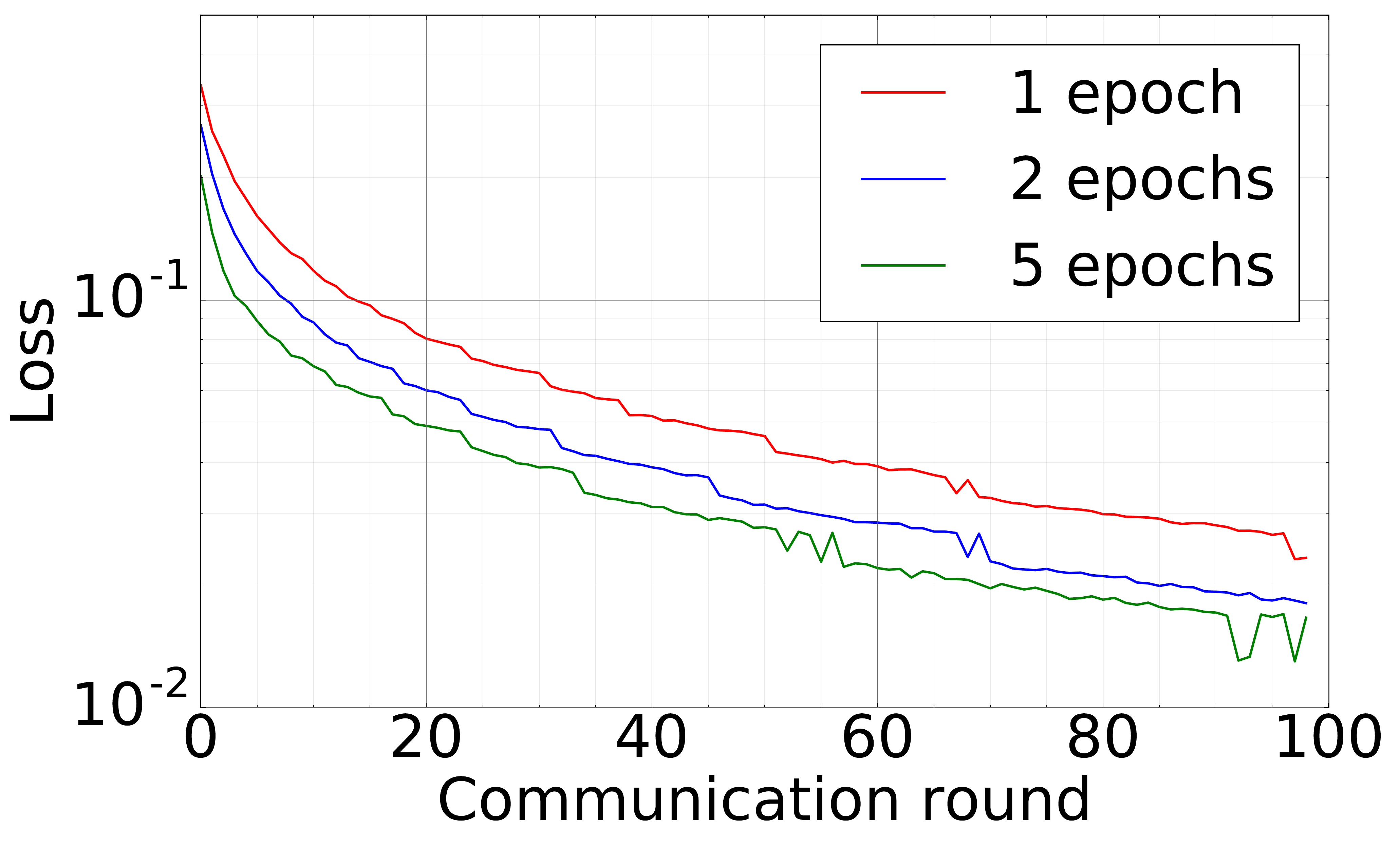}&
\hskip -0.45cm\includegraphics[width=0.32\linewidth]{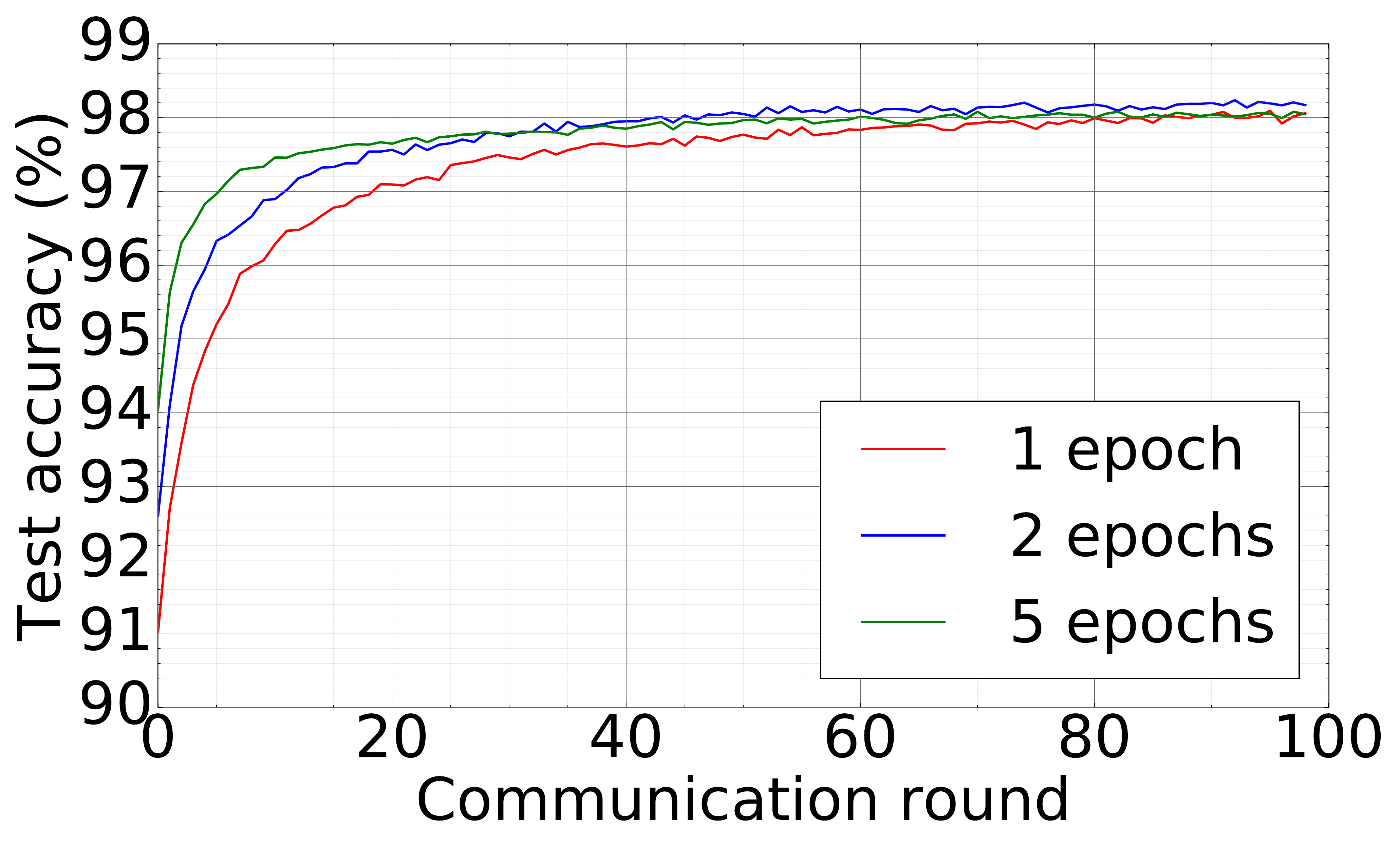}&
\hskip -0.45cm\includegraphics[width=0.32\linewidth]{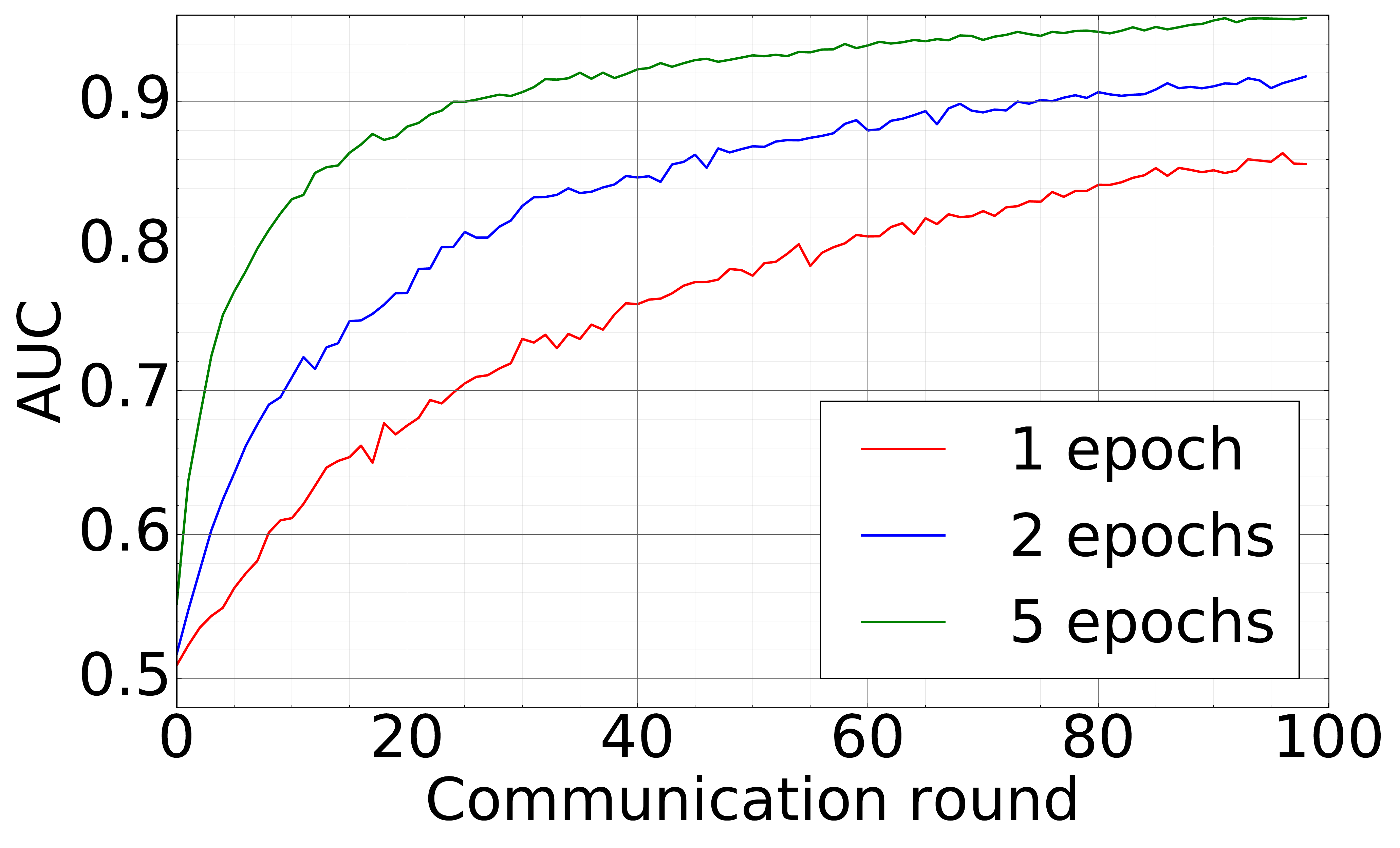}\\
{\footnotesize CR vs. Training loss} & {\footnotesize CR vs. Test acc} & {\footnotesize CR vs. AUC}\\
\end{tabular}
\vskip -0.3cm
\caption{Training 2NN for IID MNIST classification with DFedAvgM using: different communication bits but fix local epoch to one (first row) and different local epochs but fix the communication bits to 16 (second row). Different quantized DFedAvgM performs almost similar, and more local epoch can accelerate training at the cost of faster privacy leakage.
}
\label{fig:mnist-mlp-iid}
\end{figure}

\begin{figure}[t!]
\centering
\begin{tabular}{ccc}
\hskip -0.3cm\includegraphics[width=0.32\linewidth]{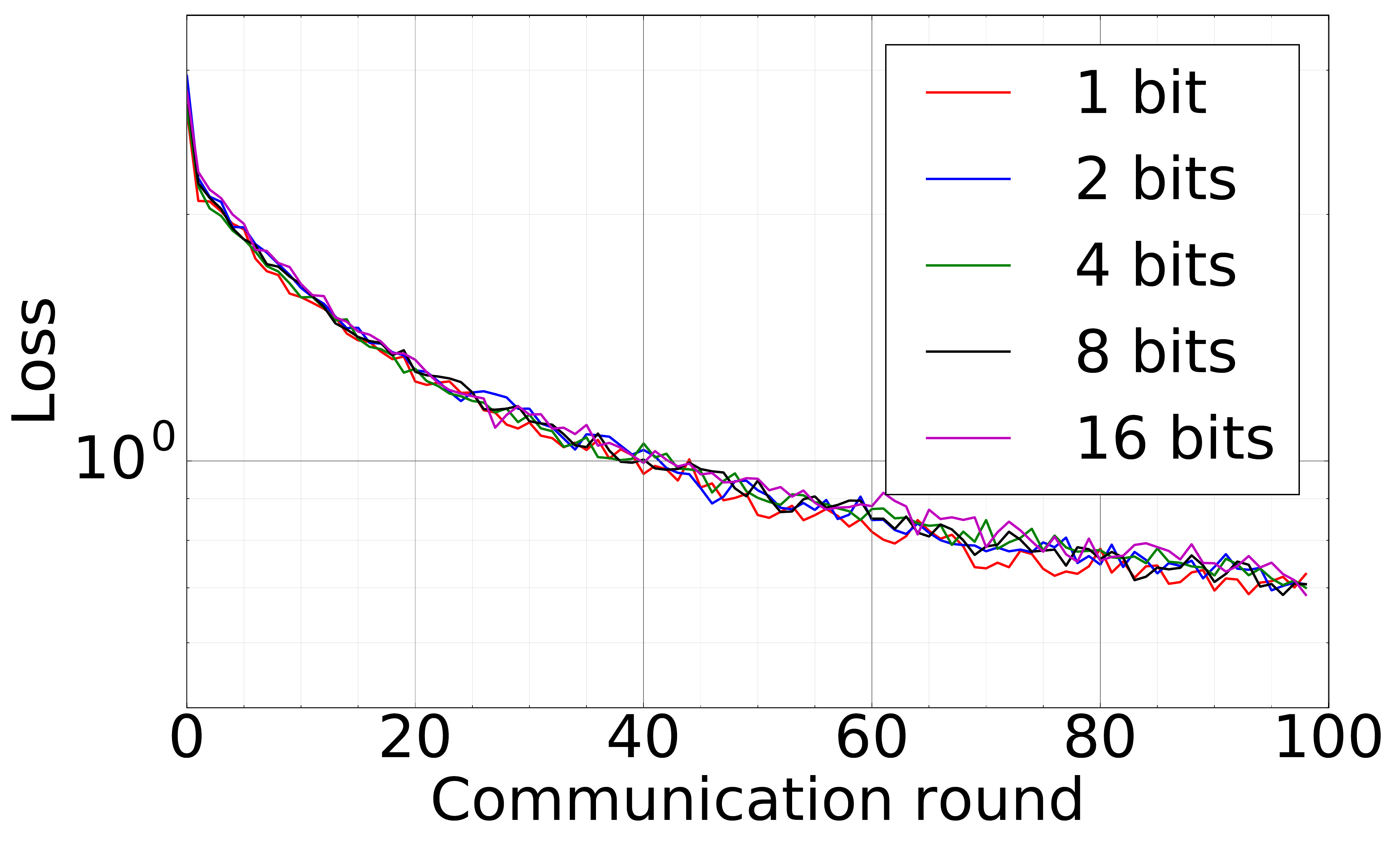}&
\hskip -0.45cm\includegraphics[width=0.32\linewidth]{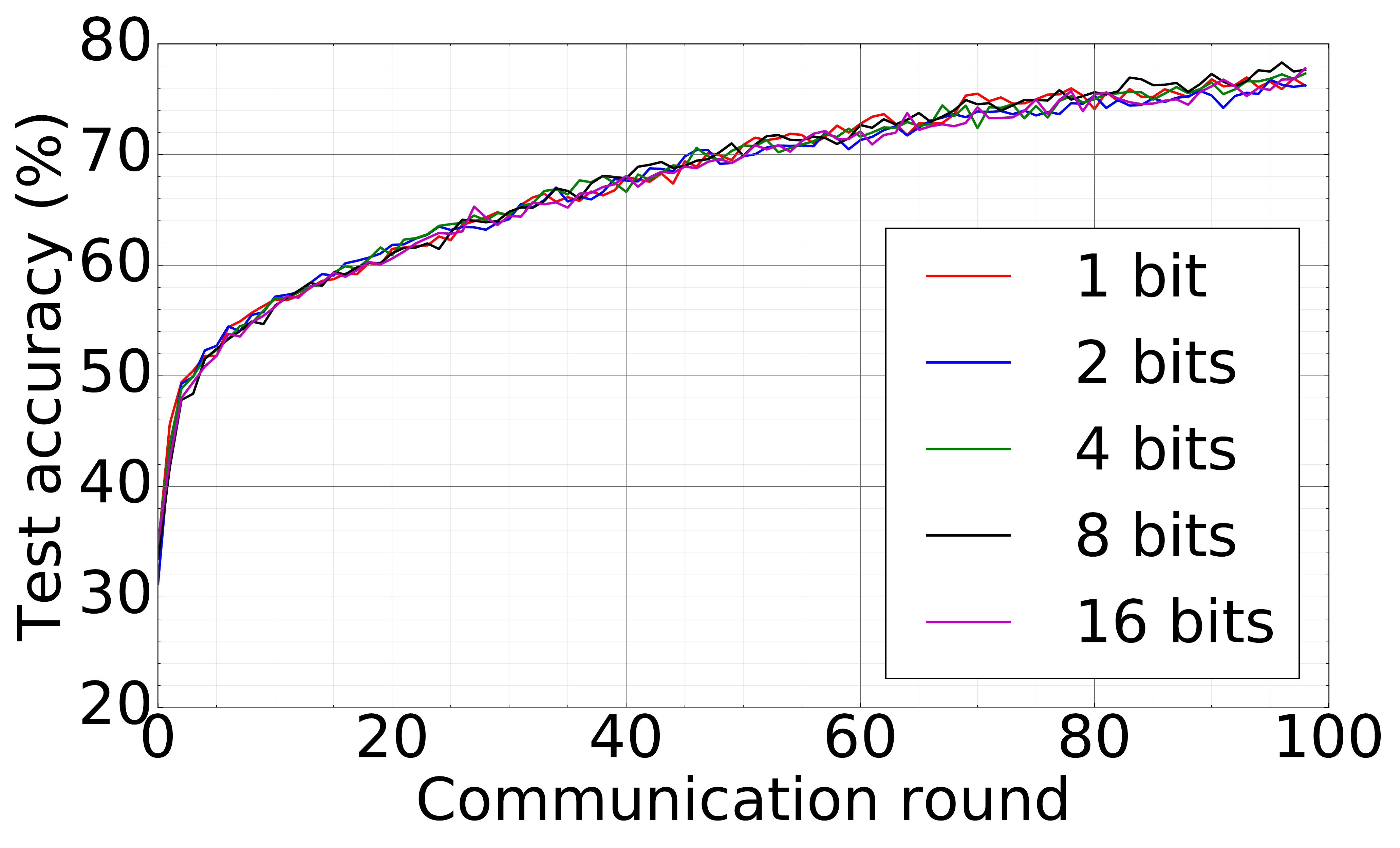}&
\hskip -0.45cm\includegraphics[width=0.32\linewidth]{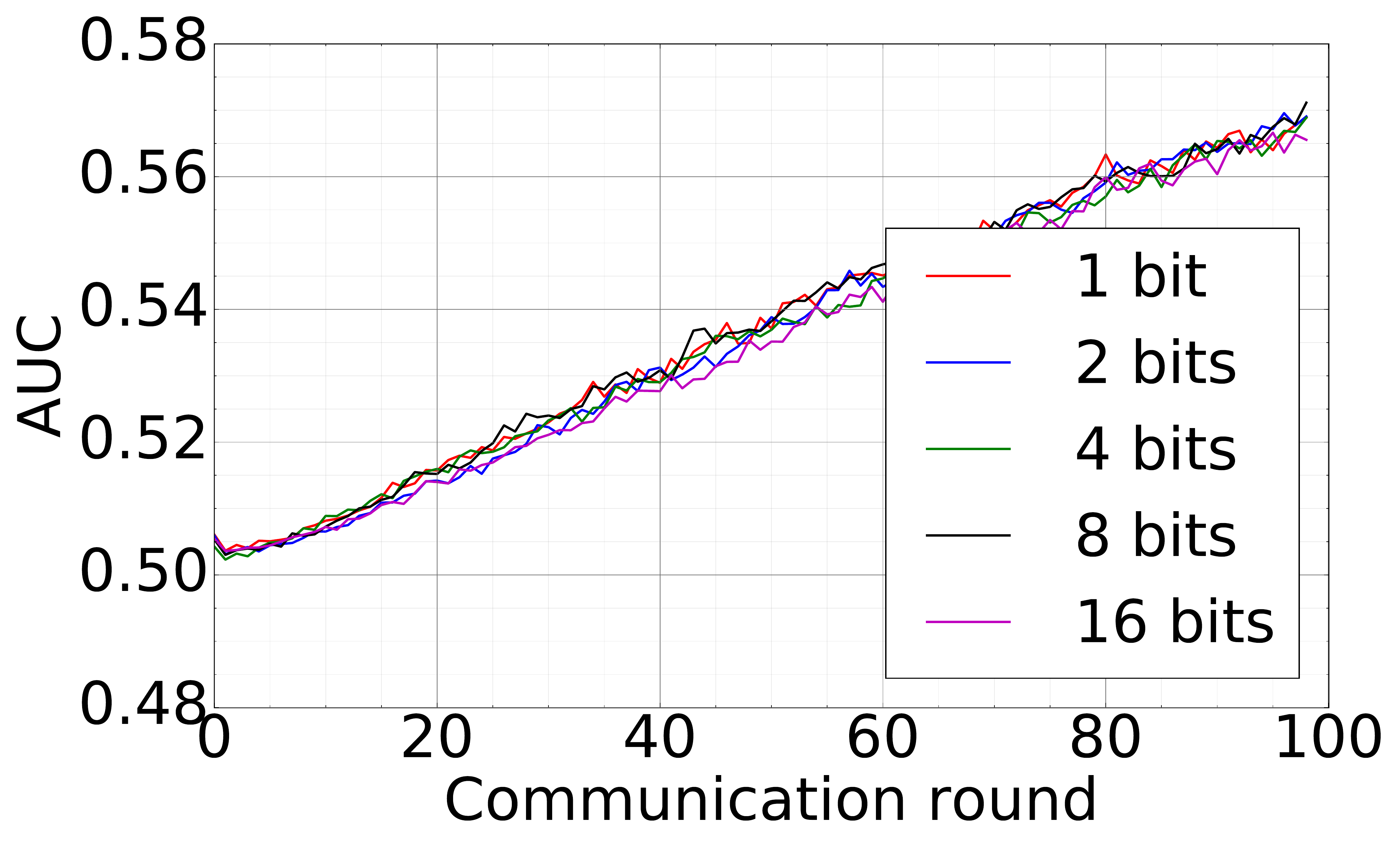}\\
{\footnotesize CR vs. Training loss} & {\footnotesize CR vs. Test acc} & {\footnotesize CR vs. AUC}\\
\hskip -0.3cm\includegraphics[width=0.32\linewidth]{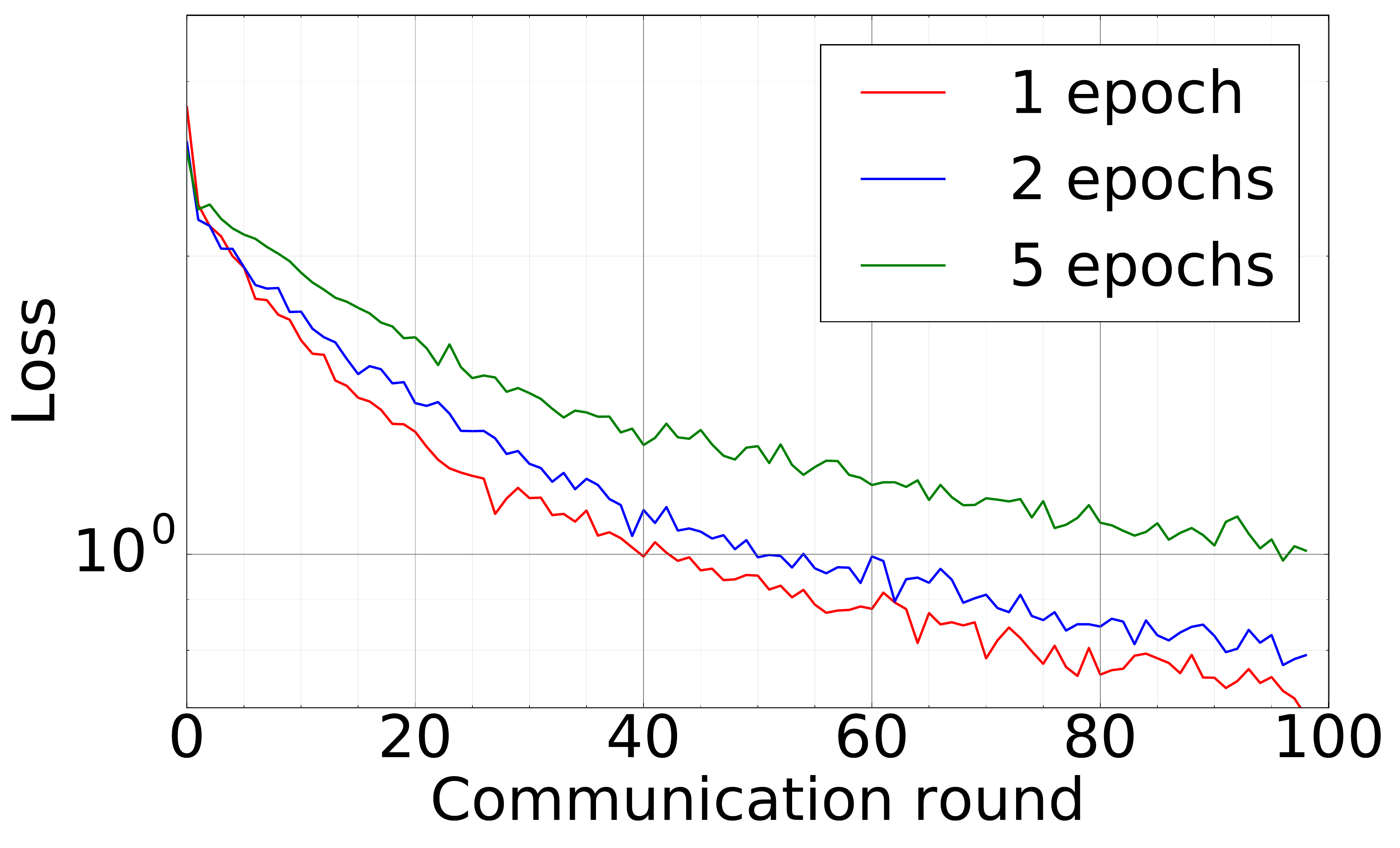}&
\hskip -0.45cm\includegraphics[width=0.32\linewidth]{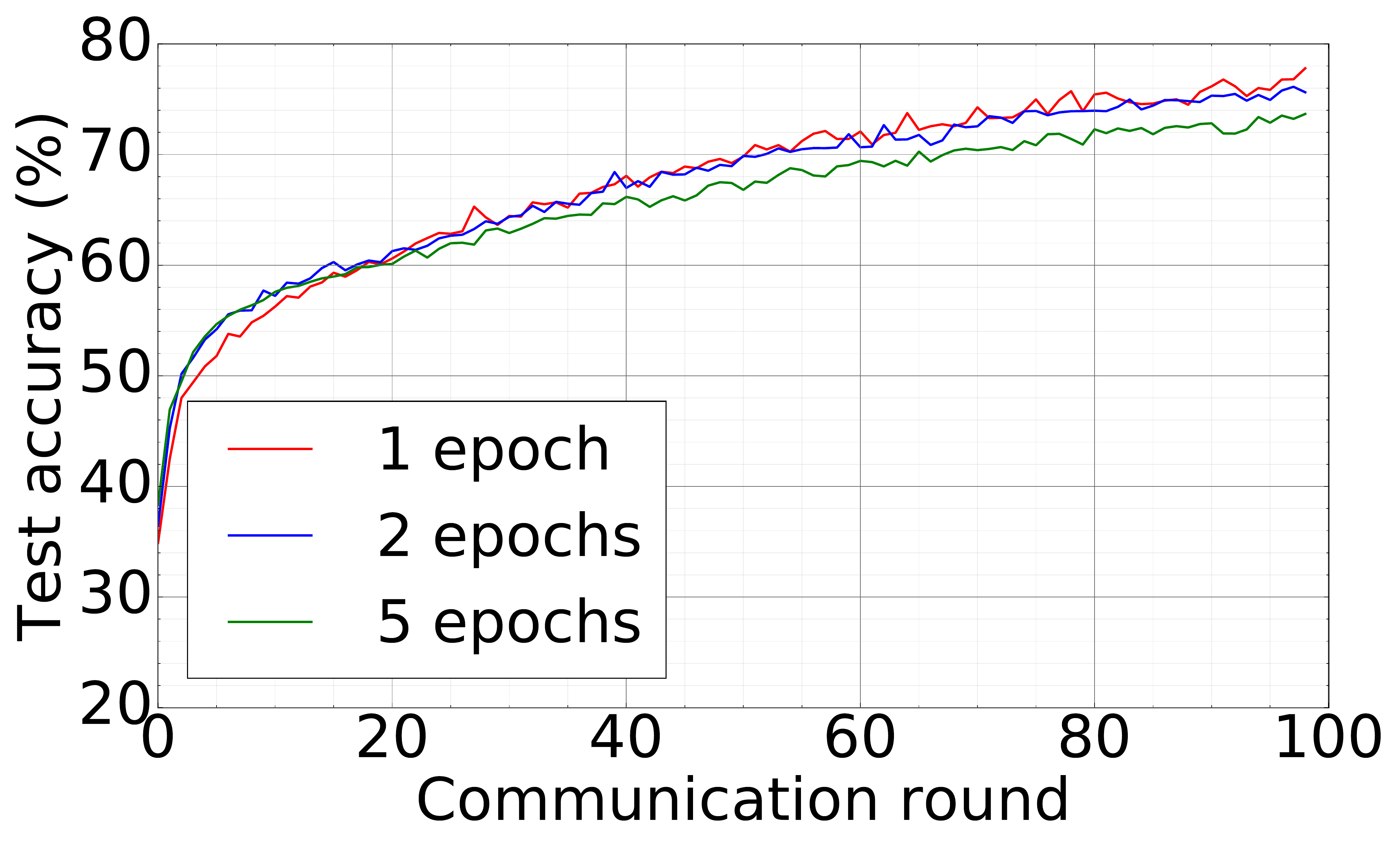}&
\hskip -0.45cm\includegraphics[width=0.32\linewidth]{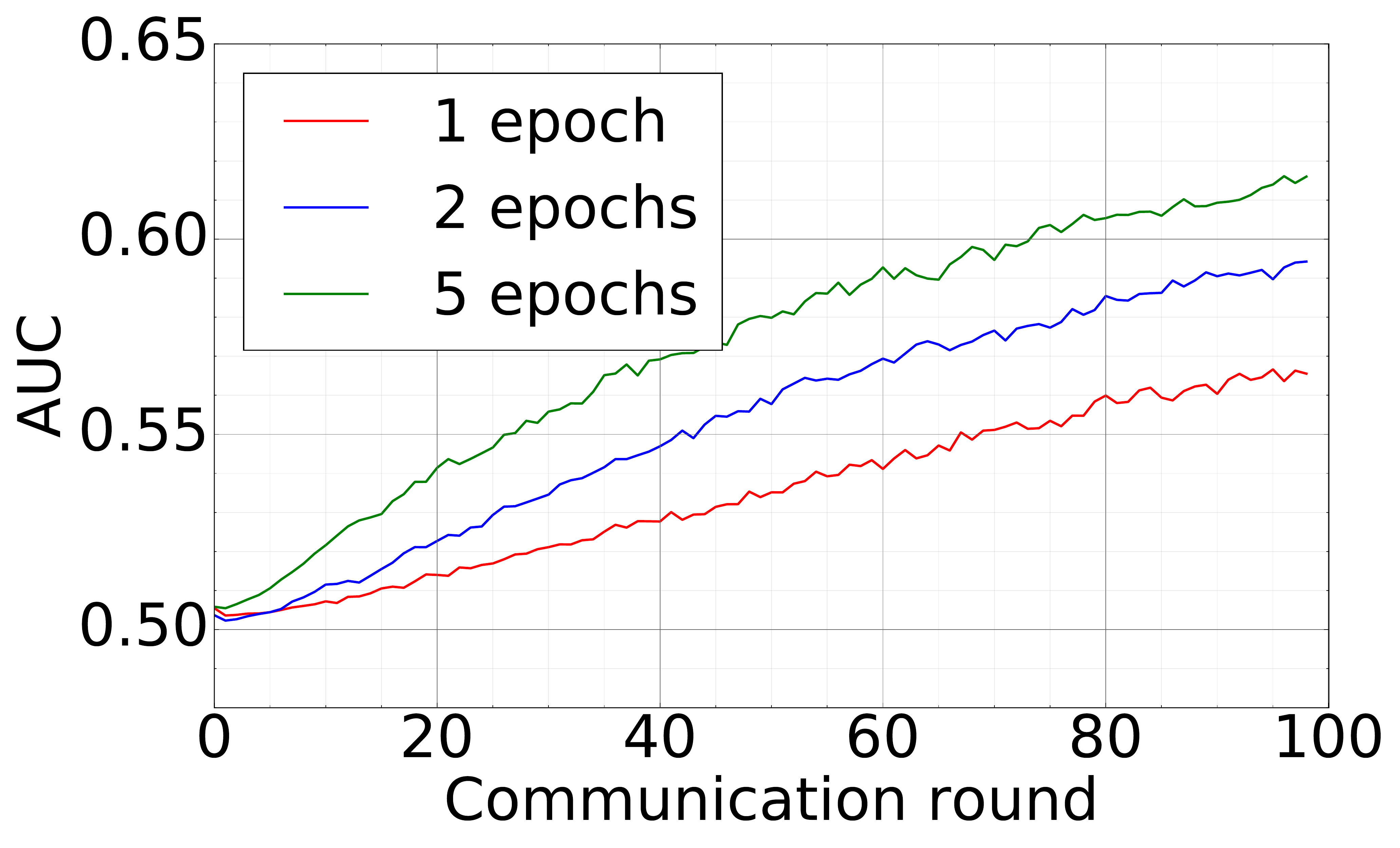}\\
{\footnotesize CR vs. Training loss} & {\footnotesize CR vs. Test acc} & {\footnotesize CR vs. AUC}\\
\end{tabular}
\vskip -0.3cm
\caption{Training 2NN for Non-IID MNIST classification with DFedAvgM using: different communication bits but fix local epoch to one (first row) and different local epochs but fix the communication bits to 16 (second row). Different quantized DFedAvgM does not lead to much difference in performance. More local epoch does not help in accelerating training or protect data privacy.
}
\label{fig:mnist-mlp-noniid}
\end{figure}

\paragraph{The efficiency of DFedAvgM.} We train two DNNs for MNIST classification using 100 clients: 1) A simple multilayer-perceptron with 2-hidden layers with 200 units each using ReLU activation (199,210 total parameters), which we refer to as 2NN. 2) A CNN with two $5\times 5$ convolution layers (the first with 32 channels, the second with 64, each followed with $2\times 2$ max pooling), a fully connected layer with 512 units and ReLU activation, and the final output layer (1,663,370 total parameters). We study two partitioning of the MNIST data over clients, i.e., IID and Non-IID. In IID setting, the data is shuffled, and then partitioned into 20 clients each receiving 3000 examples. In Non-IID, we first sort the data by digit label, divide it into 40 shards of size 1500, and assign each of 20 clients 2 shards. In training, we set the local batch size (batch size of the training data on clients) to be 50, learning rate 0.01, and momentum 0.9. Figures~\ref{fig:mnist-cnn-iid} and \ref{fig:mnist-cnn-noniid} show the results of training CNN for MNIST classification (Fig.~\ref{fig:mnist-cnn-iid}: IID and Fig.~\ref{fig:mnist-cnn-noniid} Non-IID) by DFedAvgM using different communication bits and different local epochs. These results confirm the efficiency of DFedAvgM for training DNNs; in particular, when the clients' data are IID. For both IID and Non-IID settings, the communication bits do not affect the performance of DFedAvgM; as we see that the training loss, test accuracy, and AUC under the membership inference attack are almost identical. Increasing local training epochs can accelerate training for IID setting at the cost of faster privacy leakage. However, for Non-IID, increasing local training epochs does not help DFedAvgM in either training or privacy protection. Training 2NN by DFedAvgM behaves similarly, see Figs.~\ref{fig:mnist-mlp-iid} and \ref{fig:mnist-mlp-noniid}.

\begin{figure}[t!]
\centering
\begin{tabular}{cc}
\hskip -0.1cm\includegraphics[width=0.36\linewidth]{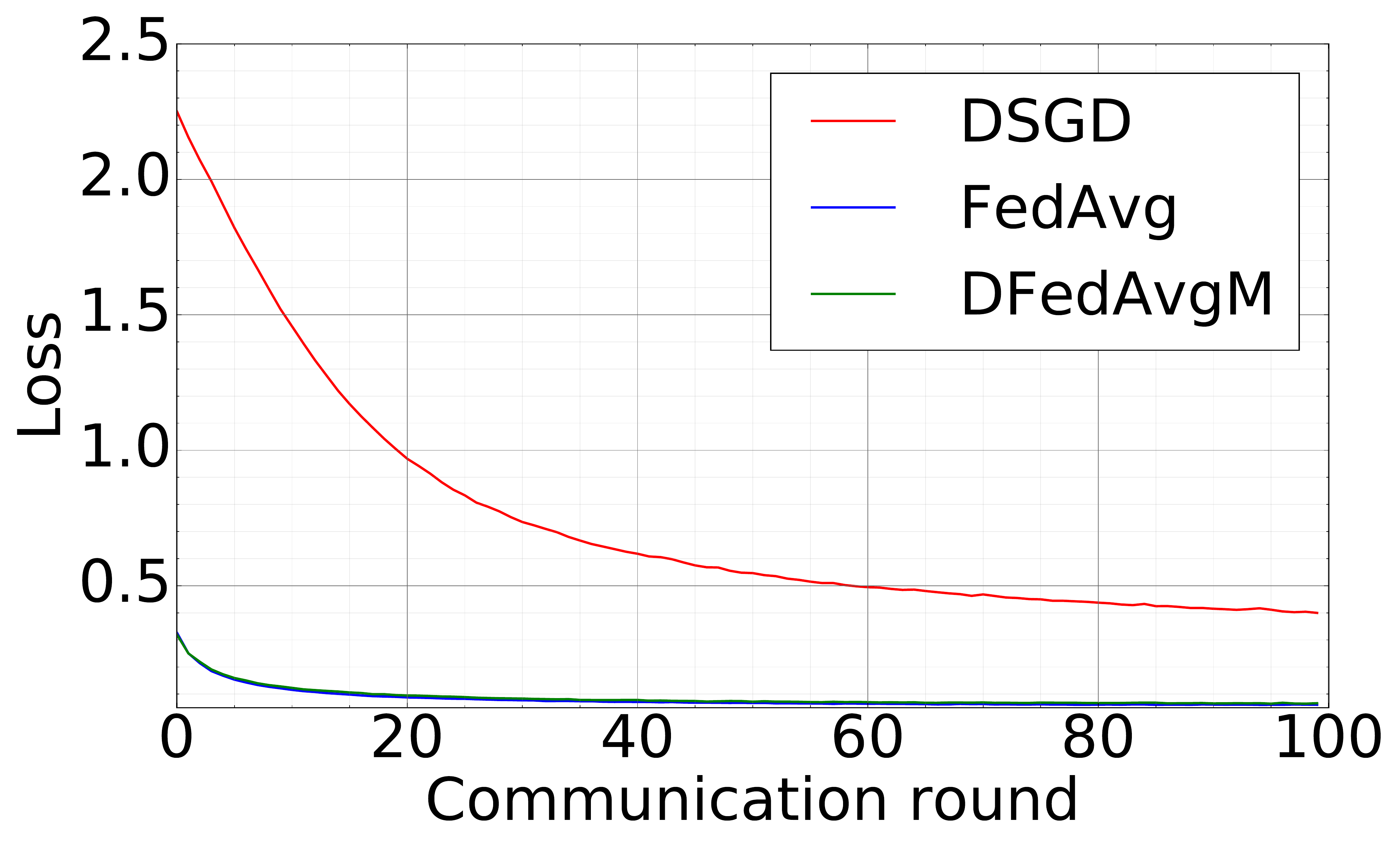}&
\hskip -0.4cm\includegraphics[width=0.36\linewidth]{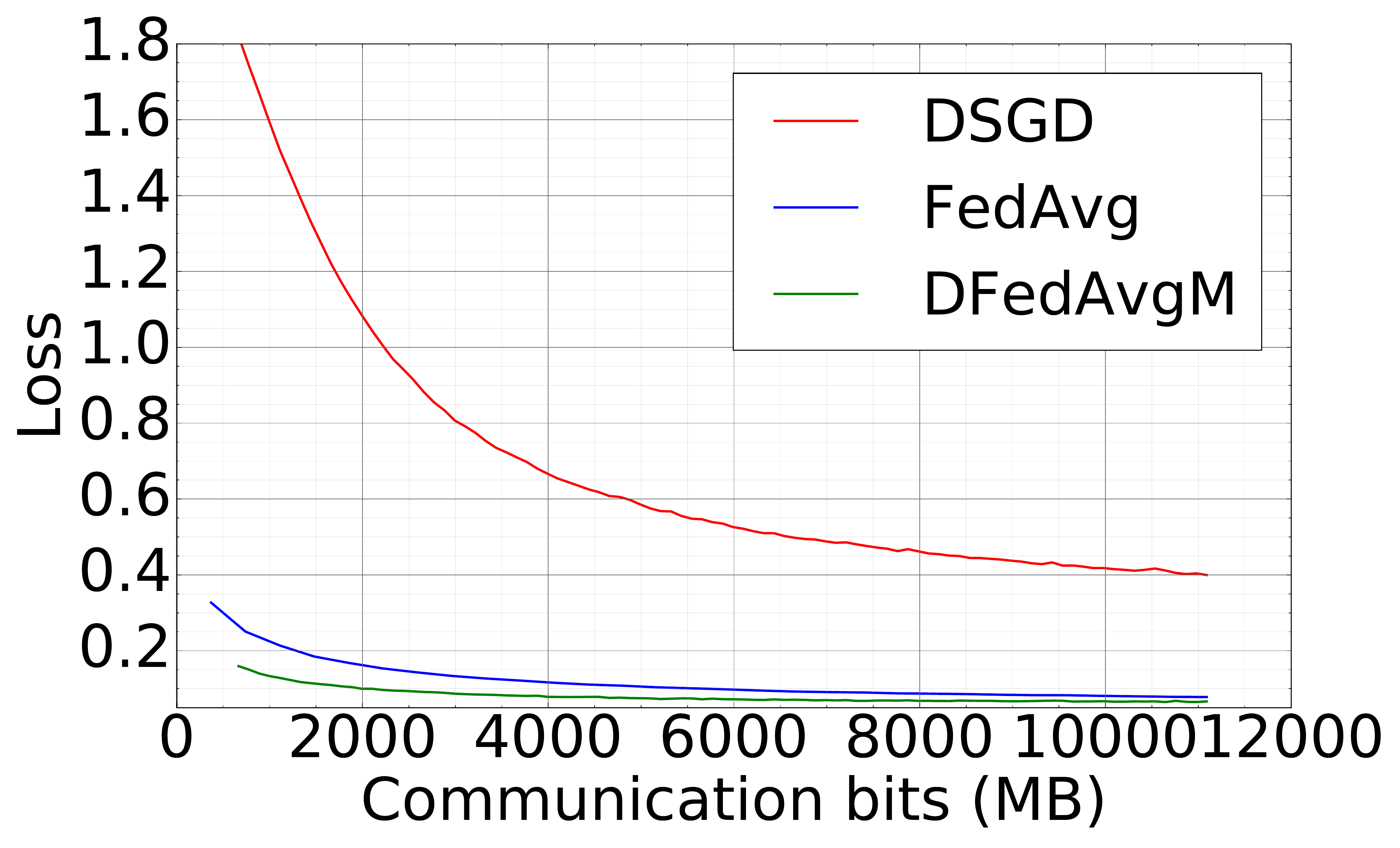}\\
{\footnotesize CR vs. Test loss}  & {\footnotesize CB vs. Test loss} \\
\hskip -0.1cm\includegraphics[width=0.36\linewidth]{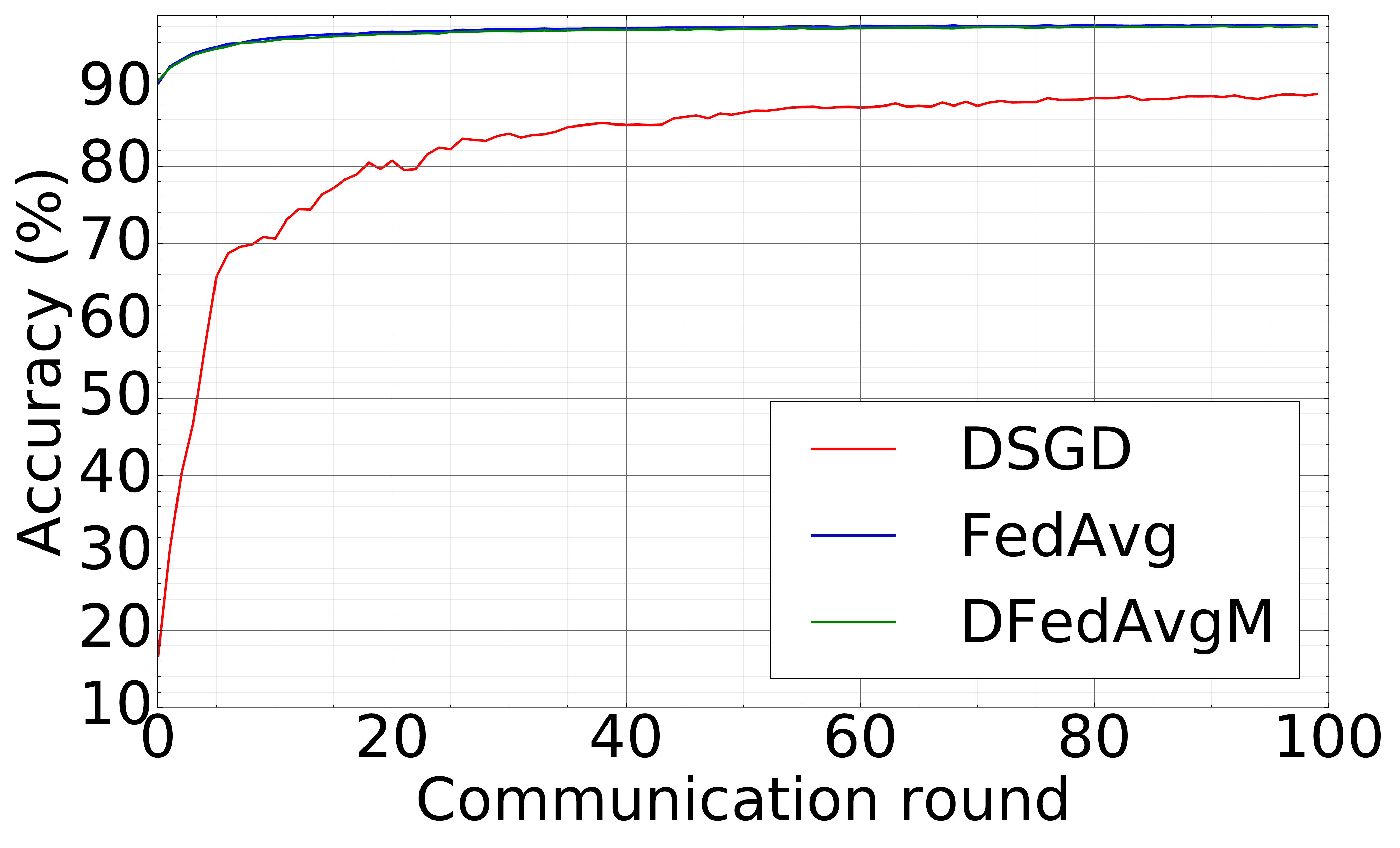}&
\hskip -0.4cm\includegraphics[width=0.36\linewidth]{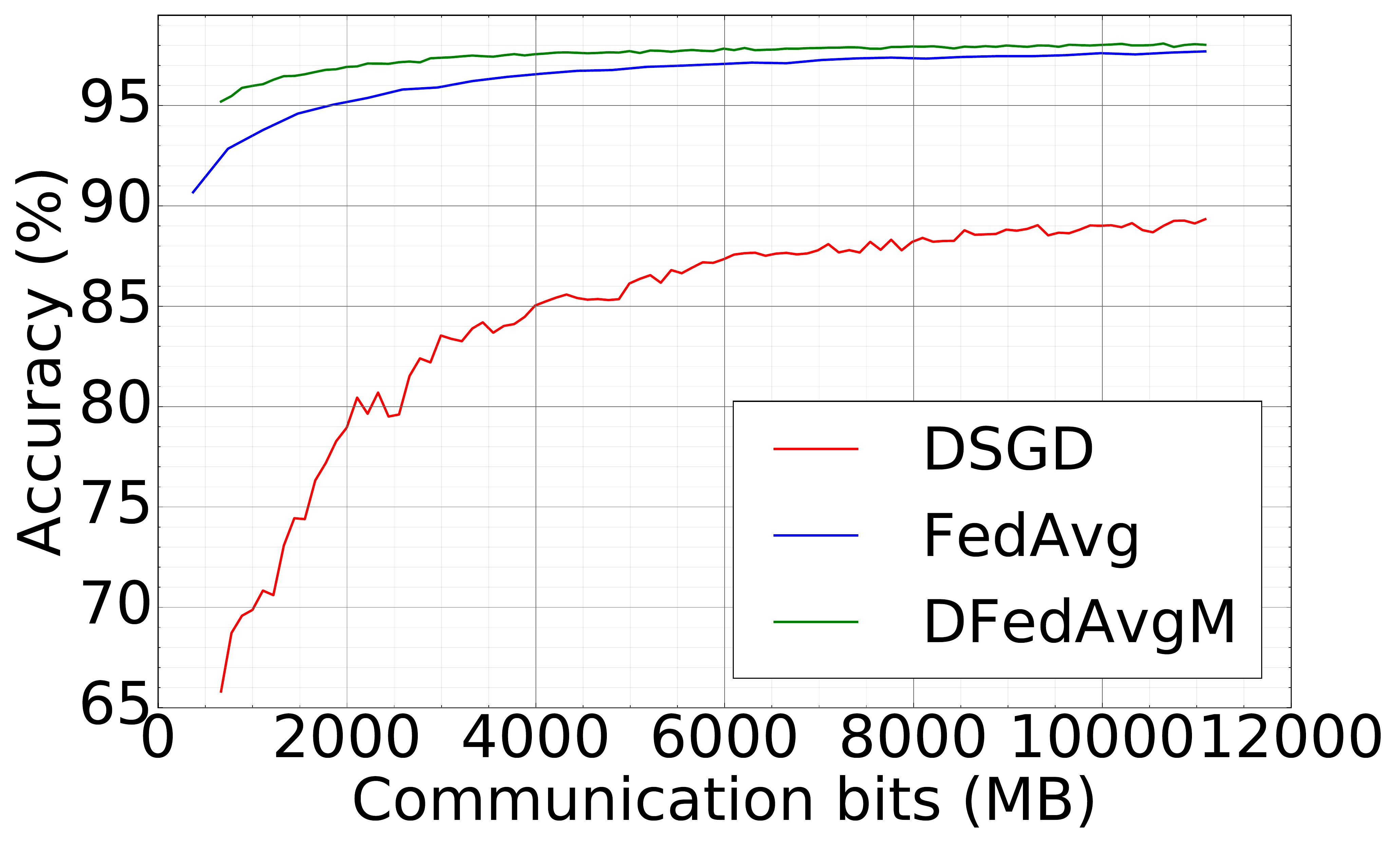}\\
{\footnotesize CR vs. Test acc} & {\footnotesize CB vs. Test acc}\\
\end{tabular}
\vskip -0.3cm
\caption{The efficiency comparison between DSGD, FedAvg, and DFedAvgM in training 2NN for MNIST classification. (a) and (c): test loss and test accuracy vs. communication round. (b) and (d): test loss and test accuracy vs. communication bits. DFedAvgM performs on par with FedAvg in terms of communication rounds, but DFedAvgM is significantly more efficient than FedAvg from the communication cost viewpoint. CR: communication round; CB: communication bits.
}
\label{fig:compare}
\end{figure}

\paragraph{Comparison between DFedAvgM, FedAvg, and DSGD.} Now, we compare the DFedAvgM, FedAvg, and DSGD in training 2NNs for MNIST classification. We use the same local batch size 50 for both FedAvg and DSGD, and the learning rates are both set to 0.1\footnote{We note that DFedAvg requires smaller learning rates than FedAvg and DSGD for numerical convergence.}. For FedAvg, we select all clients to get involved in training and communication in each round. Figure~\ref{fig:compare} compares three algorithms in terms of test loss and test accuracy for IID MNIST over communication round and communication cost. In terms of communication rounds, DFedAvgM converges as fast as FedAvg, and both are much faster than DSGD. DFedAvgM has a significant advantage over FedAvg and DSGD in communication costs. For Non-IID MNIST, training 2NN by FedAvg can achieve 96.81\% test accuracy, but both DFedAvg and DSGD cannot bypass 85\%. This disadvantage is because both DSGD and DFedAvgM only communicate with their neighbors, while the neighbors and itself may not contain enough training data to cover all possible classes. One feasible solution to resolve the issues of DFedAvgM for the Non-IID setting is by designing a new graph structure for more efficient global communication.

\subsection{LSTM for Language Modeling}
\begin{figure}[t!]
\centering
\begin{tabular}{cc}
\hskip -0.1cm\includegraphics[width=0.36\linewidth]{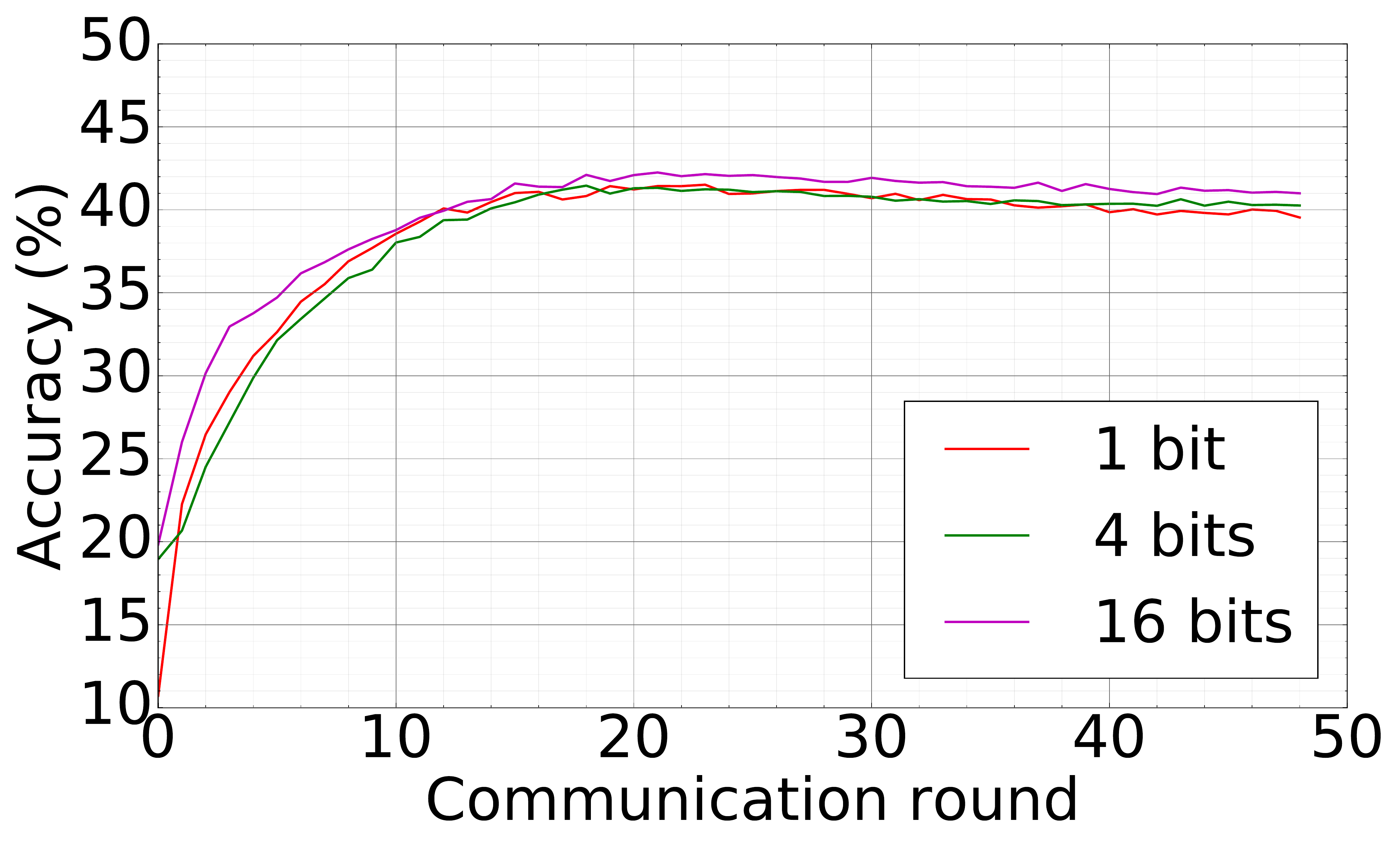}&
\hskip -0.4cm\includegraphics[width=0.36\linewidth]{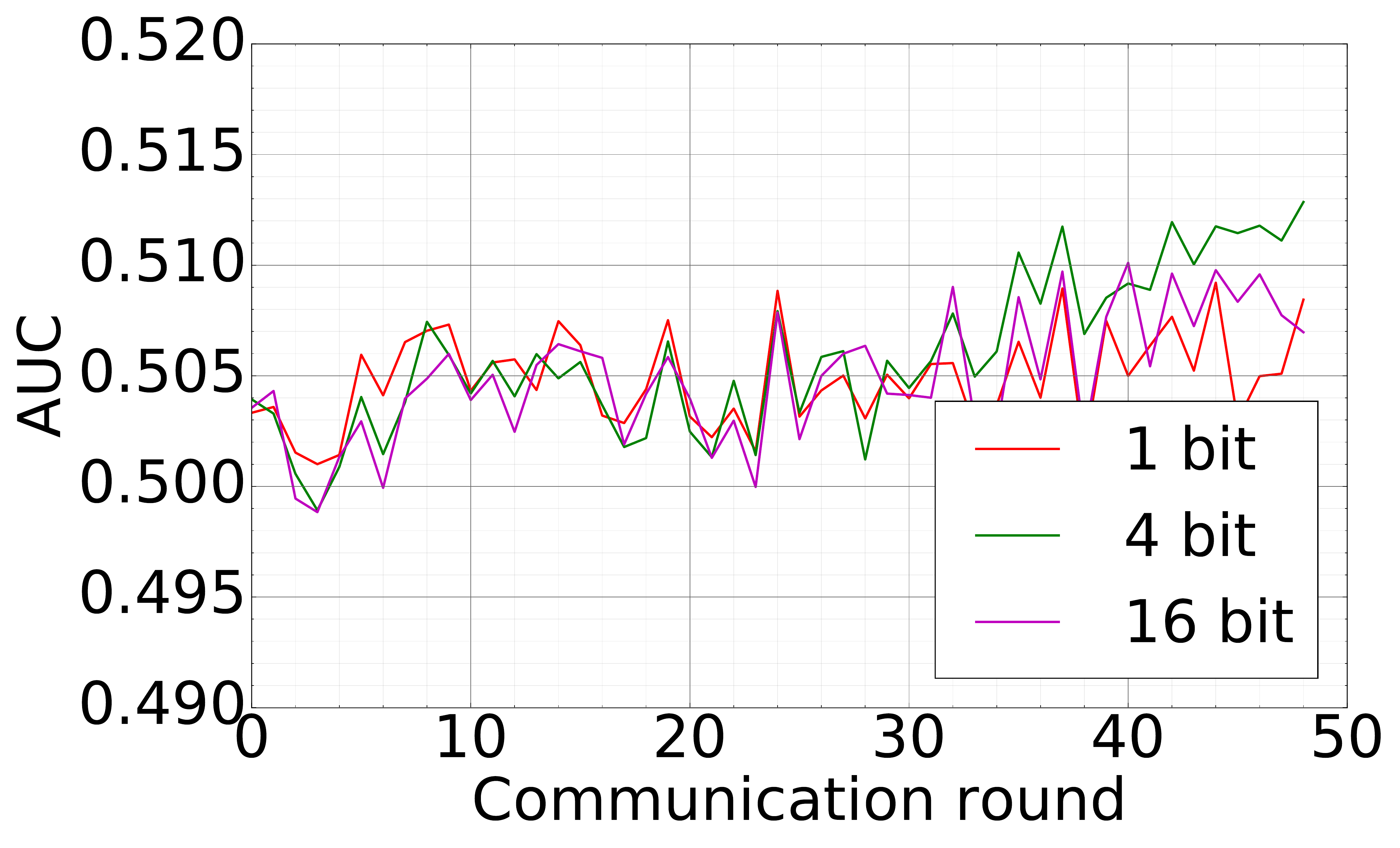}\\
{\footnotesize CR vs. Test accuracy}  & {\footnotesize CR vs. AUC} \\
\hskip -0.1cm\includegraphics[width=0.36\linewidth]{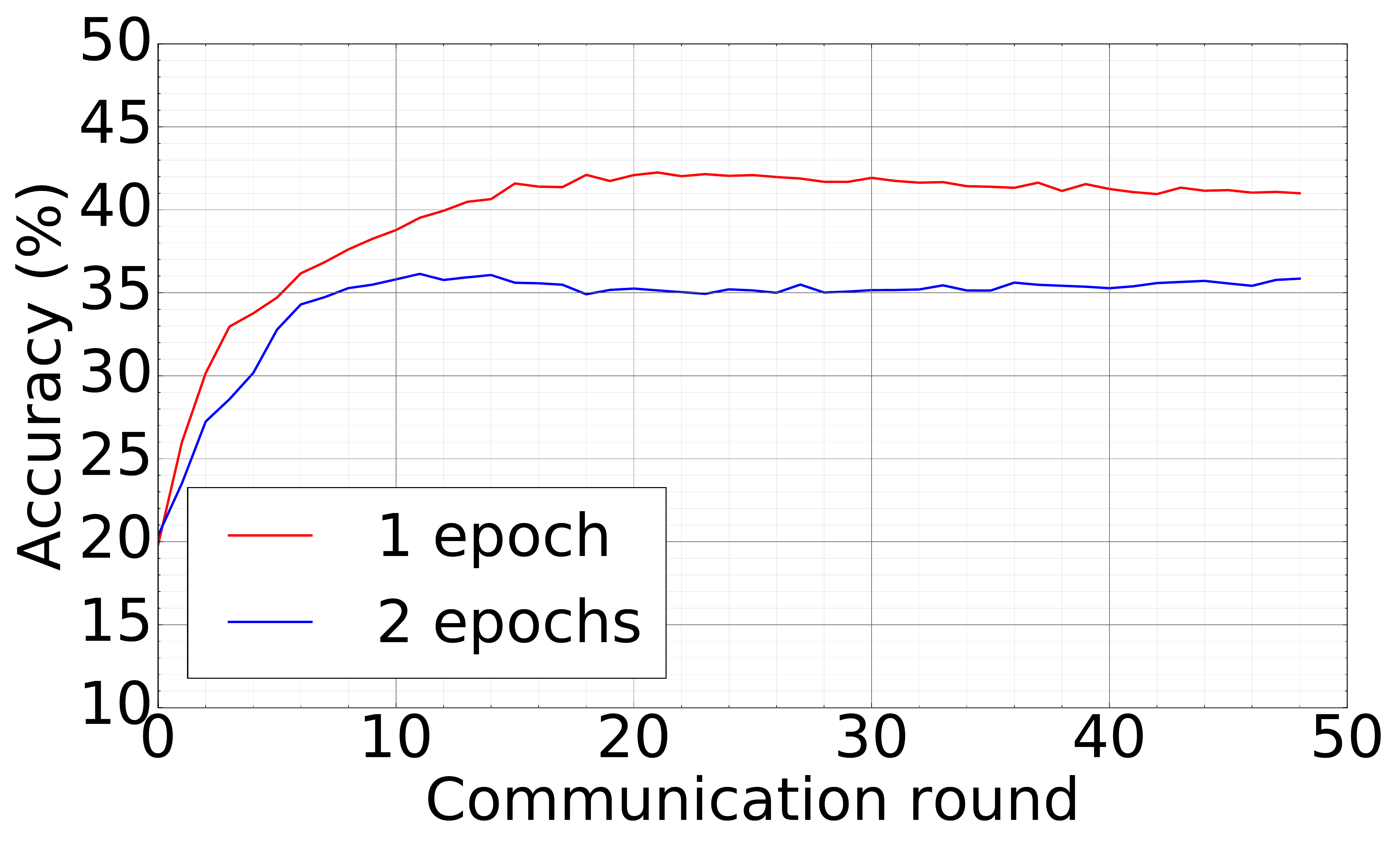}&
\hskip -0.4cm\includegraphics[width=0.36\linewidth]{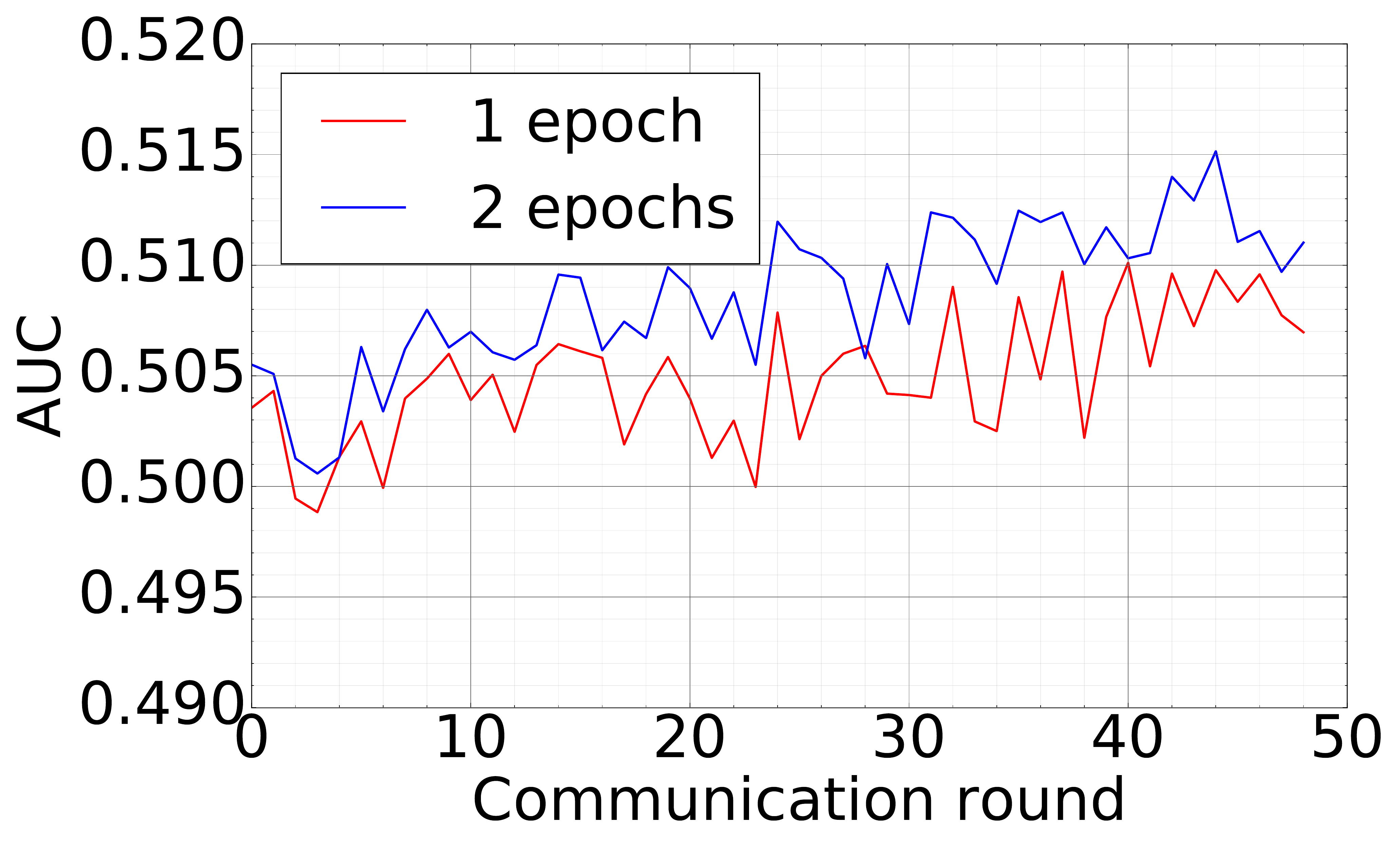}\\
{\footnotesize CR vs. Test accuracy} & {\footnotesize CR vs. AUC}\\
\end{tabular}
\vskip -0.3cm
\caption{
Training LSTM for SHAKESPEARE classification with DFedAvgM using: different communication bits but fix local epoch to one (first row) and different local epochs but fix the communication bits to 16 (second row). Using higher precision communication can slightly improve the performance.
More local epoch does not help in accelerating training or protect data privacy.
}
\label{fig:lstm-noniid}
\end{figure}

We consider the SHAKESPEARE dataset and we follow the processing as that used in \cite{mcmahan2017communication-efficient}, resulting in a dataset distributed over 1146 clients in the Non-IID fashion. On this data, we use DFedAvgM to train a stacked character-level LSTM language model, which predicts the next character after reading each character in a line. The model takes a series of characters as input and embeds each of these into a learned 8-dimensional space. The embedded characters are then processed through 2 LSTM layers, each with 256 nodes. Finally, the output of the second LSTM layer is sent to a softmax output layer with one node per character. The full model has 866,578 parameters, and we trained using an unroll length of 80 characters. We set the local batch size to 10, and we use a learning rate of 1.47, which is the same as \cite{mcmahan2017communication-efficient}. The momentum is selected to be 0.9. Figure~\ref{fig:lstm-noniid} plots the communication round vs. test accuracy and AUC under MIA for different quantization and different local epochs. These results show that 1) both the accuracy and MIA increase as training goes; 2) higher communication cost can lead to faster convergence; 3) increase local epochs deteriorate the performance of DFedAvgM.

\subsection{CIFAR10 Classification}
\begin{figure}[t!]
\centering
\begin{tabular}{ccc}
\hskip -0.3cm\includegraphics[width=0.32\linewidth]{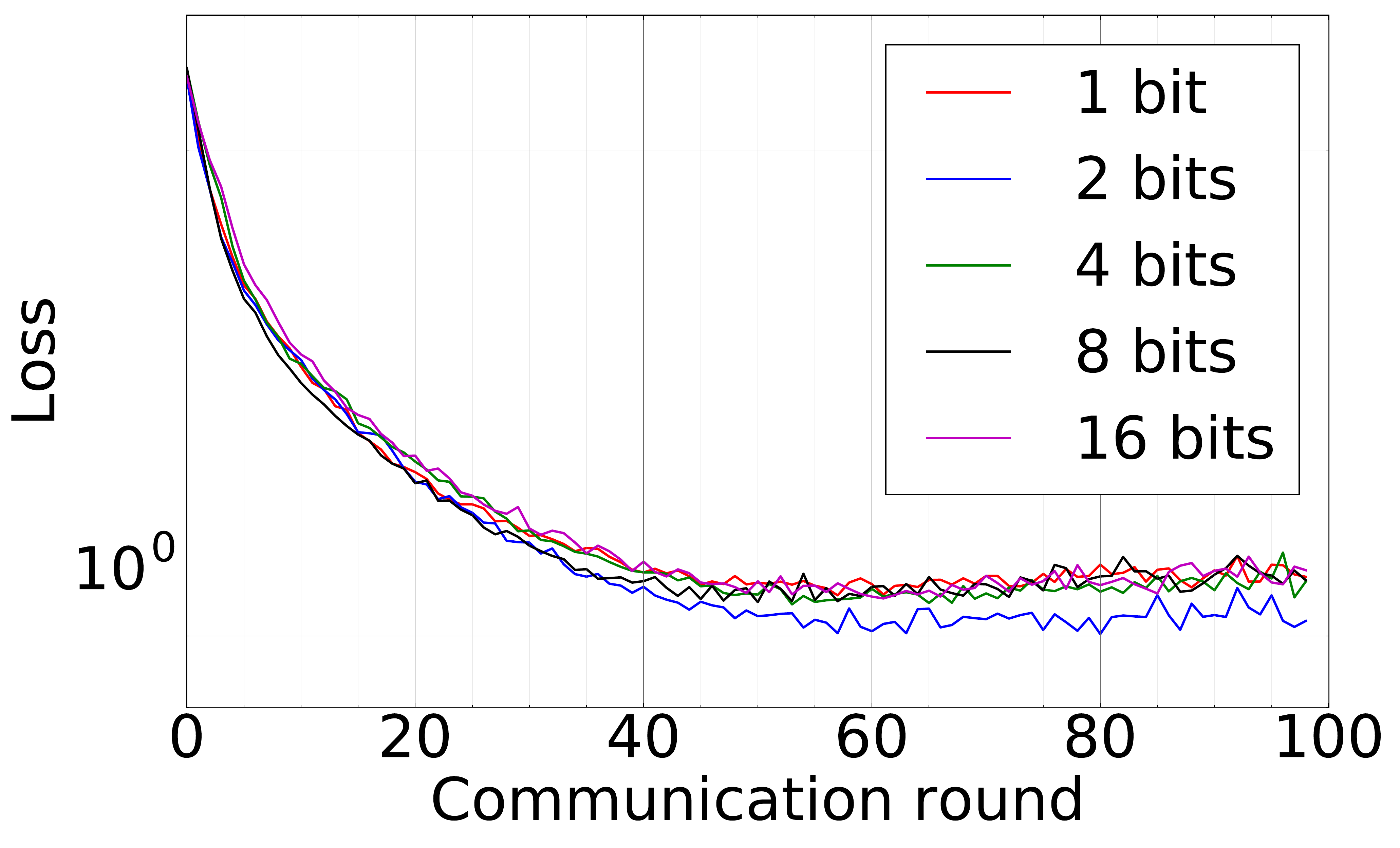}&
\hskip -0.45cm\includegraphics[width=0.32\linewidth]{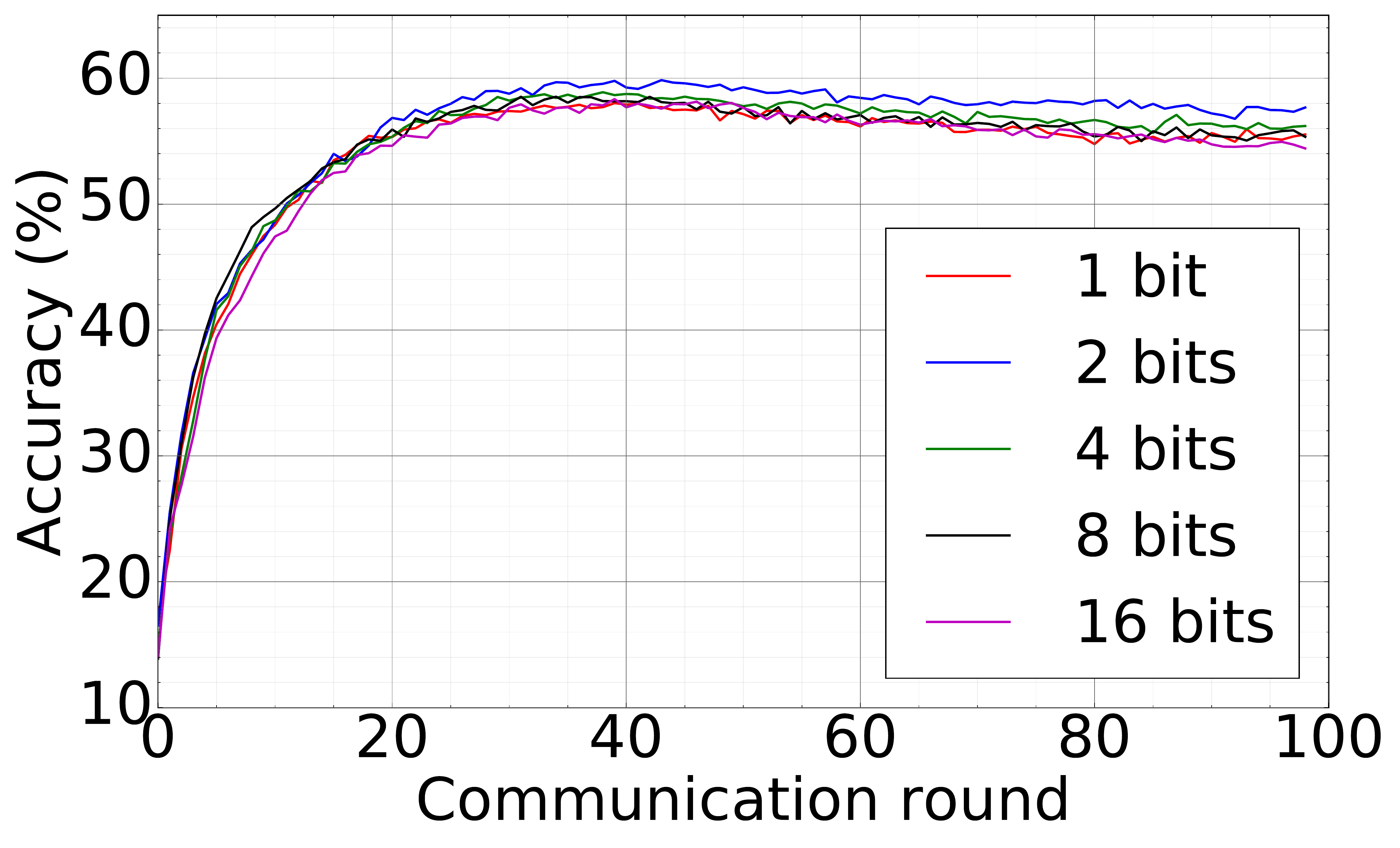}&
\hskip -0.45cm\includegraphics[width=0.32\linewidth]{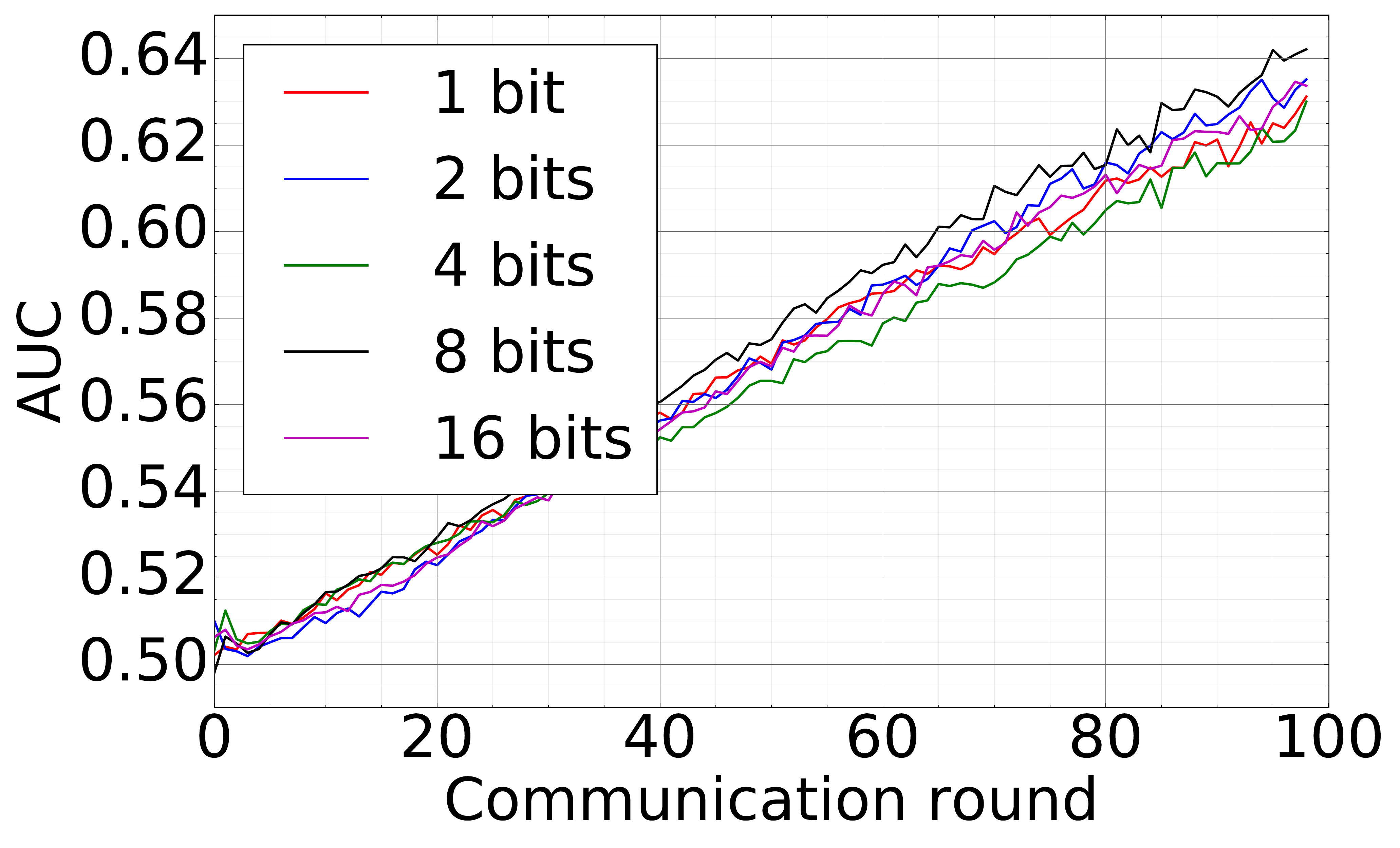}\\
{\footnotesize CR vs. Training loss} & {\footnotesize CR vs. Test acc} & {\footnotesize CR vs. AUC}\\
\hskip -0.3cm\includegraphics[width=0.32\linewidth]{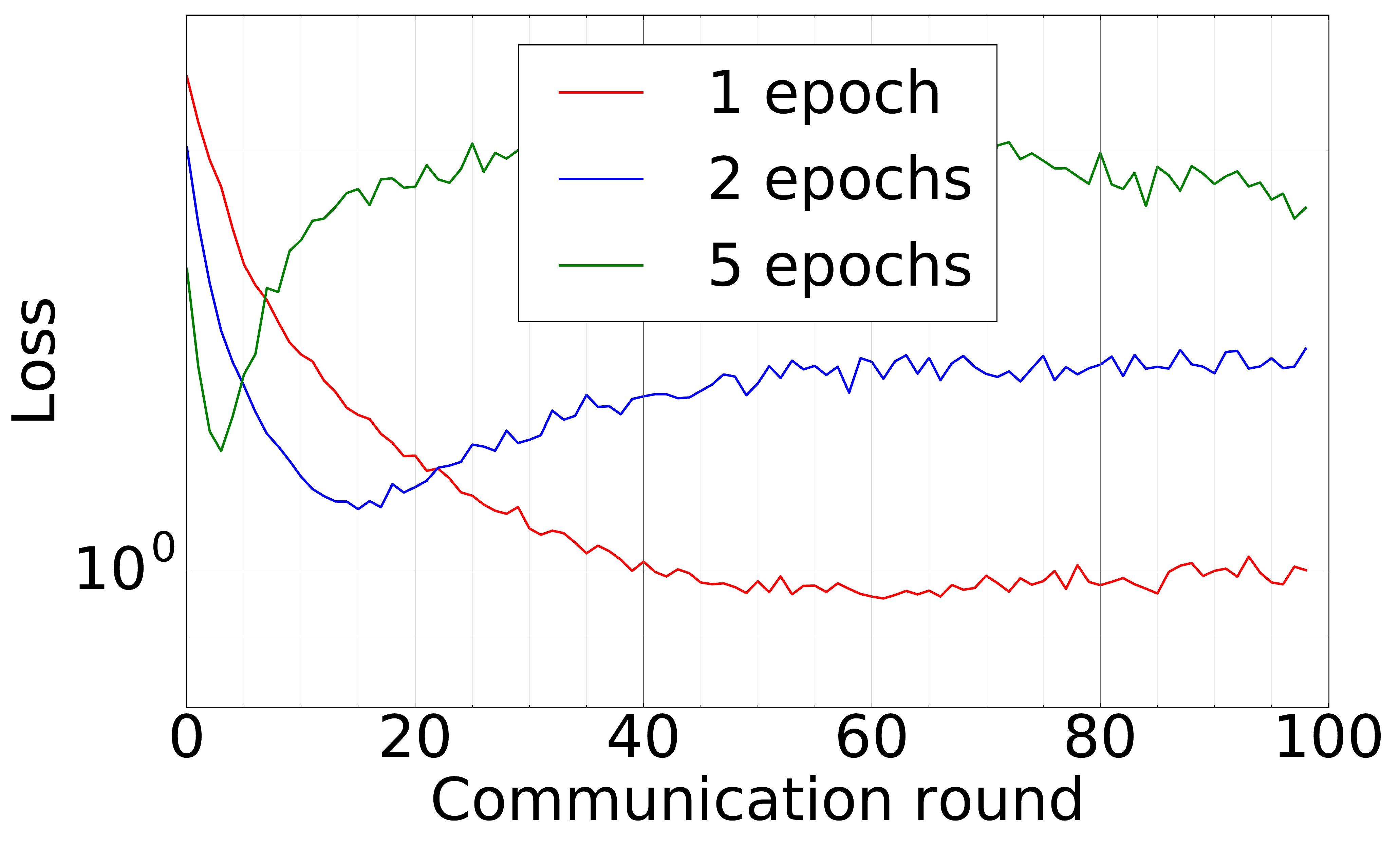}&
\hskip -0.45cm\includegraphics[width=0.32\linewidth]{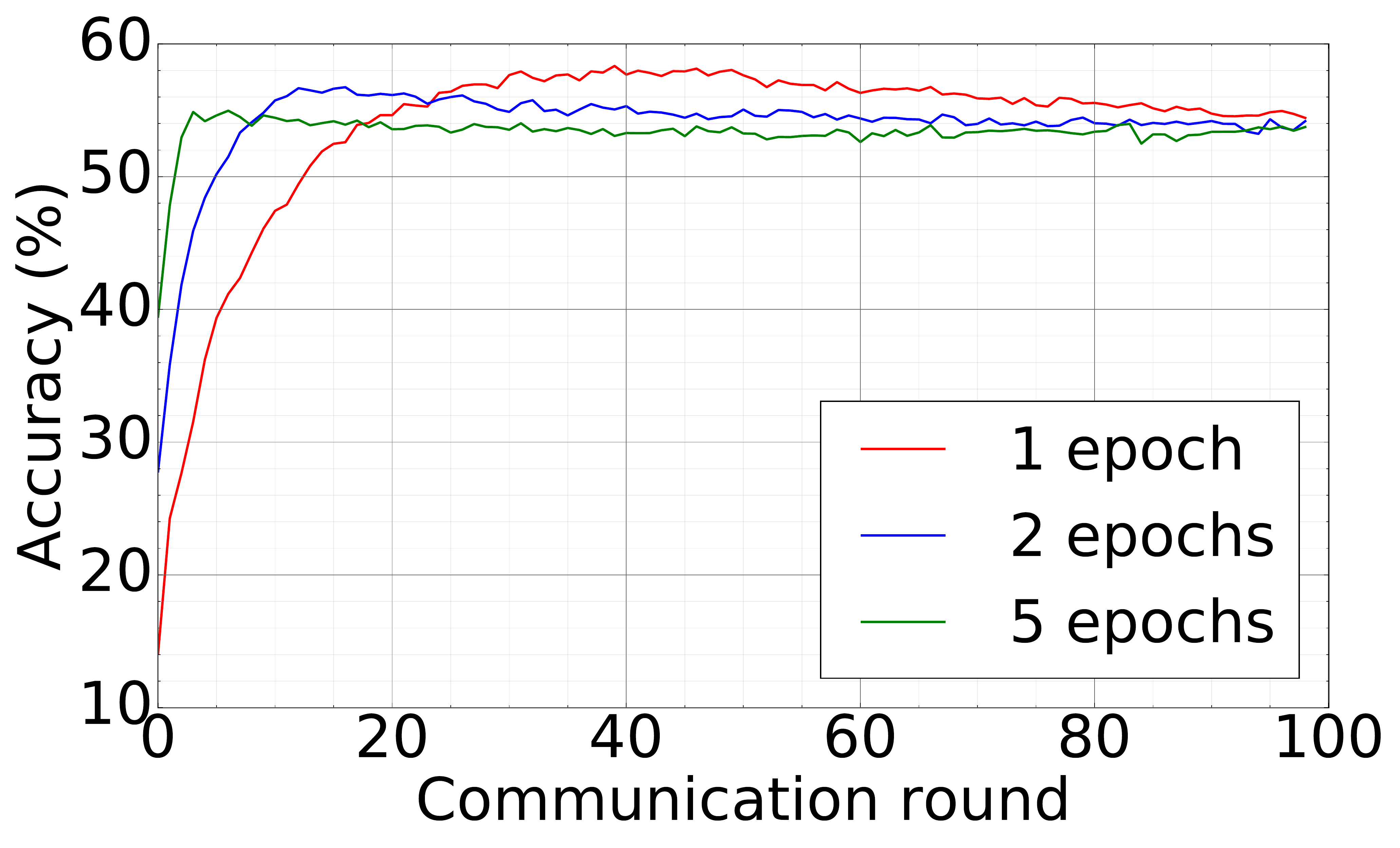}&
\hskip -0.45cm\includegraphics[width=0.32\linewidth]{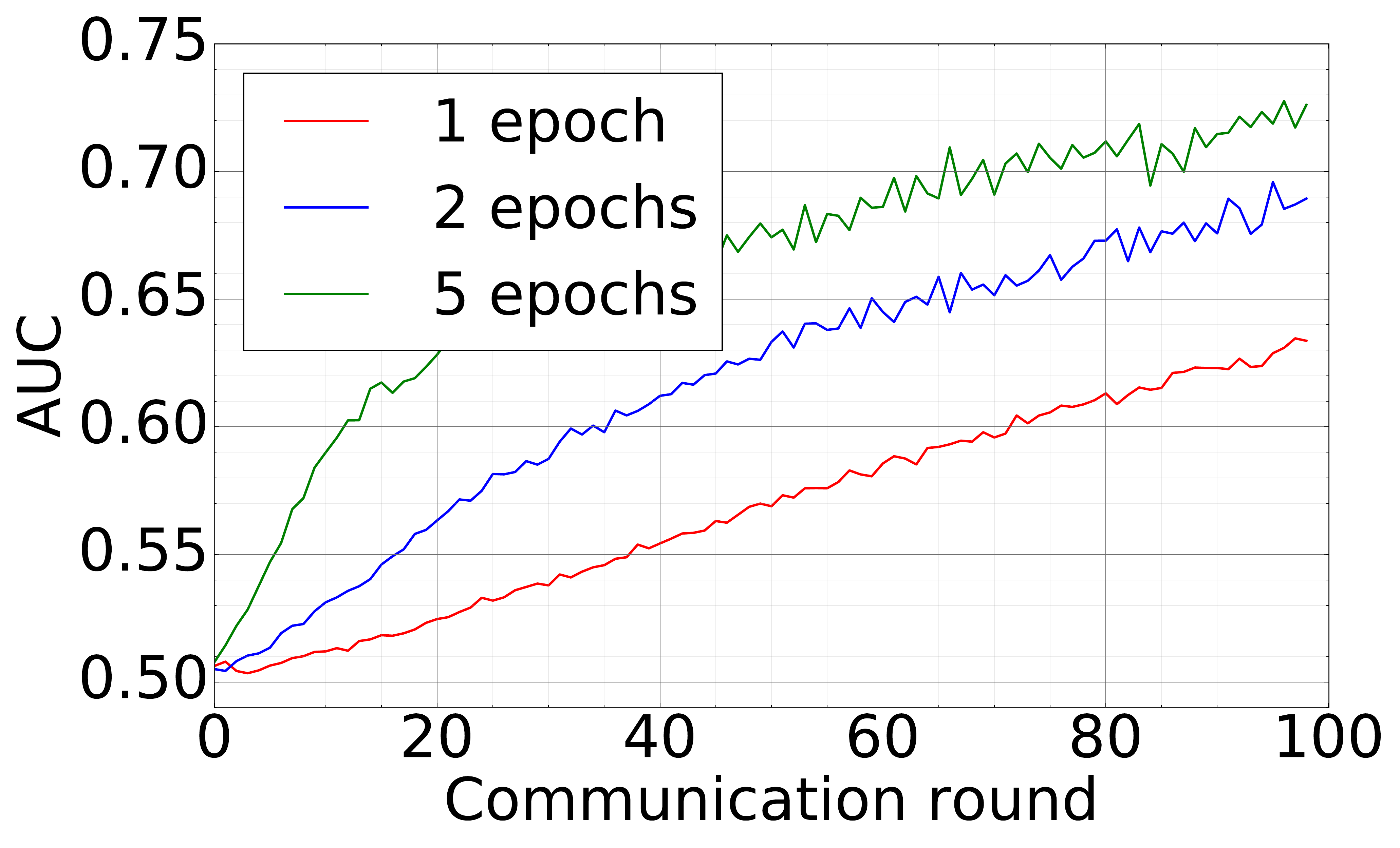}\\
{\footnotesize CR vs. Training loss} & {\footnotesize CR vs. Test acc} & {\footnotesize CR vs. AUC}\\
\end{tabular}
\vskip -0.3cm
\caption{Training CNN for IID CIFAR10 classification with DFedAvgM using: different communication bits but fix local epoch to one (first row) and different local epochs but fix the communication bits to 16 (second row). Different quantized DFedAvgM performs almost similar and more local epochs can accelerate training at the beginning but does not perform very well as training continues.
}
\label{fig:cifar-iid}
\end{figure}

Finally, we use DFedAvgM to train ResNet20 for CIFAR10 classification, which consists of 10 classes of $32\times 32$ images with three channels. There are 50,000 training and 10,000 testing examples, which we partitioned into 20 clients uniformly, and we only consider the IID setting following \cite{mcmahan2017communication-efficient}. We use the same data augmentation and DNN as that used in \cite{mcmahan2017communication-efficient}. The local batch size is set to 50, the learning rate is set to 0.01, and the momentum is set to 0.9. Figure~\ref{fig:cifar-iid} shows the communication round vs. test accuracy and AUC under MIA for different quantization and different local epochs. If the local epoch is set to 1, different communication bits does not lead to a significant difference in training loss, test accuracy, and AUC under MIA. However, for the fixed communication bits 16, increase the local epochs from 1 to 2 or to 5 will make training not even converge.



\vspace{-2mm}
\section{Concluding Remarks}\label{sec:conclusion}
In this paper, we proposed a DFedAvgM and its quantized version. There two major benefits of the DFedAvgM over the existing FedAvg: 1) In FedAvg, communication between the central parameter server and local clients is required in each communication round, and this communication will be very expensive as the number of clients is very large. On the contrary, in DFedAvgM communications are between clients which are significantly less than FedAvg. 2) In FedAvg, the central server collects the updated models from clients, and attack the central server can break the privacy of the whole system. In contrast, conceptually, it is harder to break the privacy in DFedAvgM than FedAvg. Furthermore, we established the theoretical convergence for DFedAvgM and its quantized version under general nonconvex assumptions, and we showed that the worst-case convergence rate of (quantized) DFedAvgM is the same as that of DSGD. In particular, we proved a sublinear convergence rate of (quantized) DFedAvgM when the objective functions satisfy the P{\L} condition. We perform extensive numerical experiments to verify the efficacy of DFedAvgM and its quantized version in training ML models and protect membership privacy.

%

%
%

\end{document}